\documentclass[aps,prm,amsmath,amssymb,reprint,superscriptaddress,]{revtex4-2}
\usepackage{charter,graphicx,verbatim,threeparttable,float,amssymb,gensymb }
\usepackage{upgreek}
\usepackage{booktabs}
\begin{document}

\preprint{APS/123-QED}

\title{
Evolution of highly anisotropic magnetism in the titanium-based kagome metals \textit{Ln}Ti$_3$Bi$_4$ (\textit{Ln}: La...Gd$^{3+}$, Eu$^{2+}$, Yb$^{2+}$)}

\author{Brenden R. Ortiz}
\email{ortizbr@ornl.gov}
\affiliation{Materials Science and Technology Division, Oak Ridge National Laboratory, Oak Ridge, TN 37831, USA}

\author{Hu Miao} 
\affiliation{Materials Science and Technology Division, Oak Ridge National Laboratory, Oak Ridge, TN 37831, USA}

\author{David S. Parker}
\affiliation{Materials Science and Technology Division, Oak Ridge National Laboratory, Oak Ridge, TN 37831, USA}

\author{Fazhi Yang} 
\affiliation{Materials Science and Technology Division, Oak Ridge National Laboratory, Oak Ridge, TN 37831, USA}

\author{German D. Samolyuk}
\affiliation{Materials Science and Technology Division, Oak Ridge National Laboratory, Oak Ridge, TN 37831, USA}

\author{Eleanor M. Clements} 
\affiliation{Materials Science and Technology Division, Oak Ridge National Laboratory, Oak Ridge, TN 37831, USA}

\author{Anil Rajapitamahuni} 
\affiliation{National Synchrotron Light Source II, Brookhaven National Laboratory, Upton NY 11973, USA}

\author{Turgut Yilmaz} 
\affiliation{National Synchrotron Light Source II, Brookhaven National Laboratory, Upton NY 11973, USA}

\author{Elio Vescovo} 
\affiliation{National Synchrotron Light Source II, Brookhaven National Laboratory, Upton NY 11973, USA}

\author{Jiaqiang Yan} 
\affiliation{Materials Science and Technology Division, Oak Ridge National Laboratory, Oak Ridge, TN 37831, USA}

\author{Andrew F. May} 
\affiliation{Materials Science and Technology Division, Oak Ridge National Laboratory, Oak Ridge, TN 37831, USA}

\author{Michael A. McGuire} 
\affiliation{Materials Science and Technology Division, Oak Ridge National Laboratory, Oak Ridge, TN 37831, USA}

\date{\today}

\begin{abstract}
Here we present the family of titanium-based kagome metals of the form \textit{Ln}Ti$_3$Bi$_4$ (\textit{Ln}: La...Gd$^{3+}$, Eu$^{2+}$, Yb$^{2+}$). Single crystal growth methods are presented alongside detailed magnetic and thermodynamic measurements. The orthorhombic (\textit{Fmmm}) \textit{Ln}Ti$_3$Bi$_4$ family of compounds exhibit slightly distorted titanium-based kagome nets interwoven with zig-zag lanthanide-based (\textit{Ln}) chains. Crystals are easily exfoliated parallel to the kagome sheets and angular resolved photoemission (ARPES) measurements highlight the intricacy of the electronic structure in these compounds, with Dirac points existing at the Fermi level. The magnetic properties and the associated anisotropy emerge from the quasi-1D zig-zag chains of \textit{Ln}, and impart a wide array of magnetic ground states ranging from anisotropic ferromagnetism to complex antiferromagnetism with a cascade of metamagnetic transitions. Kagome metals continue to provide a rich direction for the exploration of magnetic, topologic, and highly correlated behavior. Our work here introduces the \textit{Ln}Ti$_3$Bi$_4$ compounds to augment the continuously expanding suite of complex and interesting kagome materials.
\end{abstract}

\maketitle

\section{Introduction}

The kagome lattice has long been heralded as one of the prototypical frustrated lattices in condensed matter physics, and was historically valued for contributions in the search for insulating quantum spin liquids.\cite{balents2010spin,wulferding2010interplay,han2012fractionalized,fu2015evidence,freedman2010site} Fueled in part by the relatively recent discovery of the \textit{A}V$_3$Sb$_5$ kagome superconductors ~\cite{ortiz2019new,ortizCsV3Sb5,ortiz2020KV3Sb5,RbV3Sb5SC}, research into kagome metals has accelerated dramatically. An innate connection between the kagome motif and the electronic structure drives the manifestation of an electronic structure hosting Dirac points, flat bands, and Van Hove singularities.~\cite{park2021electronic,PhysRevB.87.115135,kiesel2013unconventional,meier2020flat}. Chemical tuning can then be used to tune the Fermi level, enabling a wide array of electronic instabilities ranging from bond density wave order ~\cite{PhysRevB.87.115135,PhysRevLett.97.147202}, charge fractionalization~\cite{PhysRevB.81.235115, PhysRevB.83.165118}, charge-density waves~\cite{PhysRevB.80.113102,yu2012chiral,kiesel2013unconventional}, and superconductivity~\cite{PhysRevB.87.115135,ko2009doped,kiesel2013unconventional}. 

The development of key material systems remains a persistent opportunity for solid state chemistry, and the discovery of new host systems can spur transformative paradigm shifts in the community. The nonmagnetic \textit{A}V$_3$Sb$_5$ (\textit{A}: K, Rb, Cs) materials are a good example, where the nonmagnetic kagome network of vanadium ions filled near the Van Hove points induces a unique intertwining of charge density wave (CDW) order and a superconducting ground state~\cite{ortizCsV3Sb5,ortiz2020KV3Sb5,RbV3Sb5SC,ortiz2021fermi,zhao2021cascade,hu2022coexistence,kang2022microscopic,jiang2021unconventional}. However, the introduction of magnetic degrees of freedom alongside the kagome network is a fertile area for exploration. While magnetic analogs of the \textit{A}V$_3$Sb$_5$ are still in development, the CoSn family of kagome compounds and it's derivatives (e.g. \textit{A}\textit{M}$_6$\textit{X}$_6$) have exemplified the diversity and complexity of mixing the magnetic sublattices with the kagome motif.\cite{PhysRevLett.127.266401,PhysRevB.103.014416,PhysRevB.104.235139,PhysRevMaterials104202,PhysRevB.103.014416,PhysRevB.106.115139,sciadv_abe2680,PhysRevLett.129.216402,Yin_2020,PhysRevMaterials.6.105001,PhysRevMaterials.6.083401,ZhangShao-ying_2001,PhysRevLett.126.246602}

We reported previously on a class of materials of the form \textit{A}\textit{M}$_3$\textit{X}$_4$ (\textit{A}: Lanthanide, Ca, \textit{M}: V, Ti, \textit{X}: Sb, Bi).\cite{ortiz2023ybv} These compounds exhibit slightly distorted \textit{M}-based kagome sublattices with zig-zag chains of \textit{A}-site ions. The potential for magnetism through choice of the \textit{A}-site provides a degree of chemical flexibility analogous to the \textit{A}\textit{M}$_6$\textit{X}$_6$ family. Reports of the phases are sporadic, with off-hand reports in exploratory chemistry papers\cite{ovchinnikov2018synthesis,ovchinnikov2019bismuth} mentioning the phases but spending little time systematically exploring the connection between chemistry and the impact on properties. Our prior discovery of the V-Sb based analogs YbV$_3$Sb$_4$ and EuV$_3$Sb$_4$\cite{ortiz2023ybv} represent some of the only explorations into the magnetic and transport properties of the wider family of compounds. Still, the \textit{A}\textit{M}$_3$\textit{X}$_4$ structures known to date are limited to (LaTi$_3$Bi$_4$, CeTi$_3$Bi$_4$, and SmTi$_3$Bi$_4$, CaV$_3$Sb$_4$, CaTi$_3$Bi$_4$, YbV$_3$Sb$_4$ and EuV$_3$Sb$_4$)~\cite{ovchinnikov2018synthesis,ovchinnikov2019bismuth,motoyama2018magnetic,ortiz2023ybv}. Considering the flexibility of the structure in adopting both V-Sb and Ti-Bi based frameworks, we aimed to provide a systematic survey of the \textit{Ln}-Ti-Bi based \textit{A}\textit{M}$_3$\textit{X}$_4$ compounds along with detailed structural and magnetic characterization.

In this work, we present the single crystal growth and characterization of the titanium-based kagome metals of the form \textit{Ln}Ti$_3$Bi$_4$ (\textit{Ln}: La...Gd$^{3+}$, Eu$^{2+}$, Yb$^{2+}$). Our study surveyed the formation of the phase along the entire lanthanide row, finding that the phase is preferentially stabilized by large lanthanide cations. The \textit{Ln}Ti$_3$Bi$_4$ provide an interesting opportunity by interweaving two interesting structural motifs: 1) the titanium-based kagome nets, and 2) the zig-zag chains of \textit{Ln}. We adopt a structure-forward approach to the characterization in this manuscript, first examining the influence of the kagome network on the electronic structure through DFT and ARPES studies, finding a rich electronic landscape with Dirac points at the Fermi level and Van Hove singularities nearby. Afterwards, we perform an in-depth suite of magnetic measurements with specific care to the underlying crystal symmetry of the lattice -- tracing the evolution of the magnetic anisotropy with detailed orientation-dependent measurements. As expected, the quasi-1D nature of the chains naturally imparts highly complex magnetism throughout the series, ranging from anisotropic ferromagnetism, potential helical phases, and complex antiferromagnetism with staged metamagnetic transitions. These observations demonstrate how the underlying crystal structure synergizes with the inherent anisotropy of the rare-earth elements to create a host of interesting ground states. Our results augment the growing suite of kagome metals by weaving together the intrinsic complexity of the kagome electronic structure with the chemical diversity offered by a magnetic sublattice on an exfoliatable single crystal platform.

\section{Experimental Methods}

\subsection{Single Crystal Synthesis}
\textit{Ln}Ti$_3$Bi$_4$ single crystals are grown through a bismuth self-flux. Elemental reagents of La (AMES), Ce (AMES), Pr (AMES), Nd (Alfa 99.8\%), Sm (AMES), Eu (AMES), Gd (AMES), Yb (Alfa 99.9\%), Ti (Alfa 99.9\% powder), and Bi (Alfa 99.999\% low-oxide shot) were combined at a 2:3:12 ratio into 2~mL Canfield crucibles fitted with a catch crucible and a porous frit.\cite{canfield2016use} The crucibles were sealed under approximately 0.7~atm of argon gas in fused silica ampoules. Each composition was heated to 1050\degree C at a rate of 200\degree C/hr. Samples were allowed to thermalize and homogenize at 1050\degree C for 12-18~h before cooling to 500\degree C at a rate of 2\degree C/hr. Excess bismuth was removed through centrifugation at 500\degree C. 

Crystals are a lustrous silver with hexagonal habit. The samples are mechanically soft and are easily scratched with a knife or wooden splint. They are layered in nature and readily exfoliate using adhesive tape. For all members of the family except EuTi$_3$Bi$_4$, the crystal size is limited by the volume of the growth vessel, and samples with side lengths up to 1~cm are common. Samples of EuTi$_3$Bi$_4$ are substantially smaller and rarely exceed 1~mm side lengths. We note that samples are moderately stable in air and tolerate common solvents and adhesives (e.g. GE Varnish, isopropyl alcohol, toluene) well. However, the samples are not indefinitely stable and will degrade, tarnish, and spall if left in humid air for several days.

\subsection{Bulk Characterization}
Single crystals of \textit{Ln}Ti$_3$Bi$_4$ were mounted on kapton loops with Paratone oil for single crystal x-ray diffraction (SCXRD). Diffraction data were collected at 100~K on a Bruker D8 Advance Quest diffractometer with a graphite monochromator using Mo K$\alpha$ radiation ($\lambda$ = 0.71073~\AA). Data integration, reduction, and structure solution was performed using the Bruker APEX3 software package. A numerical absorption correction was performed using a face-indexing algorithm. For large crystals, an additional spherical absorption correction was occasionally used as well. All atoms were refined with anisotropic thermal parameters. CIF files for all structures are included in the supplementary information\cite{ESI}. To orient and analyze the facets of the as-grown single crystals, Laue diffraction was performed on a Multiwire Back-Reflection Laue Detector. As a consistency check, facet scans were also performed using an a PANalytical X'Pert Pro MPD diffractometer (monochromated Cu K$_{\alpha 1}$ radiation) in standard Bragg-Brentano ($\theta$-2$\theta$) geometry with crystals mounted with the easy-axis perpendicular to the diffraction plane.

Magnetization measurements of \textit{Ln}Ti$_3$Bi$_4$ single crystals were performed in a 7~T Quantum Design Magnetic Property Measurement System (MPMS3) SQUID magnetometer in vibrating-sample magnetometry (VSM) mode. Samples were mounted to quartz paddles using a small quantity of GE varnish or n-grease. Angle-resolved magnetization measurements were performed on a 7~T Quantum Design Magnetic Property Measurement System (MPMSXL) equipped with a rotator stage. Supplementary high field measurements to 12~T were performed as needed in a 14~T Quantum Design Physical Property Measurement System (PPMS) equipped with the VSM option. All magnetization measurements were performed under field-cooled conditions unless specified.

Heat capacity measurements on \textit{Ln}Ti$_3$Bi$_4$ single crystals between 300~K and 1.8~K were performed in a Quantum Design 9~T Dynacool Physical Property Measurement System (PPMS) equipped with the heat capacity option. Both LaTi$_3$Bi$_4$ and YbTi$_3$Bi$_4$ were measured as nonmagnetic reference samples, with LaTi$_3$Bi$_4$ used as the reference for the trivalent rare-earth \textit{Ln}Ti$_3$Bi$_4$ compounds and YbTi$_3$Bi$_4$ as the reference for divalent EuTi$_3$Bi$_4$. 

Electronic resistivity measurements on YbTi$_3$Bi$_4$ and LaTi$_3$Bi$_4$ were performed in a Quantum Design 9~T Dynacool Physical Property Measurement System (PPMS). Single crystals were mounted to a sheet of sapphire and then subsequently the sapphire plate was then adhered to the sample puck stage.  GE varnish was used to ensure electrical isolation and thermal contact. Samples were then exfoliated and contacts established using silver paint (DuPont cp4929N-100) and platinum wire (Alfa, 0.05~mm Premion 99.995\%). We used a dc current of 1~mA to measure the resistivity under zero-field conditions.

ARPES experiments are performed on single crystals SmTi$_3$Bi$_4$. While the effect of the magnetic order is not the focus of this manuscript, samples of SmTi$_3$Bi$_4$ has the highest magnetic transition temperature, which enables future comparisons above and below the onset of ferromagnetic order. The samples are cleaved in-situ in a vacuum with pressures $<3\times10^{-11}$~torr. The experiment is performed at beam line 21-ID-1 at the NSLS-II. The measurements are taken with synchrotron light source and a Scienta-Omicron DA30 electron analyzer. The total energy resolution of the ARPES measurement is approximately 12~meV. The sample stage is maintained at 30~K (above the ferromagnetic transition in SmTi$_3$Bi$_4$) throughout the experiment. 

\subsection{Electronic Structure Calculations}
In order to understand the electronic structure of these compounds, We have conducted first principles calculations of LaTi$_3$Bi$_4$, using the linearized augmented Plane-wave density functional theory code WIEN2K \cite{blaha2001}, within the Generalized gradient approximation\cite{perdew1996}. The experimentally derived structure was used as the basis for calculations. Sphere radii of 2.48, 2.50, and 2.50 Bohr were employed, for Ti, La and Bi, respectively. An RK$_{max}$ of 9.0 was employed; being the product of the smallest sphere radius (Ti), and the largest plane-wave vector. Spin-orbit coupling was not included, and no internal coordinate relaxation was conducted. An 8$\times$8$\times$8 k-mesh, comprising 95 points in the irreducible Brillouin zone, was employed. We present here non-spin-polarized calculations.

Note that we have also conducted calculations of the corresponding Cerium and Neodymium compounds,
With the main differences being the presence and location, in these latter compounds,
of the 4$f$-electron derived bands.

\section{Results \& Discussion}

\subsection{Crystal Structure and Phase Stability}

\begin{figure}
\includegraphics[width=\linewidth]{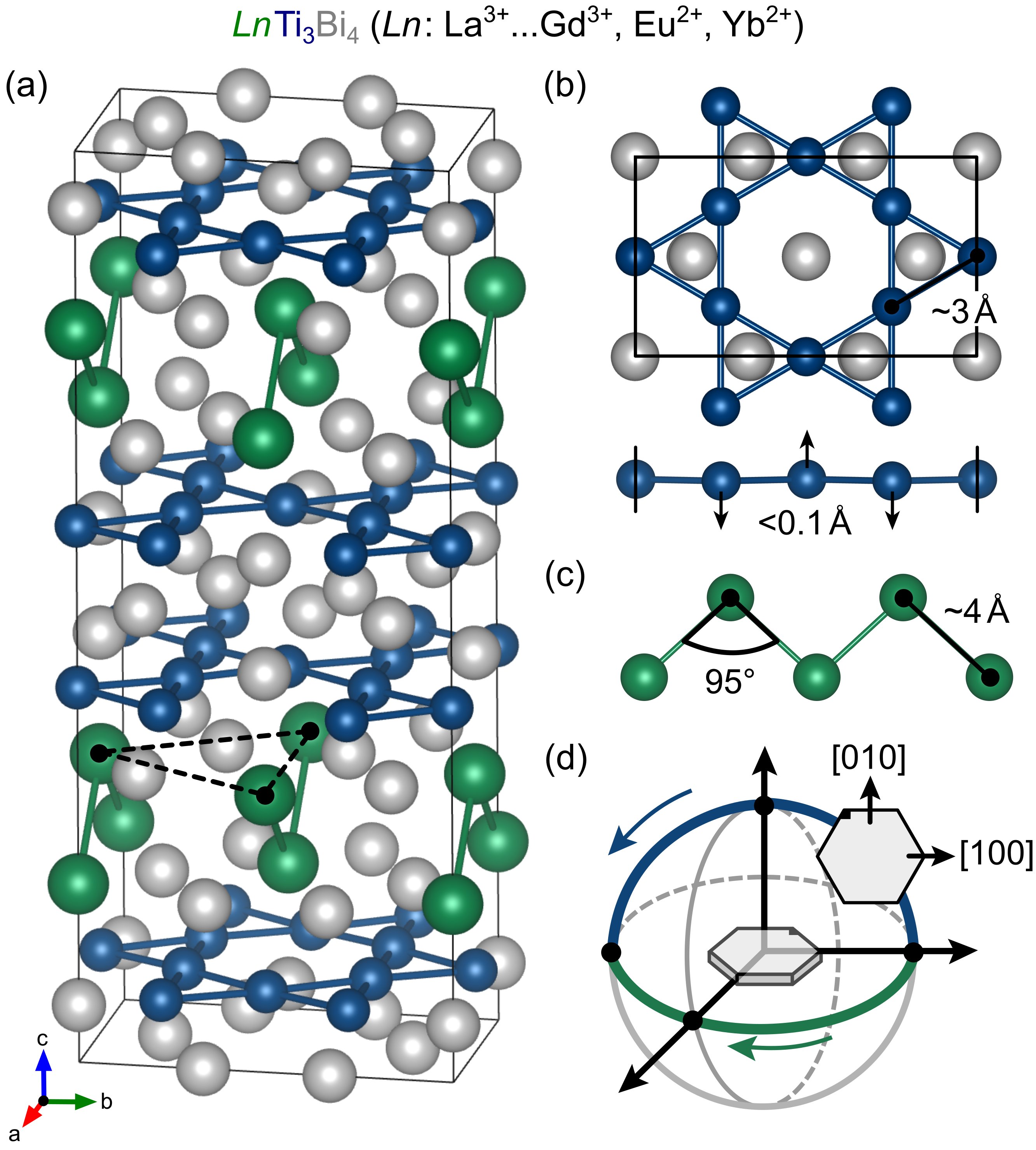}
\caption{(a) Unit cell of the \textit{Ln}Ti$_3$Bi$_4$ structure, with Ti--Ti and \textit{Ln}--\textit{Ln} bonds drawn to highlight the kagome and zig-zag chains. (b) The kagome sublattice is very slightly distorted due to the reduced symmetry (\textit{Fmmm}) of the unit cell. (c) The \textit{Ln} sublattice is best visualized as zig-zag chains parallel to the \textit{a}-axis. with an intra-chain and inter-chain (black dashed) \textit{Ln}--\textit{Ln} distance of 4~\AA~ and 6~\AA, respectively. (d) To reflect the two-fold symmetry of the unit cell, samples are oriented along the easy-axis and then rotated through a fixed magnetic field in both the out-of-plane (blue) and in-plane (green) directions.}
\label{fig:1}
\end{figure}

Unlike the small, generally ``simpler'' unit cells of the AV$_3$Sb$_5$, CoSn, and AM$_6$X$_6$ kagome prototypes, the \textit{Ln}\textit{M}$_3$\textit{X}$_4$ compounds are substantially more complex. Figure \ref{fig:1}(a) illustrates the overall crystal structure with Ti--Ti and \textit{Ln}-\textit{Ln} bonds drawn to highlight the two atomic sublattices of note: 1) the Ti-based kagome, and the 2) \textit{Ln}-based zig-zag chains. The overall symmetry of the unit cell is \textit{Fmmm}, necessitated by the quasi-1D (two-fold) nature of the \textit{Ln-Ln} chains. Concurrently, the kagome lattice is slightly distorted.  Figure \ref{fig:1}(b) highlights one of the kagome layers and the slight ($<$0.1~\AA) out-of-plane buckling. We note that the \textit{A}\textit{M}$_3$\textit{X}$_4$ actually contains elements from the CoSn and AM$_6$X$_6$ kagome prototypes. If we consider stacking along the \textit{c}-axis, the \textit{A}\textit{M}$_3$\textit{X}$_4$ structure consists of $X_4$--$M_3$--$AX_2$--[$AX_2$--$M_3$--$X_4$--$M_3$--$AX_2$]--$AX_2$--$AX_2$--$M_3$--$X_4$ layers. The bracketed segment of the stacking represents the same motif as the HfFe$_6$Ge$_6$ prototype structure. There are two sets of paired [$AX_2$--$M_3$--$X_4$--$M_3$--$AX_2$] kagome layers per unit cell, and they are offset from one another, yielding the larger \textit{c}-axis.

Figure \ref{fig:1}(c) highlights the \textit{Ln}-based zig-zag chains that run parallel to the \textit{a}-axis. The nearest \textit{Ln}-\textit{Ln} interaction in the zig-zag chain (intra-chain) is approximately 4~\AA. Treated as a single object, the chains are relatively well isolated, with the associated planes separated by approximately 5~\AA. However, the nearest inter-chain \textit{Ln}-\textit{Ln} distance is even larger (approximately 6~\AA) as each adjacent chain is inverted. One could alternatively picture the rare-earth sublattices as two stacks of offset triangular lattices, however this depiction is misleading. The ``triangular'' lattice interactions are redundant with the inter-chain \textit{Ln}-\textit{Ln} interaction (dashed lines in Figure \ref{fig:1}(a)) and disregard the much closer nearest-neighbor inter-chain interactions. Viewing the structure as stacked triangular lattices also disguises the reduced symmetry, which has a dramatic effect on the magnetic properties.

Due to the reduced symmetry of the unit cell, and the quasi-1D nature of the \textit{Ln} zig-zag chains, care must be taken to fully capture the magnetic properties of \textit{Ln}Ti$_3$Bi$_4$ single crystals. In our previous report on EuV$_3$Sb$_4$, we highlighted the difference between the out-of-plane (\textit{H}$\parallel$\textit{c}) and in-plane (\textit{H}$\perp$\textit{c}) results. Due to the spin-only nature of Eu$^{2+}$ and the exceedingly small crystal size of EuV$_3$Sb$_4$, this approximation was considered sufficient for an initial study. However, in this work we have endeavored to provide a more comprehensive mapping of the magnetic properties of the \textit{Ln}Ti$_3$Bi$_4$ family, particularly with regards to the underlying 2-fold symmetry imparted by the \textit{Fmmm} space group.

Almost all crystals show substantial magnetic anisotropy between the in-plane and out-of-plane directions. For all results except for Gd and Eu, the easy-axis direction corresponds to the [010] direction. Gd possesses two directions of interest (the [100] and [001]), and Eu is the only compound to show an out-of-plane [001] easy-axis. It is essential to note that the pseudo-hexagonal crystal habit in the \textit{Ln}Ti$_3$Bi$_4$ family disguises the reduced symmetry of the unit cell. The (010) plane in crystals of \textit{Ln}Ti$_3$Bi$_4$ always corresponds to two of the natural hexagonal facets (the (010) and the 180-degree equivalent facet), and (100) always corresponds to two of the orthogonal hexagonal corners. \textit{Important note}: Resist the temptation to treat the cell as pseudo-hexagonal, as adjacent hexagonal facets are \textit{not} equivalent. Two of the key directions have been marked in Figure \ref{fig:1}(d) for reference. 

Figure \ref{fig:1}(d) illustrates a simple diagnostic strategy for collecting magnetization data on single crystals of \textit{Ln}Ti$_3$Bi$_4$ compounds. The crystal is first oriented with the magnetic field parallel to the in-plane easy-axis. For compounds with an out-of-plane easy-axis (e.g. (001) in EuTi$_3$Bi$_4$), or for those with multiple directions of interest (e.g. (100) and (001) in GdTi$_3$Bi$_4$) whichever in-plane direction exhibits the largest magnetization is chosen. Magnetization results are then collected by rotating the crystal within a fixed magnetic field. Three scans are performed: 1) first a 180\degree rotation in-plane (Figure \ref{fig:1}(d) green trace) surveys the in-plane magnetization, 2) the crystal is then returned to the easy-axis starting position and rotated out-of-plane through 180\degree (Figure \ref{fig:1}(d) blue trace), 3) Isolated orientations of interest are selected and scanned in more detail, if necessary.

The difficulty of orienting each crystal is abated by the naturally large size and obvious faceting in single crystals of \textit{Ln}Ti$_3$Bi$_4$ grown from a bismuth self-flux (see Methods section). All crystals exhibit a bright silver luster, are mechanically quite soft, and are easily exfoliated. The divalent EuTi$_3$Bi$_4$ and YbTi$_3$Bi$_4$ are notably much softer than the rest of the series. The samples appear to tolerate air, water, common adhesives, and solvents (e.g. GE varnish, ethanol, toluene). However, samples are not indefinitely stable in air, and exfoliated surfaces will tarnish in the course of a day if left exposed to humid air. To this end, we note an unusual property of the Ti$_3$Bi$_4$-based \textit{Ln}\textit{M}$_3$\textit{X}$_4$, where crystals exposed to air for an extended time appear to swell along the \textit{c}-axis and eventually spall, layer by layer. As such, efforts were made to minimize exposure to air, water, and solvents throughout the course of our measurements. Out of precaution, if samples needed to be exposed for extended periods of time, the crystals were exfoliated and then coated in a thin layer of n-grease to serve as a passivating layer.

\begin{figure}
\includegraphics[width=\linewidth]{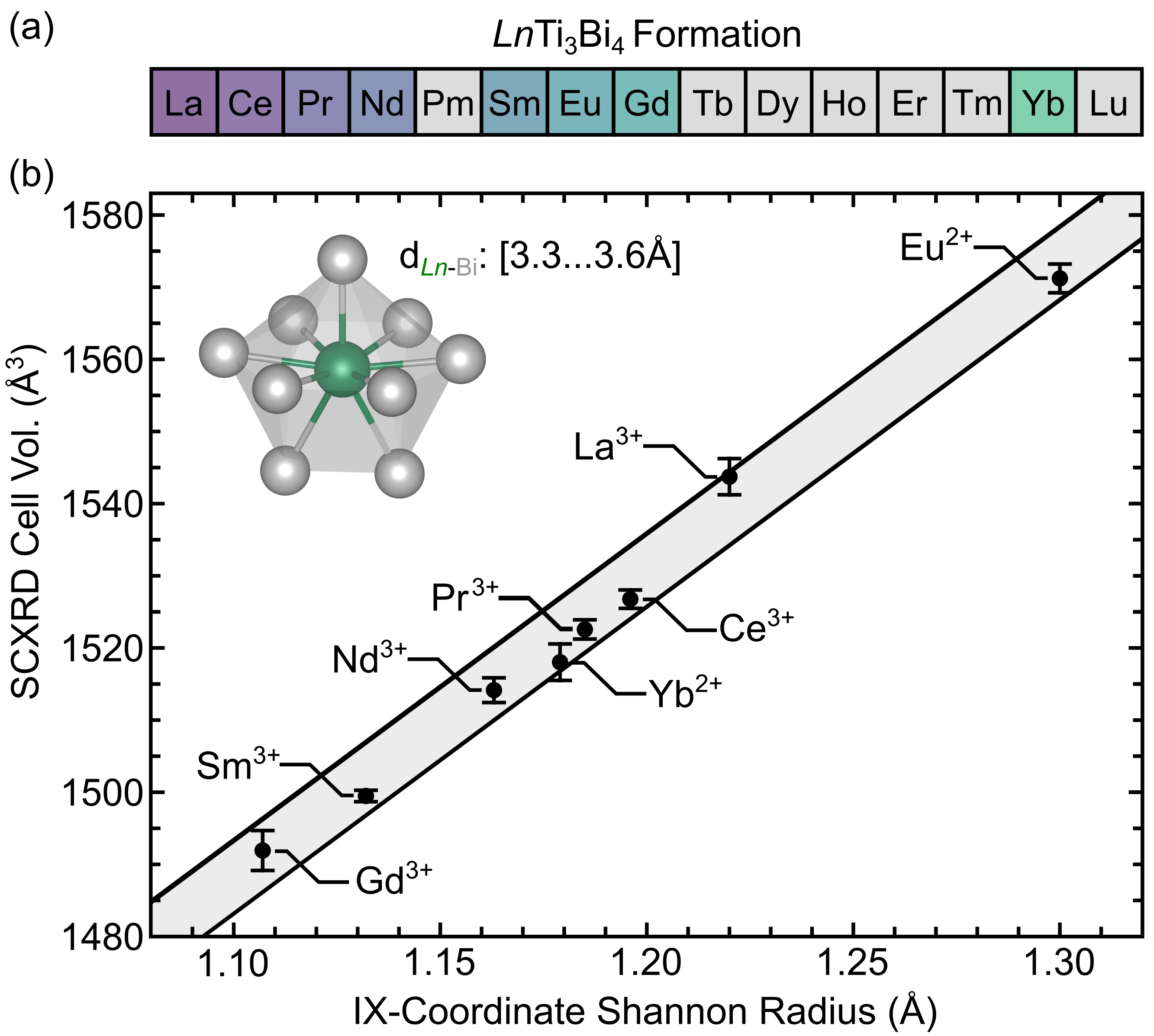}
\caption{(a) Simple schematic showing the phase stability of the \textit{Ln}Ti$_3$Bi$_4$ family across the lanthanide row. Recall that the notable outlier (Yb) is divalent and possesses an ionic radius closer to that of Pr$^{3+}$. (b) Presuming a small spread of nearest-neighbor \textit{Ln}--Bi distances, the coordination of \textit{Ln} is approximately 9-coordinate. The unit cell volume from single crystal x-ray diffraction correlates linearly with the 9-coordinate Shannon ionic radius. The gray shading is a guide to the eye based on the linear regression of the cell volume.}
\label{fig:2}
\end{figure}

\begin{figure*}[t]
\includegraphics[width=1\textwidth]{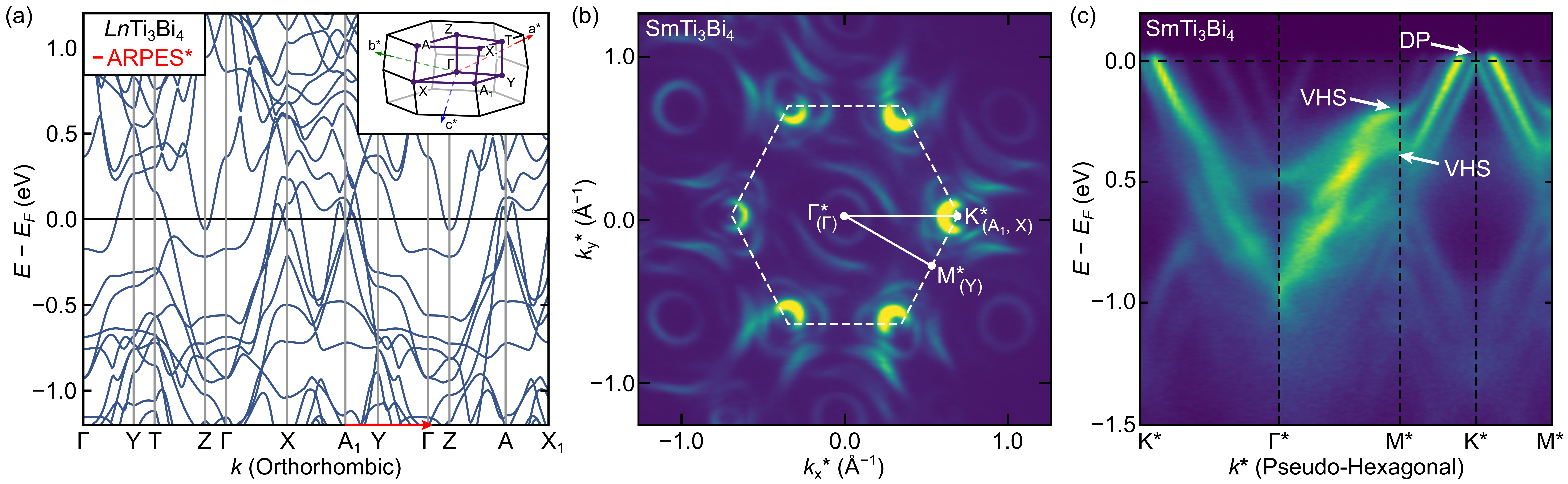}
\caption{(a) Representative electronic band diagram of the \textit{Ln}Ti$_3$Bi$_4$ family. The \textit{Ln} atoms do not contribute substantially to the electronic density of states near the Fermi level. To first order, we expect that the electronic structure will not change (b) Fermi surface mapping of SmTi$_3$Bi$_4$ measured by linear horizontal photon polarization. The surface high-symmetry points, $K^*$ and $M^*$, are defined in the pseudo-hexagonal symmetry (asterisks). Approximately equivalent points in the orthorhombic zone are annotated. (c) Linecuts through the ARPES data shows the characteristic electronic structure of the titanium-based kagome sublattice, including van Hove singularities (VHS) and Dirac points (DP). Due to the slight distortion, the $K^*\rightarrow M^*\rightarrow \Gamma^*$ is approximately equivalent to $A_1\rightarrow Y^*\rightarrow \Gamma^*$ (highlighted in red on the DFT derived band diagram.}
\label{fig:arpes}
\end{figure*}

The mechanical properties of the \textit{Ln}Ti$_3$Bi$_4$ compounds make single crystal diffraction (SCXRD) a challenging endeavor. Due to the high absorption and plate-like geometry, absorption corrections are absolutely essential. Large sample sizes are prohibitive, further exacerbated by the soft, layered nature of the samples. Attempts to cut samples often destroy the crystal quality. Such issues were noted previously \cite{ovchinnikov2019bismuth} and impeded the initial structural identification of the \textit{Ln}\textit{M}$_3$\textit{X}$_4$ compounds. Careful selection of small, well-faceted crystals from fast growths (to limit crystal coarsening), coupled with face-indexing absorption corrections allowed us to solve the entire \textit{Ln}Ti$_3$Bi$_4$ series through single crystal diffraction. The resulting CIF files are included in the supplementary information\cite{ESI}. 

Figure \ref{fig:2}(a) provides a summary of the chemical and structural data within the \textit{Ln}Ti$_3$Bi$_4$ family. The periodic table highlights the stability of the \textit{Ln}Ti$_3$Bi$_4$ phase relative to other adjacent phases. Common secondary phases observed in the flux growths are the binary bismide \textit{Ln}Bi$_2$ and the ternary compounds \textit{Ln}$_3$TiBi$_5$. Larger lanthanides appear to stabilize the structure. Recall that Yb adopts the Yb$^{2+}$ state in YbTi$_3$Bi$_4$, and thus possesses an ionic radii similar to Pr$^{3+}$. As a simple exercise, we assume that the \textit{Ln} atoms are well described by an ionic model in \textit{Ln}Ti$_3$Bi$_4$. Considering a range of nearest neighbor \textit{Ln}-Bi bonds (approximately 3.3\AA~ to 3.6\AA), the coordination environment of \textit{Ln} is approximately 9-coordinate. Figure \ref{fig:2}(b) shows the unit cell volume from SCXRD plotted against the 9-coordinate Shannon ionic radius. Within the error of SCXRD, the compounds obey a roughly linear relationship between the Shannon radius and the unit cell volume, as expected. As a conceptually pleasing aside, the CaTi$_3$Bi$_4$ is the only non-lanthanide Ti$_3$Bi$_4$-based compound known,~\cite{ovchinnikov2019bismuth} and Ca$^{2+}$ possesses a 9-coordinate Shannon radius of 1.18\AA~ which agrees with the stability field found here. Considering the preference for large cation radii, a potential curiosity would be the exploration of actinide variants of the \textit{Ln}Ti$_3$Bi$_4$ structure. 

\subsection{Electronic Structure}

Due to the orthorhombic structure, one needs to determine how the the small distortion on the kagome network effects the ``hallmark'' features expected of a kagome metal (Van Hove singularities, Dirac points, flat bands). To first order, density of states calculations estimate that the orbitals near the Fermi level are dominated by contributions from the titanium and bismuth-based orbitals. As such, we provide a ``representative band diagram'' near the Fermi level based on density functional theory (DFT) calculations on LaTi$_3$Bi$_4$. The strictly nonmagnetic nature of LaTi$_3$Bi$_4$ simplifies the discussion and will be a good first-order approximation for the \textit{Ln}Ti$_3$Bi$_4$ family above the magnetic transition temperatures. 

Figure \ref{fig:arpes}(a) demonstrates the DFT-GGA calculated electronic structure of the \textit{Ln}Ti$_3$Bi$_4$ family near the Fermi level. In keeping with the slight distortion of the normally hexagonal kagome motif and the small deviation of the orthorhombic \textit{b/a} lattice parameter ratio from the nominal $\sqrt{3}$ applicable to the hexagonal lattice, the electronic structure at the $X$, $X_1$, and $A_1$ points is rather similar near the Fermi level, with hole bands present around each point. These points, of course, would be equivalent if the hexagonal symmetry was not perturbed. 

Several points of interest can be seen immediately: 1) Dirac-like crossings near $X$ and $A_1$, 2) saddle point (Van Hove singularity) features at $Y$, and 3) kagome-like flat bands scattered between 0.5-0.75~eV below $E_\text{F}$. A schematic of the face-centered orthorhombic Brilluoin zone is shown as an inset to Figure \ref{fig:arpes}(a). Clearly the core elements of the kagome-based electronic structure persist in the calculated band diagram. The proximity of these features to the Fermi level is promising, and cements these materials as prime candidates for ARPES/STM studies.

Experimentally, samples of \textit{Ln}Ti$_3$Bi$_4$ are easily exfoliated and readily available in large sizes, so we performed a suite of preliminary ARPES measurements on crystals of SmTi$_3$Bi$_4$ above the ferromagnetic transition temperature. To first order, we suspect that the \textit{Ln}Ti$_3$Bi$_4$ family will show similar ARPES results. The specifics of other samples (e.g. GdTi$_3$Bi$_4$, LaTi$_3$Bi$_4$, and YbTi$_3$Bi$_4$) are slated to be published elsewhere. Figure \ref{fig:arpes}(b) demonstrates the Fermi surface mapping of SmTi$_3$Bi$_4$ as measured by linearly polarized ARPES. The data were taken at a photon energy of 123~eV. We have superimposed the pseudo-hexagonal 2D Brillouin zone on the Fermi surface plot, highlighting the $M^*$, $K^*$, and $\Gamma^*$ high-symmetry points. Please note the asterisk(*) labels, which denote a pseudo-hexagonal interpretation of the unit cell. 

A ``conversion'' between the ARPES projection and the high-symmetry paths from DFT can be valuable for qualitative comparisons. As discussed before, many of the features in the DFT calculation are only slightly perturbed from the nominal pseudo-hexagonal interpretation. At the limit where the distortion vanishes, the in-equivalent $A_1$ and $X$ points collapse to $K^*$ and Similarly, $Y$ becomes $M^*$. This provides a more transparent way to compare line cuts through the ARPES data to the predicted DFT electronic structure. 

Figure \ref{fig:arpes}(c) demonstrates a selection of linecuts through the ARPES data in the framework of the pseudo-hexagonal projection. We see several Dirac-like points and potential Van Hove singularities (saddle points) near the experimental Fermi level. The Dirac points are clearest at $K^*$, and likely correspond to the features seen near $A_1$ and $X$ in the bulk band structure. The saddle point-like features arise when examining $K^*\rightarrow M^*\rightarrow \Gamma^*$, which is approximately equivalent to $A_1\rightarrow Y^*\rightarrow \Gamma^*$ (highlighted in red on Figure \ref{fig:arpes}(a)). The astonishing robustness of the qualitative results, even when comparing between different members of the \textit{Ln}Ti$_3$Bi$_4$ family -- \textit{and} the translation between the orthorhombic and pseudo-hexagonal interpretations is a testament to the robustness of the kagome motif in controlling the electronic structure of these systems.

While the remainder of this manuscript will primarily focus on the magnetic properties of the \textit{Ln}Ti$_3$Bi$_4$ family, we believe that electrical transport measurements will be a valued tool for future studies. The intermixing of the unique electronic structure and the complex magnetism is precisely why the metallic kagome magnets continue to interest the community.

\subsection{Magnetic and Thermodynamic Properties}
The magnetic properties of the \textit{Ln}Ti$_3$Bi$_4$ family are relatively complex, particularly when reviewing the high-dimensional space of rotation angle, magnetic field, and temperature. As a result, the compounds are divided into three loose classifications: 1) antiferromagnetic/metamagnetic, 2) ferromagnetic, and 3) nonmagnetic. This categorization is a broad simplification for the sake of organization and facile comparison. We acknowledge that magneto-transport would be extremely interesting in this family of materials -- but considering the number of compounds, we leave this for a future study where measurements can be specialized and tuned based on this foundational work. We begin with the most complex and interesting of the compounds, the complex antiferromagnetism and cascade of metamagnetic transitions in GdTi$_3$Bi$_4$.

\subsubsection{Antiferromagnetic GdTi$_3$Bi$_4$}

\begin{figure*}[t]
\includegraphics[width=1\textwidth]{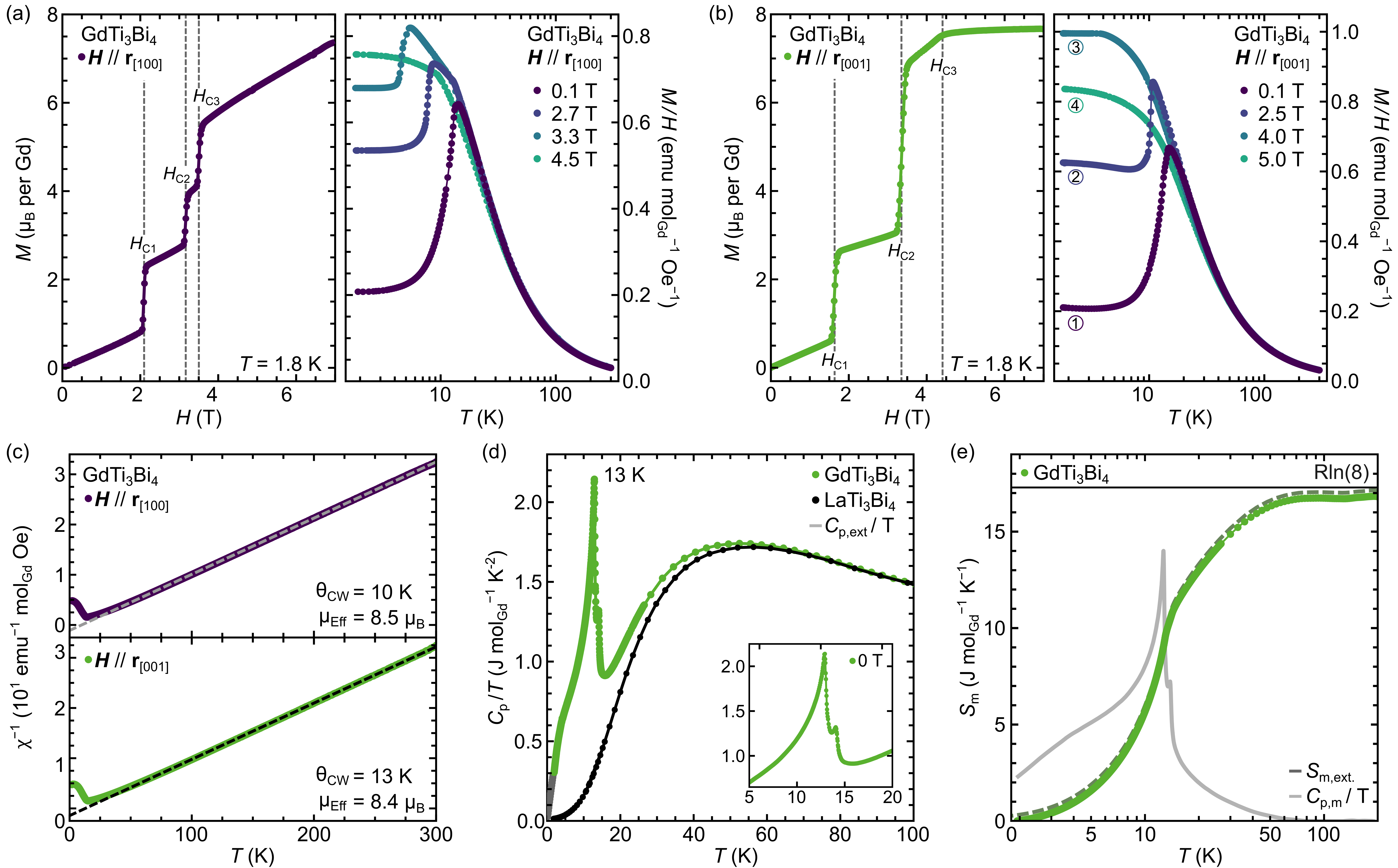}
\caption{(a) Isothermal magnetization for GdTi$_3$Bi$_4$ at 1.8~K demonstrating successive metamagnetic transitions when $H\parallel r_{[100]}$. Select fields were chosen to examine the temperature-dependent magnetization between each metamagnetic transition. (b) Similar results, except for $H\parallel r_{[001]}$. (c) The inverse susceptibility performed at low-fields (100~Oe) yields similar results for both orientations. The effective paramagnetic moment agrees well with that expected of Gd$^{3+}$, though both the isothermal magnetization and Curie-Weiss moment are enhanced slightly. (d) Heat capacity for GdTi$_3$Bi$_4$ showing the zero-field antiferromagnetic transition alongside the nonmagnetic LaTi$_3$Bi$_4$ analog. (e) The integrated entropy approaches the expected $R\ln 8$ for Gd$^{3+}$ by 200~K. The magnetic heat capacity (not to scale) is superimposed in grey for easy reference to the transition temperature.}
\label{fig:3}
\end{figure*}

\begin{figure*}[t]
\includegraphics[width=1\textwidth]{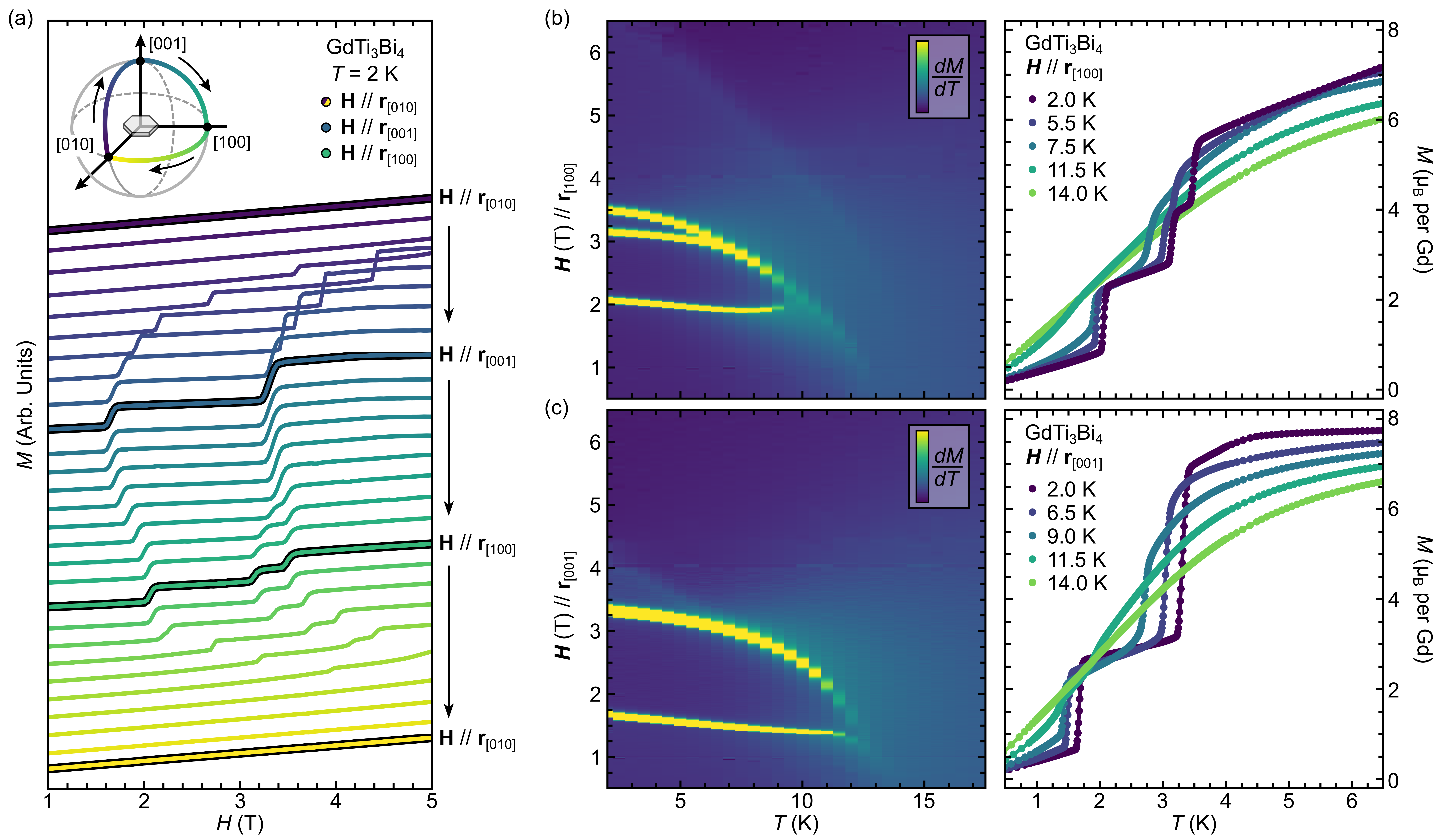}
\caption{(a) Motivated by the complex metamagnetic transitions observed when $H\parallel r_{[100]}$ and $H\parallel r_{[001]}$, we designed a closed loop to probe the orientation dependence of the metamagnetism in GdTi$_3$Bi$_4$. Here we demonstrate isothermal magnetization traces collected at 5 degree increments between the three orthogonal directions. Results clearly demonstrate that the metamagnetism observed at $H\parallel r_{[001]}$ morphs continuously into the response observed when $H\parallel r_{[100]}$. Further, we observe that orthogonal to both metamagnetic directions ($H\parallel r_{[010]}$) exhibits featureless, linear magnetization. (b,c) Full temperature-field phase diagrams of the metamagnetism in GdTi$_3$Bi$_4$ along the special directions  $H\parallel r_{[100]}$ and $H\parallel r_{[001]}$. The phase pockets are in excellent agreement with the critical fields identified earlier. Select isothermal cuts through the data sets are shown to help illustrate the shifts between different regimes.}
\label{fig:4}
\end{figure*}

\begin{figure*}[t]
\includegraphics[width=1\textwidth]{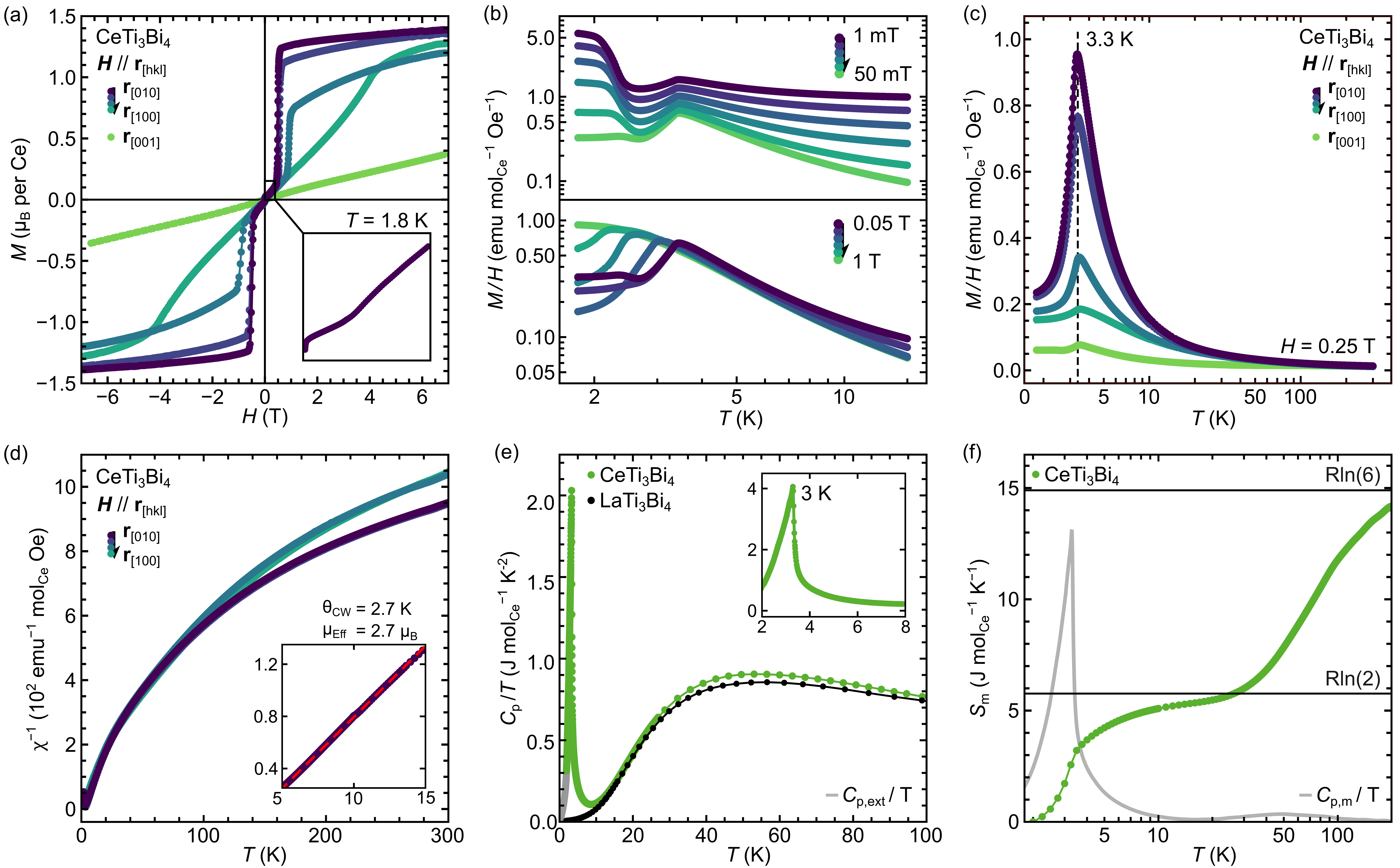}
\caption{(a) Isothermal magnetization for CeTi$_3$Bi$_4$ at 1.8~K demonstrating a single metamagnetic transition which is sharpest along the $H\parallel r_{[010]}$ orientation. The inset highlights the low-field behavior, which exhibits a initial sharp onset followed by a slope change. (b) Temperature-dependent magnetization over the low-field (top) and high-field (bottom) regimes. A clear plateau in the magnetization is noted at low fields, which is quenched by application of fields higher than 50~mT. (c) The orientation dependence of the magnetization at a moderate field of 0.25~T shows the largest response along $H\parallel r_{[010]}$. (d) Curie-Weiss analysis performed at a low-temperature limited range recovers a paramagnetic moment $\mu_\text{Eff}=2.7\mu_\text{B}$, in good agreement with a Ce$^{3+}$ ion. (e) Heat capacity results show a clear lambda anomaly at 3~K and the integrated entropy (f) approaches the expected $R\ln 6$ for Ce$^{3+}$ by 200~K. The extracted magnetic heat capacity superimposed in gray (not to scale) for easy reference.}
\label{fig:5}
\end{figure*}

\begin{figure}
\includegraphics[width=\linewidth]{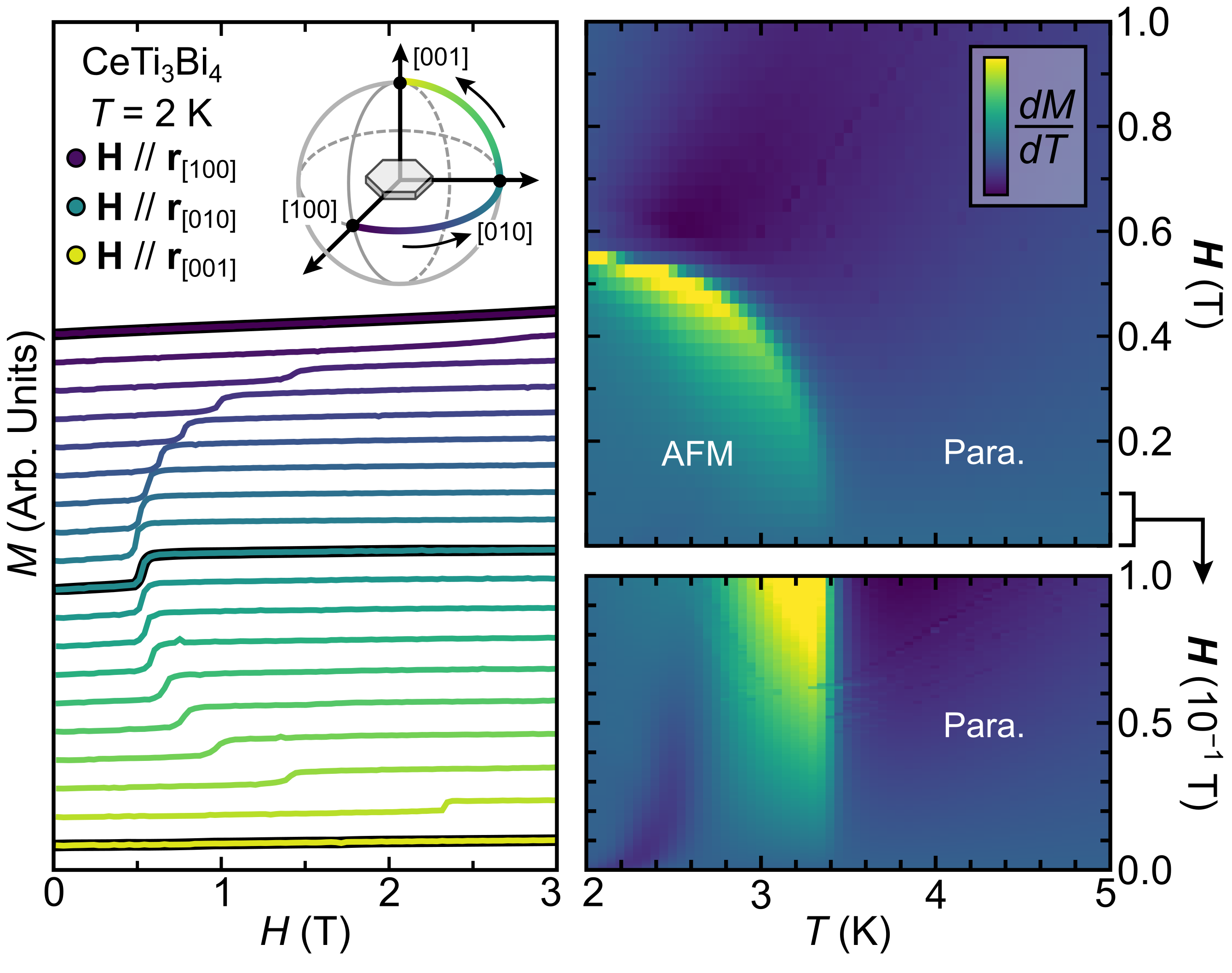}
\caption{CeTi$_3$Bi$_4$ possesses a primary metamagnetic transition when $H\parallel r_{[010]}$. To examine the evolution of the transition as a function of angle, we rotated the sample between the [010], [100], and [001] directions. The transition is suppressed as we rotate away from the [010] direction. The phase pocket (right) formed by the primary transition is obvious in the derivative. The low-field, low-temperature plateau observed in the magnetization creates an additional pocket in the low-field data (right, bottom).}
\label{fig:6}
\end{figure}

\begin{figure*}[t]
\includegraphics[width=1\textwidth]{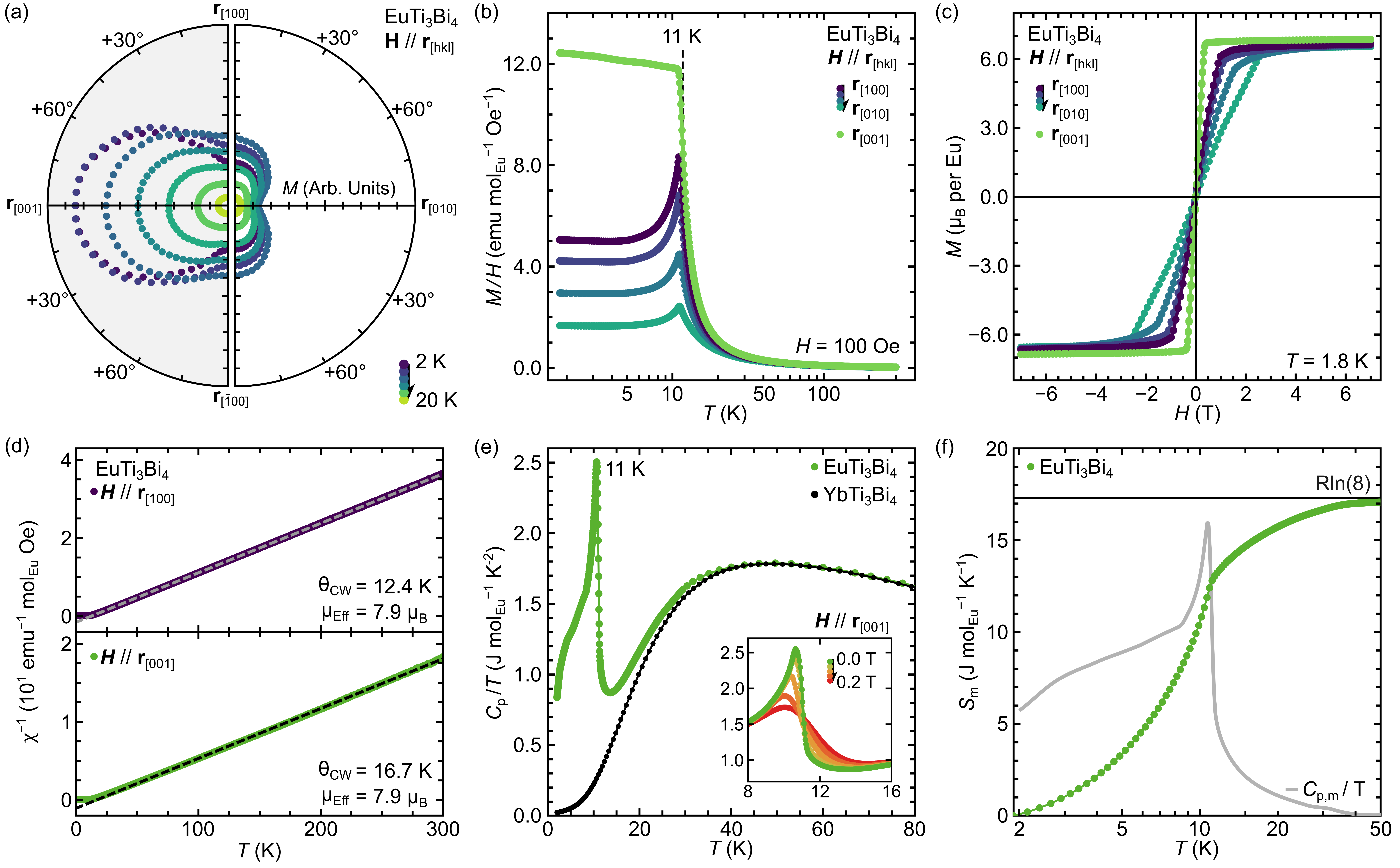}
\caption{(a) Polar magnetization plot for EuTi$_3$Bi$_4$ demonstrating the magnetic anisotropy below and through the magnetic transition. EuTi$_3$Bi$_4$ is the sole exception in the LnTi$_3$Bi$_4$ ferromagnets, where the easy-axis lies in the [001] out-of-plane direction. (b) Temperature-dependent magnetization highlight the ferromagnetic transition along the [001] easy-axis. There is a cusp in the in-plane magnetization for all directions. (c) Isothermal magnetization of EuTi$_3$Bi$_4$ shows rapid (soft) magnetization when $H\parallel r_{[001]}$ . The saturation magnetization is in agreement with that expected for Eu$^{2+}$ ($gJ=7\mu_\text{B}$). (d) Curie-Weiss analysis in both the in-plane and out-of-plane directions are similar and in agreement with the expected Eu$^{2+}$ free ion. (e) Heat capacity results show strong lambda-like anomaly at 11~K consistent with magnetization results. (f) Integrated entropy captures the full Eu$^{2+}$ $R\ln 8$ entropy by 50~K, in agreement with other results.}
\label{fig:7}
\end{figure*}

GdTi$_3$Bi$_4$ is a previously unknown member of the \textit{Ln}Ti$_3$Bi$_4$ family that exhibits a rich and complex set of magnetic properties. When performing the diagnostic set of orientation-dependent magnetization measurements, we observed multiple metamagnetic transitions when $H\parallel r_{[001]}$ and $H\parallel r_{[100]}$. Figure \ref{fig:3}(a,left) highlights a 1.8~K isothermal magnetization measurement up to 7~T where $H\parallel r_{[100]}$. Three distinct metamagnetic transitions can be observed at critical fields of $H_\text{C1}\approx2$~T, $H_\text{C2}\approx3$~T, and $H_\text{C3}\approx3.5$~T.

To investigate further, Figure \ref{fig:3}(a,right) shows temperature-dependent magnetization traces performed at fields between each successive metamagnetic transition. At the lowest field, the system is a clear antiferromagnet with a $T_\text{N}\approx13$~K. The two following fields of 2.7~T and 3.3~T loosely resemble the low-field limit, though the crash in the susceptibility associated with onset of antiferromagnetic order is substantially diminished. This suggests that $H_\text{C1}$, and $H_\text{C2}$ may indicate two intermediate spin-flop transitions where the resulting magnetic order still has an antialigned component. Increased fields after $H_\text{C3}$ appear to induce a largely field polarized state, though the moment doesn't saturate by 7~T, suggesting $H_\text{C3}$ is a final spin-flop and the subsequent linear magnetization is the final rotation to a fully field-polarized state. 

Orthogonal to the first orientation, we find another set of metamagnetic transitions when $H\parallel r_{[001]}$. Figure \ref{fig:3}(b,left) demonstrates the 1.8~K isothermal magnetization in this direction. There are two well-defined critical fields $H_\text{C1}\approx1.5$~T and $H_\text{C2}\approx3.5$~T. A third transition around $H_\text{C3}\approx4.5$~T marks the crossover to a fully field polarized state. At full saturation the magnetization is approximately 7.5$\mu_\text{B}$, which compares favorably with the expected $gJ=7\mu_\text{B}$ for Gd$^{3+}$. The excess magnetization ($\approx$10\%) above the expected $gJ$ may be related to massing errors or shape effects, particularly considering the large moment of Gd$^{3+}$. Note that demagnetization effects will influence results most strongly when $H\parallel r_{[001]}$ due to the plate-like crystal habit. However, as the effect of demagnetization will be to reduce the effective field (e.g. $H_{applied} < H_{internal}$), the effect will not change the qualitative nature of the plots. The orientation with $H\parallel r_{[001]}$ will still saturate substantially faster than $H\parallel r_{[100]}$.

As before, Figure \ref{fig:3}(b,right) examines the intermediate states between each metamagnetic transition. In this orientation, the low-field limit (0.1~T) exhibits the same antiferromagnetic order as before, and the subsequent $H_\text{C1}$ appears like a spin-flop that preserves some component of the antiferromagnetic alignment in the intermediate field regime (2.5~T). The next spin-flop transition at $H_\text{C2}$ appears to destroy most of the antiferromagnetism. From $H_\text{C2}$ the isothermal magnetization increases linearly as moments gradually rotate to the fully polarized state past $H_\text{C3}$.  

Figure \ref{fig:3}(c) examines the inverse susceptibility and the resulting Curie-Weiss analysis for the two primary orientations with $H\parallel r_{[100]}$ and $H\parallel r_{[001]}$ under $H=100$~Oe. In both cases, the effective paramagnetic moment $\mu_\text{Eff}$ is approximately 8.5$\mu_\text{B}$, which agrees well with the expected moment from Gd$^{3+}$ (7.9$\mu_\text{B}$). Our Curie-Weiss analysis and isothermal magnetization measurements are largely consistent, though the persistent enhancement above the expected results for Gd$^{3+}$ suggest a systematic error (e.g. massing, shape factor). However, one can turn to the specific heat to ensure that the enhancement does not result from something more exotic or unexpected.

Figure \ref{fig:3}(d,e) show the specific heat for GdTi$_3$Bi$_4$ and the resulting entropy analysis. Note that all heat capacity results are performed with $H\parallel r_{[001]}$. This corresponds to the hexagonal plate laying flat on the heat capacity stage, ensuring the best thermal contact and minimal orientation error. Regardless, we are largely examining the properties of the zero-field state. Figure \ref{fig:3}(d) examines $C_\text{p}/T$ for GdTi$_3$Bi$_4$ alongside the nonmagnetic analog LaTi$_3$Bi$_4$ (discussed in detail later). Besides a standard correction for the differences in the molar masses (which is on the order of 1\%) there are no additional scaling factors applied to the data. A strong lambda anomaly is noted at 13~K, in good agreement with the temperature-dependent magnetization at 0.1~T. The inset of Figure \ref{fig:3}(c) highlights that the lambda anomaly is actually a pair of transitions. The field-dependence of the transitions in GdTi$_3$Bi$_4$ can be found in the supplementary information\cite{ESI}. The double peak has been reproduced between multiple samples and is believed to be intrinsic to GdTi$_3$Bi$_4$ at this time. The integrated entropy (Figure \ref{fig:3}(e)) approaches the full $R\ln 8$ expected of Gd$^{3+}$ by 200~K. The magnetic contribution to the heat capacity $C_\text{p,m}/T$ is shown superimposed on the integrated entropy is not to scale and is provided for easy reference only.

GdTi$_3$Bi$_4$ provides a good opportunity to reflect on the in-plane anisotropy and the ambiguity of measurements performed solely with $H\parallel c$ and H$\perp c$ in the \textit{Ln}Ti$_3$Bi$_4$ family. Metamagnetic transitions exist in both the $H\parallel r_{[001]}$ and $H\parallel r_{[100]}$ measurements. However, there is no metamagnetic response along the $H\parallel r_{[010]}$ direction. If the simpler set of $H\parallel c$ and H$\perp c$ measurements were made, a qualitatively different picture of GdTi$_3$Bi$_4$ would emerge. Oriented appropriately, the metamagnetic transitions are extremely sharp and well-defined.

To exemplify this effect, we have performed a series of field-temperature phase diagrams for the metamagnetic transitions in GdTi$_3$Bi$_4$. Figure \ref{fig:4} summarizes these results. Let us first examine Figure \ref{fig:4}(a), which shows a series of isothermal magnetization measurements performed as the sample is rotated between  $H\parallel r_{[001]}$, $r_{[001]}$, and $r_{[001]}$. To help orient the reader, we have included a schematic showing the rotation paths used in this study. A closed loop that bridged the two metamagnetic transitions and the non-metamagnetic directions was constructed as the diagnostic path. Every 5 degrees of rotation, we then conducted an isothermal magnetization sweep at 2~K. These results are stitched together in the resulting waterfall plot. The results are to scale, though they have been offset in the y-direction for visual clarity. Several keystone traces are highlighted at specific orientations. In this representation there are two key results: 1) the two metamagnetic transitions blend into each other continuously, and 2) when the field is oriented with $H\parallel r_{[010]}$, the isothermal magnetization is mundane and featureless. 

Returning to the primary metamagnetic transitions, we then constructed temperature-field phase diagrams. Isothermal magnetization measurements were performed at approximately 1~K increments to produce a dense grid of $M(T,H)$ data. Figure \ref{fig:4}(b) demonstrates the phase diagram for $H\parallel r_{[100]}$. Phase boundaries are highlighted through a simple temperature derivative of the magnetization data. The three metamagnetic transitions at $H_\text{C1}$, $H_\text{C2}$, and $H_\text{C3}$ create the three pockets clearly observed in the phase diagram. There is another weak feature at high fields which is evident in the derivative, and can be observed when examining several isothermal cuts through the data (Figure \ref{fig:4}(b, right)). Analogous results can be seen for the orthogonal orientation $H\parallel r_{[001]}$, where the three pockets correspond to the three $H_\text{C1}$, $H_\text{C2}$, and $H_\text{C3}$ transitions identified in the isothermal magnetization from Figure \ref{fig:3}(b). 

Altogether, our results suggest that GdTi$_3$Bi$_4$ possesses an antiferromagnetic ground state that is extremely susceptible to field-induced metamagnetism. The angle-temperature-field diagram is complex, with multiple orientation-dependent spin-flop transitions. The transitions likely correspond to staged destruction of the antiferromagnetic order. Considering the potential analogies to other Gd$^{3+}$ systems, GdTi$_3$Bi$_4$ may be an excellent Skyrmion candidate material.\cite{yasui2020imaging,kurumaji2019skyrmion,hirschberger2019skyrmion} 

\subsubsection{Antiferromagnetic CeTi$_3$Bi$_4$}

The case of CeTi$_3$Bi$_4$ is a bit unusual. The material was previously identified alongside Ce$_3$TiBi$_5$, and the initial study performed a basic suite of magnetic and thermodynamic measurements\cite{motoyama2018magnetic}. However, only the $H\parallel c$ and $H \perp c$ orientations were investigated, and no isothermal magnetization traces were shown. As part of our systematic study, we performed a thorough survey of the magnetic properties of CeTi$_3$Bi$_4$, identifying that it also exhibits metamagnetism and some unusual low-field behavior. 

Figure \ref{fig:5}(a) highlights the isothermal magnetization for CeTi$_3$Bi$_4$ at 1.8~K. Unlike GdTi$_3$Bi$_4$ the metamagnetic response is clearest along a single direction $H\parallel r_{[010]}$. Rotations away from this orientation generally degrade the sharpness of the transition. CeTi$_3$Bi$_4$ appears to exhibit a primary metamagnetic transition around 1~T that marks the rapid saturation of the moment to approximately 1.5$\mu_\text{B}$. This is approximately 70\% of the expected $gJ$ for Ce$^{3+}$. It is possible that another transition exists at higher fields, though we were not able to observe one up to 12~T. Altogether this suggests that the metamagnetism in CeTi$_3$Bi$_4$ is a spin-flip transition from the AFM ordered state to the field-polarized state.

From the full-scale plot shown in Figure \ref{fig:5}(a), there is nothing overtly unusual about the low-field magnetization. However, a closer inspection (Figure \ref{fig:5}(a, inset)) shows an initial sharp rise, followed by a subsequent change in slope. Figure \ref{fig:5}(b, top) investigates the temperature-dependent magnetization over two field regimes: 1) a low-field range highlighting the plateau, and 2) a mid-field range up to the metamagnetic transition. At fields ranging from 1--50~mT, the magnetization shows a clear plateau around 2~K. This field range corresponds to the initial sharp rise seen in Figure \ref{fig:5}(a, inset). The 2~K plateau is strongest for fields where $H<1000$~Oe and is rapidly suppressed with increasing fields. At moderate fields around 0.25~T, the plateau is completely suppressed. While the low-field features were not noted previously, the data collected around 0.25~T is in agreement with the singular antiferromagnetic transition reported in the prior study.\cite{motoyama2018magnetic} While observed in multiple exfoliated samples, additional care needs to be taken to omit the possibility of a hidden impurity phase.

Figure \ref{fig:5}(c) highlights the orientation dependence of the magnetization at a moderate (0.25~T) field. This field is sufficient to quench out the low-field behavior, which provides the most direct comparison with existing literature. From our results, we can approximate that the prior literature oriented their sample with an angle of between 60--90\degree relative to the [010] easy-axis. The qualitative behavior is similar, showing a single antiferromagnetic transition at 3.3~K. Subsequent analysis of the inverse susceptibility (Figure \ref{fig:5}(d)) yields an effective paramagnetic moment of 2.7$\mu_\text{B}$. This compares favorably with that expected from Ce$^{3+}$, $\mu_\text{Eff} = 2.53\mu_\text{B}$.

We can also examine the specific heat as a further consistency check. Figure \ref{fig:5}(d,e) show the specific heat and resulting entropy analysis for CeTi$_3$Bi$_4$ oriented with $H\parallel r_{[001]}$. A sharp lambda anomaly is noted at 3~K, consistent with the magnetization analysis. This peak does not shift with field up to 7~T, consistent with the isothermal magnetization when $H\parallel r_{[001]}$ direction. Perhaps owing to crystal-field effects and the complex coordination environment, some of the total entropy is spread throughout a broad peak in the intermediate temperature regime. Integrating up to and slightly past the primary antiferromagnetic transition (T$<$10~K) recovers approximately $R\ln 2$, suggesting a ground state doublet. Integrating over the remaining entropy contributions from T$>$10~K nearly recovers the full Ce$^{3+}$ $R\ln 6$ entropy by 200~K.

As CeTi$_3$Bi$_4$ shows only a single primary direction of interest, we performed an abbreviated set of rotation-dependent measurements across the metamagnetic transition. The path is shown in Figure \ref{fig:6}(left). While the metamagnetic transition exists throughout a wide range of angles, it broadens and shifts towards higher fields as the crystal is rotated away from $H\parallel r_{[010]}$. Along the two orthogonal directions $H\parallel r_{[100]}$ and $H\parallel r_{[001]}$, the isothermal magnetization is nearly linear over the full field range.

The simpler angular dependence of CeTi$_3$Bi$_4$ necessitates only a single temperature-field phase diagram along $H\parallel r_{[010]}$. This diagram is shown in Figure \ref{fig:6}(right). However, the added complexity of the low-field behavior in the isothermal magnetization was the impetus for a high-resolution field sweep at low fields (\textit{H}$<$0.1~T). Over the broad field range, a single phase boundary is evident, separating the antiferromagnetic and paramagnetic regimes. However, within the low-field data, a division in the low-field regime can be observed. This pocket is associated with the plateau in the temperature-dependent magnetization shown in Figure \ref{fig:5}(b,top), and will require further investigation to understand. 

At first glance CeTi$_3$Bi$_4$ seems like a simpler analog of GdTi$_3$Bi$_4$, demonstrating a single metamagnetic transition and a single preferred magnetization axis. However, there are subtle complexities in the low-field regime, that will require further investigation. The potential for a well-defined doublet ground state is also intriguing, and additional magnetotransport measurements would be a fruitful endeavor.

\subsubsection{Ferromagnetic EuTi$_3$Bi$_4$}

We now turn to examine the ferromagnetic members of the \textit{Ln}Ti$_3$Bi$_4$ family, starting with EuTi$_3$Bi$_4$. The preference for Eu to adopt the Eu$^{2+}$ state in both the Ti$_3$Bi$_4$- and V$_3$Sb$_4$-based families imparts strong spin-only magnetism. We previously reported on the magnetic properties of the V$_3$Sb$_4$ based analog EuV$_3$Sb$_4$, finding weakly anisotropic ferromagnetism with an unusual cusp in the $H\parallel c$ direction.\cite{ortiz2023ybv} Crystals of EuV$_3$Sb$_4$ were exceedingly small and difficult to work with, and while EuTi$_3$Bi$_4$ are the smallest and most difficult to grow member of the $Ln$Ti$_3$Bi$_4$ family, they are nearly an order of magnitude larger and more massive than samples of EuV$_3$Sb$_4$.

Figure \ref{fig:7}(a) provides a compact visualization of the magnetic anisotropy in the ferromagnetic \textit{Ln}Ti$_3$Bi$_4$ compounds. Such a visualization was not straightforward in the antiferromagnetic compounds due to the strong field-dependence (metamagnetism) and phase competition. The easy-axis direction will always correspond to the top of the plot. Clockwise rotations correspond to in-plane orientations, and counter-clockwise rotations correspond to out-of-plane orientations. We have rotated through 180 degrees to match the symmetry of the crystal system (2-fold rotation axis). The data has not been symmetrized, though the 2~K data sets have been normalized to each other at the $H\parallel r_{[010]}$ to remove small errors in orientation caused by remounting the sample. A series of temperature contours are shown that span from the base temperature of 2~K to midway across the ferromagnetic transition, providing a sense of the temperature-dependence as well.

With the full suite of rotation-dependent data, we can see that EuTi$_3$Bi$_4$ is anything but isotropic. The magnetization response between the minimum in-plane orientation $H\parallel r_{[010]}$ and the maximum out-of-plane orientation $H\parallel r_{[001]}$ differs by nearly an order of magnitude. Considering demagnetization effects, which would serve to reduce the internal field in the $H\parallel r_{[001]}$ measurements, the anisotropy would be even more pronounced. Some unusual temperature-dependence can be noted in the $theta$ plane as well, with isothermal contours crossing each other. Figure \ref{fig:7}(b) shows the temperature dependence of the magnetization along several select orientations. All in-plane rotations exhibit a sharp cusp followed by a drop and a subsequent plateau in the magnetization. Contrast this with the out-of-plane ($H\parallel r_{[001]}$) direction, where the magnetization rises and plateaus as one would expect for a ferromagnet. Curiously, the magnetization dependence seems to be reversed from that observed in EuV$_3$Sb$_4$\cite{ortiz2023ybv}.

Figure \ref{fig:7}(c) demonstrates the isothermal magnetization of EuTi$_3$Bi$_4$ over the same set of orientations. The saturation magnetization is nearly 7$\mu_\text{B}$, which agrees nicely with the expected $gJ=7.0\mu_\text{B}$ for spin-only Eu$^{2+}$. Minimal changes in the saturation magnetization are observed with rotations away from the easy-axis orientation. Analysis of the inverse susceptibility (Figure \ref{fig:7}(c)) shows minimal dependence on the orientation and results in a paramagnetic moment of 7.9$\mu_\text{B}$, in excellent agreement with the expected $\mu_\text{Eff}=7.93\mu_\text{B}$ expected of Eu$^{2+}$. Though the crystals of EuTi$_3$Bi$_4$ are small ($<$1~mm) we still caution that demagnetization effects will be strong for the $H\parallel r_{[001]}$ direction due to the plate-like nature of the crystals. However, the effect would be to \textit{reduce} the effective internal field, which would only serve to sharpen the $H\parallel r_{[001]}$ magnetization and accentuate the "soft-ness" of the magnetization along the easy-axis.

Specific heat measurements and the resulting entropy analysis for EuTi$_3$Bi$_4$ are shown in Figure \ref{fig:7}(e,f). The astute observer will note that the nonmagnetic standard has switched from LaTi$_3$Bi$_4$ to YbTi$_3$Bi$_4$. Use of LaTi$_3$Bi$_4$ as the nonmagnetic standard dramatically undersubtracts the phonon background and results an unreasonably high integrated entropy. Switching to YbTi$_3$Bi$_4$ completely removes any issues associated with the nonmagnetic subtraction. Interestingly, the difference between LaTi$_3$Bi$_4$ to YbTi$_3$Bi$_4$ can be noted even in the mechanical properties of the crystals. Divalent YbTi$_3$Bi$_4$ and EuTi$_3$Bi$_4$ are \textit{substantially} softer than the rest of the rare-earth series. In the end, it is conceptually pleasing to use the divalent YbTi$_3$Bi$_4$ as the standard for EuTi$_3$Bi$_4$ -- and the trivalent LaTi$_3$Bi$_4$ for the rest of the series.

Specific heat measurements shown in Figure \ref{fig:7}(e) demonstrate a single lambda anomaly at 11~K, in good agreement with the magnetization results. A broad peak is noted on the low-temperature side of the anomaly, which is often observed in Eu$^{2+}$ containing compounds. The inset of Figure \ref{fig:7}(e) shows the field evolution of the heat capacity peak. Similar to the behavior noted in EuV$_3$Sb$_4$, application of fields causes the peak to broaden and shifts the peak down in temperature.\cite{ortiz2023ybv} The integrated entropy (Figure \ref{fig:7}(f)) approaches $R\ln 8$ by 50~K as expected for Eu$^{2+}$.

\begin{figure*}[t]
\includegraphics[width=1\textwidth]{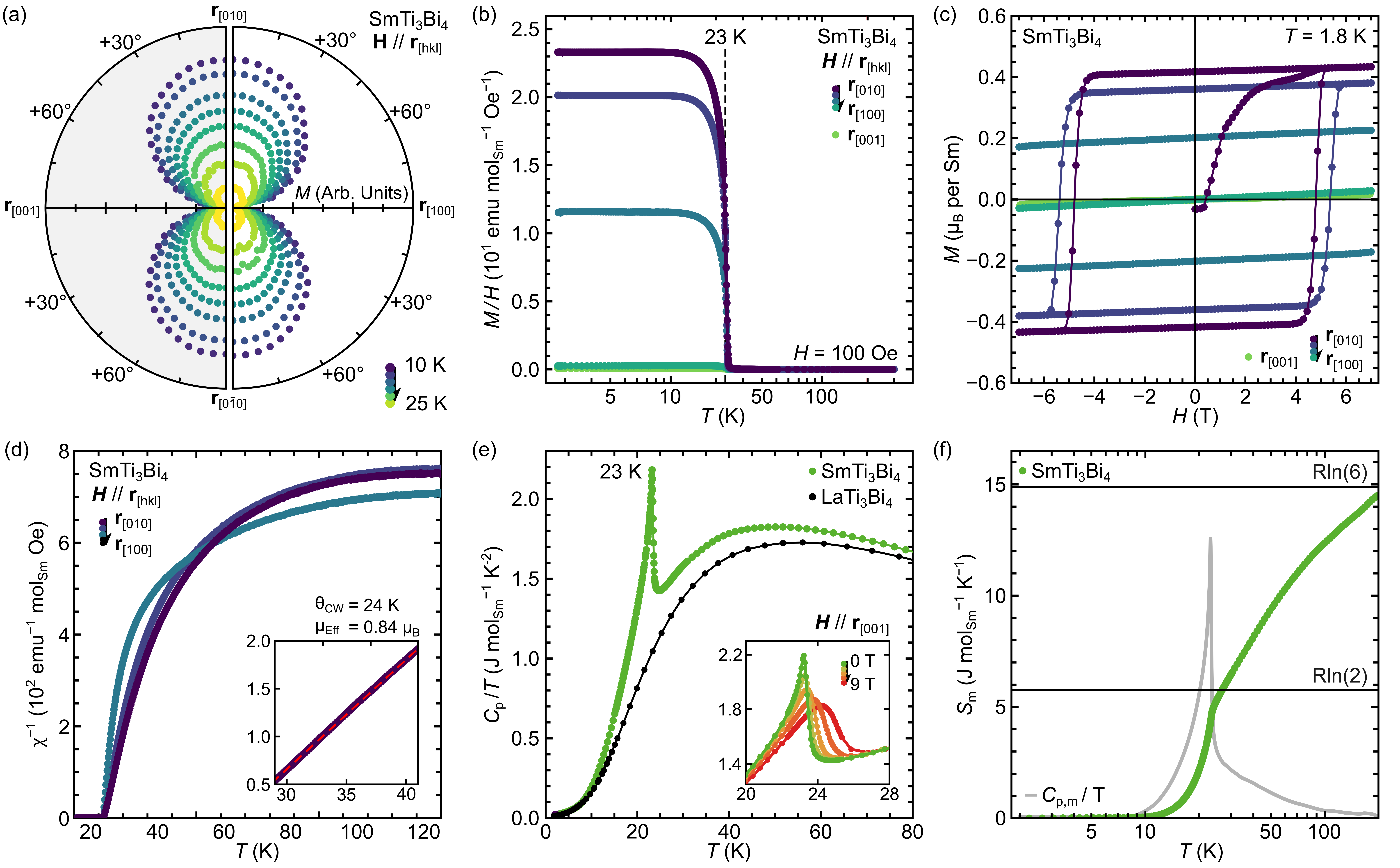}
\caption{(a) Polar magnetization plot for SmTi$_3$Bi$_4$ demonstrating the magnetic anisotropy below and through the magnetic transition. SmTi$_3$Bi$_4$ exhbits strong in-plane anisotropy with a [010] easy-axis. Nearly no magnetic response is observed when $H\parallel r_{[100]}$ or $H\parallel r_{[001]}$. (b) The temperature-dependent magnetization results highlight the ferromagnetic transition at 23~K and show the strongest response when $H\parallel r_{[010]}$. (c) Isothermal magnetization indicates that SmTi$_3$Bi$_4$ is a very hard magnet. The samples investigated here show large coercive fields nearly 5~T at 1.8~K when $H\parallel r_{[010]}$ easy-axis. A discussion of the saturation magnetization can be found in the text body. (d) Curie-Weiss analysis over a limited temperature range produce a $\theta_\text{CW}=+24$~K in agreement with the ferromagnetic transition. (e) Heat capacity results show strong lambda-like anomaly at 23~K consistent with other results. (f) Entropy release for SmTi$_3$Bi$_4$ is gradual, though the release at the magnetic transition is approximately $R\ln 2$, suggesting a magnetic doublet ground state.}
\label{fig:9}
\end{figure*}

\begin{figure*}[t]
\includegraphics[width=1\textwidth]{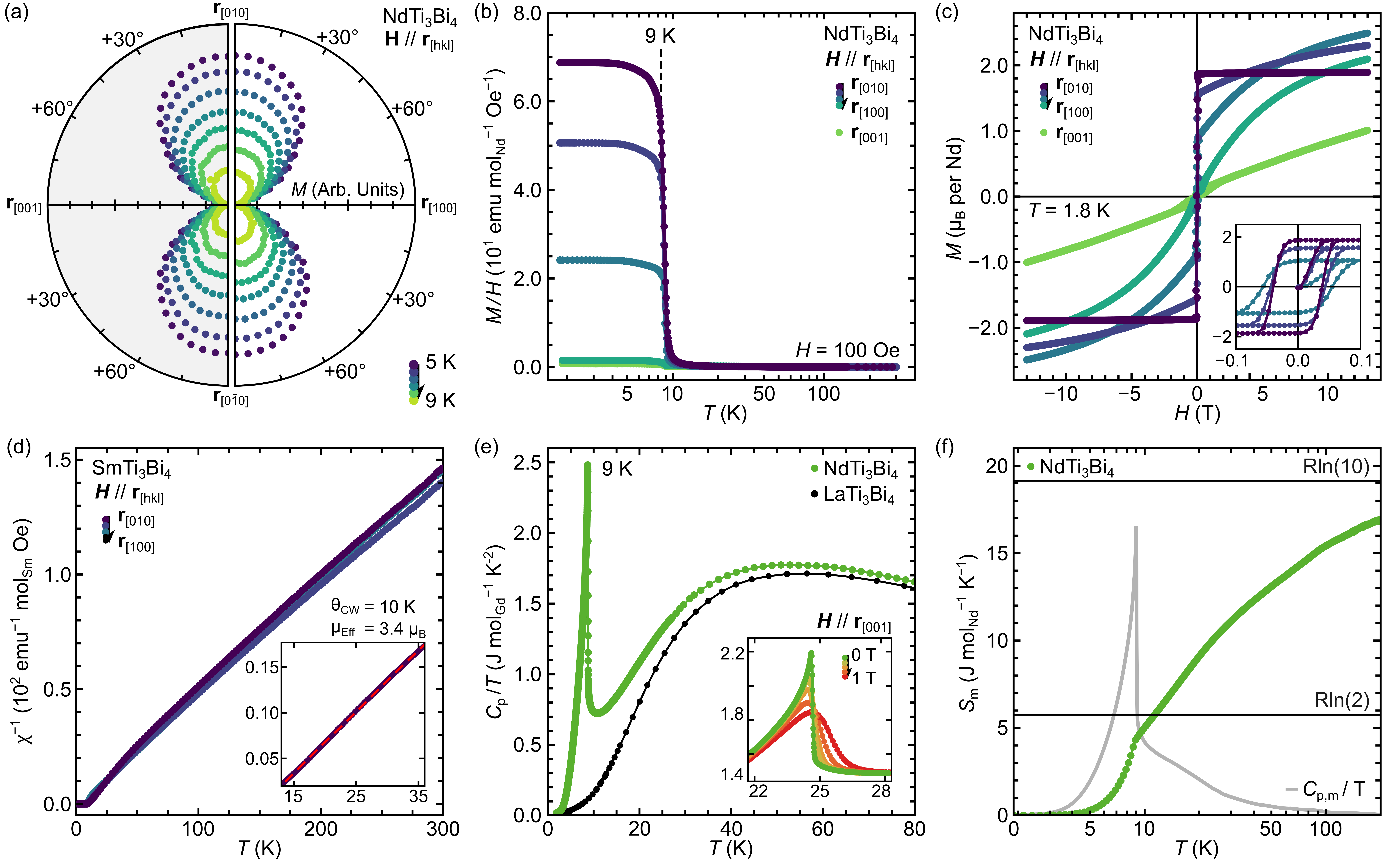}
\caption{(a) Polar magnetization plot for NdTi$_3$Bi$_4$ demonstrating the magnetic anisotropy below and through the magnetic transition. NdTi$_3$Bi$_4$ exhbits strong in-plane anisotropy with a [010] easy-axis. Nearly no magnetic response is observed when $H \parallel r_\text{[100]}$ or $H \parallel r_\text{[001]}$. (b) Temperature-dependent magnetization measurements highlight the ferromagnetic transition at 9~K. (c) Isothermal magnetization indicates that NdTi$_3$Bi$_4$ is a ``soft'' magnet that saturates by 500~Oe when $H \parallel r_\text{[010]}$. The saturation magnetization along the [010] easy-axis does not reach the expected $gJ=3.27\mu_\text{B}$ by 12~T. Intermediate orientations between the [010] and [100] directions continue to exhibit increased magnetization beyond that observed when $H \parallel r_\text{[010]}$. (d) Curie-Weiss analysis over a limited temperature range produce a $\theta_\text{CW}=+10$~K in agreement with the ferromagnetic transition (see additional discussion in text). (e) Heat capacity results show strong lambda-like anomaly at 9~K consistent with other results. (f) Entropy release for NdTi$_3$Bi$_4$ is extended over a wide temperature range, though the release at the magnetic transition is approximately $R\ln 2$, suggesting a magnetic doublet ground state.}
\label{fig:10}
\end{figure*} 

\subsubsection{Ferromagnetic SmTi$_3$Bi$_4$}

Up until this point, all the \textit{Ln}Ti$_3$Bi$_4$ compounds examined in this manuscript have exhibited substantial differences between the in-plane and out-of-plane anisotropies. SmTi$_3$Bi$_4$ remains highly anisotropic, though the rotational dependence is significantly easier to visualize. Figure \ref{fig:9}(a) demonstrates the polar magnetization plot for SmTi$_3$Bi$_4$ collected at 100~Oe. The strong magnetization observed when $H\parallel r_{[010]}$ is immediately evident, and any deviations from this orientation rapidly reduce the magnetic response. 

This is another example to stress the importance of the rotation-dependent measurements in \textit{Ln}Ti$_3$Bi$_4$ compounds. Suppose that measurements were performed out-of-plane [100] and in-plane (but unluckily along [100]).  SmTi$_3$Bi$_4$ exhibits near zero magnetization in the [100] and [001] directions, and the resulting analysis would be decidedly incorrect. The pseudo-hexagonal symmetry of the physical crystal habit is deceptive. The quasi-1D zig-zag chains of \textit{Ln} impart the clear 2-fold symmetry of the magnetic response, in agreement with the orthorhombic structure.

Figure \ref{fig:9}(b) demonstrates the temperature-dependence of the magnetization for a subset of the rotation angles shown in the polar plot. The system is clearly ferromagnetic, with a clear transition at 23~K. The transition temperature is largely unaffected by the sample rotation, but the [010] easy-axis in-plane anisotropy is clear. Figure \ref{fig:9}(c) highlights the isothermal magnetization for the same rotations. The results again indicate strong ferromagnetic order. The sample measured here demonstrates an intrinsic coercivity of approximately 5~T at 1.8~K. Orientations that rotate away from $H\parallel r_{[010]}$ easy-axis rapidly suppress the saturation magnetization and increase the coercivity. A sample rotated by about 60\degree away from the $H\parallel r_{[010]}$ towards $H\parallel r_{[100]}$ cannot be demagnetized by 7~T.

The saturation magnetization reached by 7~T at 1.8~K in Figure \ref{fig:9}(c) is slightly above 0.4$\mu_\text{B}$. This is substantially below that of the Sm free-ion ($gJ = 0.71\mu_\text{B}$). However, the Sm$^{3+}$ is notorious for deviations from the facile free-ion approximation. Van Vleck originally noted the strong temperature-independent contributions to Sm$^{3+}$ magnetization associated with the second-order Zeeman effect.\cite{white1961sign,malik1974crystal} Crystal field effects are also particularly important in Sm-containing systems, and excited levels can be readily admixed into the Sm ground state. Even within a single chemical series, the saturation magnetization can vary drastically. For example, SmNi, SmNi$_2$, SmNi$_3$, and SmNi$_5$ exhibit saturation moments of 0.23, 0.25, 0.33, and 0.70$\mu_\text{B}$, respectively.\cite{malik1974crystal}

The same caution must be used when approaching the Curie-Weiss analysis of Sm-containing compounds. Figure \ref{fig:9}(d) demonstrates the temperature dependence of the inverse susceptibility for a select set of in-plane rotations. For SmTi$_3$Bi$_4$ we do not provide the Curie-Weiss analysis for $H\parallel r_{[100]}$ or $H\parallel r_{[001]}$ directions because the normal magnetic response tends to zero and the inverse-susceptibility rapidly produces nonphysical results. Due to the aforementioned Van Vleck contribution to the susceptibility and the influence of crystal field effects, it is not possible to fit the high-temperature magnetization. While the Curie-Weiss analysis is formally a high-temperature approximation, the best linear regime possible was a small range from 20-40~K (see Figure \ref{fig:9}(d, inset)). Within these approximations we find that $\theta_\text{CW}=24$~K indicating primarly ferromagnetic interactions, in excellent agreement with the ferromagnetic transition at 23~K. The Curie-Weiss paramagnetic moment is approximately 0.84$\mu_\text{B}$, in agreement with the expected 0.84$\mu_\text{B}$ from Sm$^{3+}$, though we stress the limited temperature fitting range and intrinsic nuances of the Sm ion.

We finally turn to the thermodynamic properties of SmTi$_3$Bi$_4$. Figure \ref{fig:9}(e,f) show the heat capacity and resulting integrated magnetic entropy analysis. The heat capacity shows a strong lambda-like anomaly at 23~K, in agreement with Curie-Weiss and temperature-dependent magnetization results. The transition is broadened and shifted towards higher temperatures with the application of magnetic fields (see Figure \ref{fig:9}(e,inset)). The high fields required to shift the peak are a limitation of the geometry of the heat capacity measurement ($H \parallel r_\text{[001]}$) which is orthogonal to the [010] easy-axis in SmTi$_3$Bi$_4$. Note the residual heat capacity caused by the build-up of magnetic fluctuations immediately above the lambda-anomaly in $C_\text{p}/T$. This is reflected most obviously in the magnetic entropy plot Figure \ref{fig:9}(f).  We can see that the magnetic transition releases nearly $R\ln 2$ of entropy, suggesting that the ground state is indeed a doublet. However, the full Sm entropy of $R\ln 6$ is nearly recovered by 200~K owing to the higher temperature fluctuations.

\subsubsection{Ferromagnetic NdTi$_3$Bi$_4$} 

Like SmTi$_3$Bi$_4$, NdTi$_3$Bi$_4$ is a [010] easy-axis ferromagnet. Figure \ref{fig:10}(a) shows the polar magnetization plot, highlighting the strong magnetic response along [010] that diminishes rapidly with rotation towards [100] or [001] directions. Figure \ref{fig:10}(b) shows the temperature-dependent magnetization and highlights the ferromagnetic transition at 9~K. Like SmTi$_3$Bi$_4$, the magnetic anisotropy is strong and results in near zero magnetization when $H \parallel r_\text{[001]}$ or $H \parallel r_\text{[100]}$.

However, unlike SmTi$_3$Bi$_4$, NdTi$_3$Bi$_4$ is an exceptionally \textit{soft} ferromagnet. Figure \ref{fig:10}(c) shows several isothermal magnetization traces for various sample rotations. The orientation-dependence of the saturation magnetization is a bit unusual, however. When $H \parallel r_\text{[010]}$, the magnetization rapidly saturates by approximately 500~Oe to approximately 2$\mu_\text{B}$, which is substantially below that expected from the Nd$^{3+}$ free ion approximation of $gJ=3.27\mu_\text{B}$. As we rotate towards $H \parallel r_\text{[100]}$, the rate of saturation decreases but the ultimate saturation increased. This effect is appears largest when the field is directed at 60\degree from the [010] towards the [100] direction. For the intermediate orientations, the magnetization reaches approximately 2.5$\mu_\text{B}$ by 12~T and continues to increase. This suggests that there may be a small antiferromagnetic contribution in the ground state which is more strongly perturbed as the applied field tends towards the [100] direction. Subsequently one could expect an additional metamagnetic transition in NdTi$_3$Bi$_4$ oriented with $H \parallel r_\text{[010]}$, though one was not observed up to 12~T.

The inverse susceptibility and resulting Curie-Weiss analysis in NdTi$_3$Bi$_4$ are shown in Figure \ref{fig:10}(d). Care must be taken when picking the appropriate regime for the Curie-Weiss fits. There are two linear regimes, one which ranges from 9--40~K, and one from 50-300~K. The fit shown in Figure \ref{fig:10}(c,inset) is over the low-temperature regime, and results in a $\theta_\text{CW}=+10$~K and an effective paramagnetic moment of approximately 3.4$\mu_\text{B}$, in agreement with the expected 3.61$\mu_\text{B}$ for Nd$^{3+}$. The fit over the higher temperature regime results in a $\theta_\text{CW}=-8.4$~K and a effective paramagnetic moment of approximately 4.1$\mu_\text{B}$.

For additional clarity, we can turn to the thermodynamic properties. Figure \ref{fig:10}(e,f) show the heat capacity and resulting integrated entropy analysis for NdTi$_3$Bi$_4$. The heat capacity data (Figure \ref{fig:10}(e)) exhibits a strong lambda-like anomaly at 9~K, in agreement with the magnetization results. The figure inset demonstrates that the transition is broadened and shifted upwards in temperature with the application of a modest field. Recall that due to the geometrical constraint of the heat capacity measurement, the field is always directed with $H \parallel r_\text{[001]}$. The resulting entropy integration shows a release of approximately $R\ln 2$ at the lambda anomaly, consistent with a ground state doublet. However, extended magnetic fluctuations above the transition temperature are substantial, resulting in 80\% of the expected $R\ln 10$ being recovered by 200~K.

At first inspection, SmTi$_3$Bi$_4$ and NdTi$_3$Bi$_4$ look quite similar, with strong ferromagnetism and a [010] easy axis. Heat capacity results also suggest that both exhibit ground state doublets. However, NdTi$_3$Bi$_4$ is a very soft ferromagnet and some nuances exist in the isothermal magnetization that suggest a slightly more complex magnetic ground state. Considering that Nd does not suffer from the same difficulties as Sm in neutron diffraction, it may serve as an excellent candidate for single-crystal neutron studies.

\subsubsection{Non-Kramers PrTi$_3$Bi$_4$}

\begin{figure*}[t]
\includegraphics[width=1\textwidth]{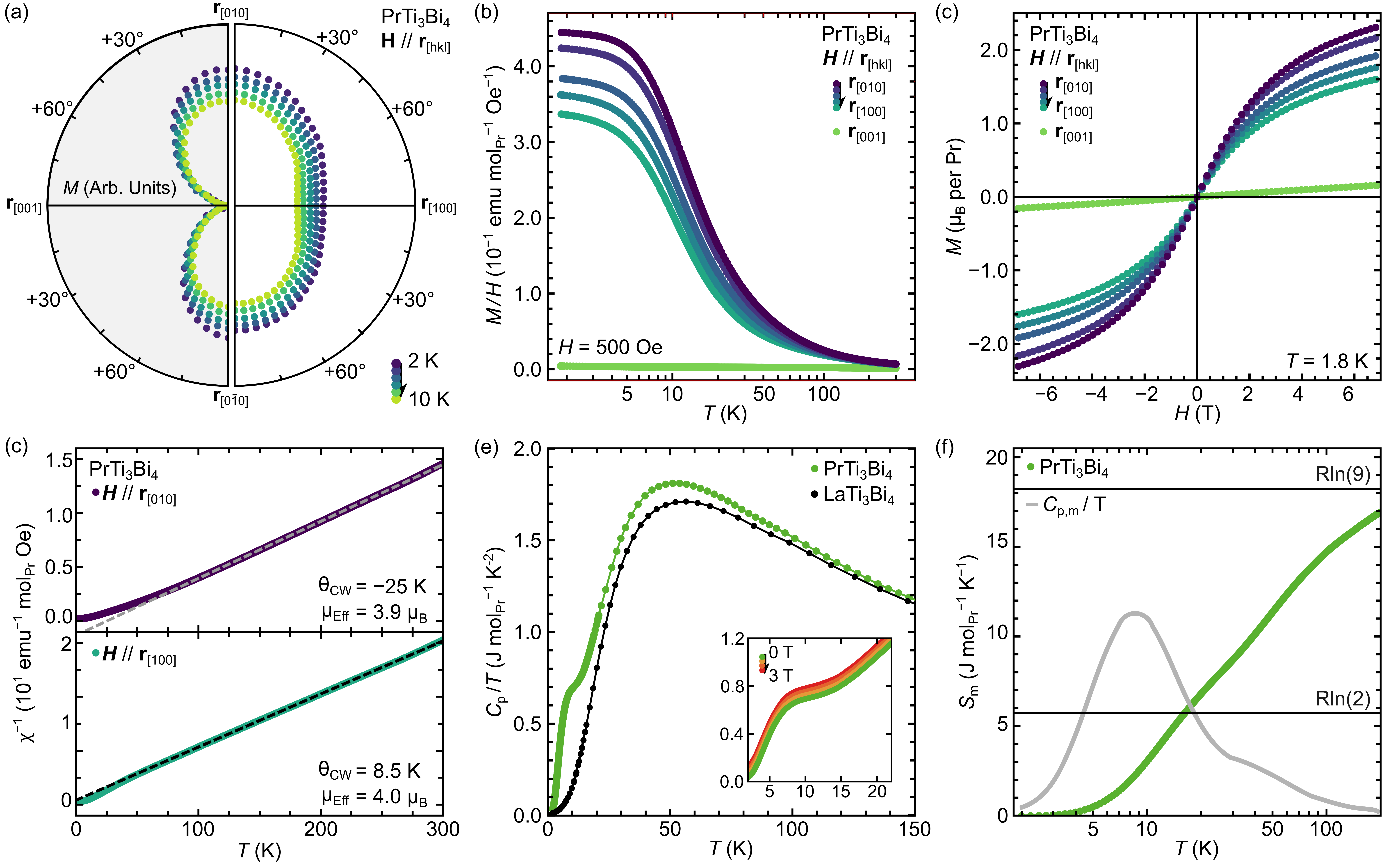}
\caption{(a) Polar magnetization plot for PrTi$_3$Bi$_4$ demonstrating the magnetic anisotropy. Similar to other compounds, there is a strong preference for in-plane magnetization. Measurements oriented such that $H \parallel r_\text{[001]}$ exhibit near zero magnetic response. The anisotropy in-plane is less significant, however, with all in-plane orientations demonstrating similar behavior. (b) No obvious magnetic transition to a long-range ordered state can be seen in the temperature-dependent magnetization. (c) Isothermal magnetization reaches 2.3$\mu_\text{B}$, approximately 70\% of the expected $gJ = 3.2\mu_\text{B}$ for Pr$^{3+}$ by 7~T. (d) Curie-Weiss analysis for fields directed along the $H \parallel r_\text{[100]}$ and $H \parallel r_\text{[010]}$ directions exhibit similar paramagnetic moments, though the Curie-Weiss temperatures vary dramatically. (e) Heat capacity results show a broad magnetic peak centered around 8.7~K and extensive magnetic fluctuations existing to high temperatures. (f) Entropy release for PrTi$_3$Bi$_4$ is extended over a wide temperature range, though the approximate entropy contribution from the broad 8.7~K peaks is approximately $R\ln 2$}
\label{fig:11}
\end{figure*} 

\begin{figure*}[t]
\includegraphics[width=1\textwidth]{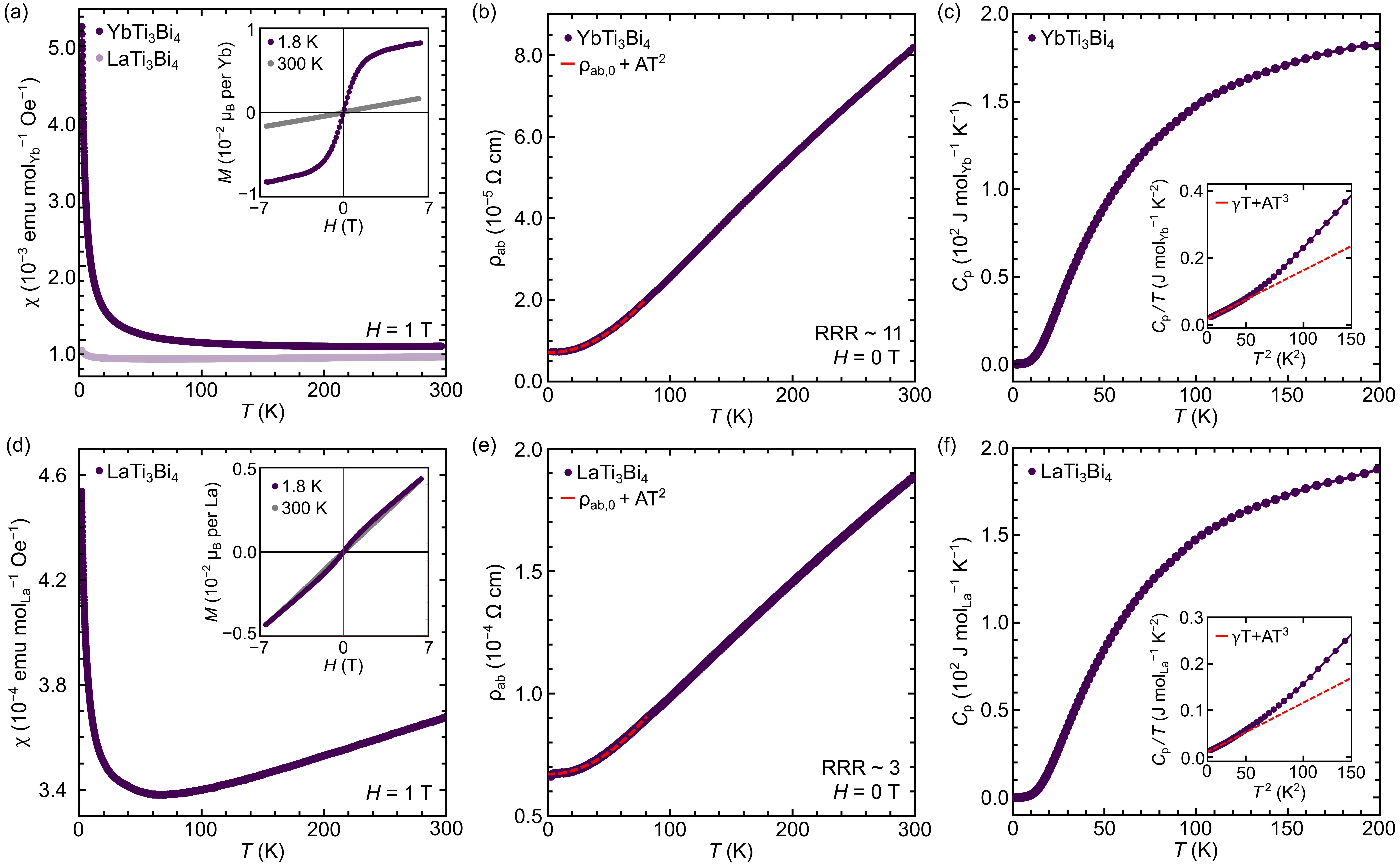}
\caption{(a,d) Magnetization results unambiguously demonstrate the nonmagnetic nature of for YbTi$_3$Bi$_4$ and LaTi$_3$Bi$_4$. The insets show the isothermal magnetization at 1.8~K and 300~K. Note the scale of the insets -- both compounds exhibit $<$0.01$\mu_\text{B}$ per \textit{Ln} atom. (b,e) Electronic resistivity shows properties consistent with a metal. Measurements were done to screen the systems for superconductivity and potential quantum oscillations, though neither were observed at this point in time. (c,f) Heat capacity results for YbTi$_3$Bi$_4$ and LaTi$_3$Bi$_4$ are similar, though not identical. As lattice standards, the best fidelity was noted when the valence of the rare-earth was matched between the magnetic system and the nonmagnetic reference.}
\label{fig:12}
\end{figure*}
The cusp in the magnetization for the in-plane orientations remains a point of interest in EuTi$_3$Bi$_4$. As such, we ventured to explore the field-dependence and orientation-dependence of this feature. An additional set of measurements explorint the field-dependence and orientation-dependence can be found in the ESI.\cite{ESI} Rotations in-plane largely scale the entire signal, with the weakest magnetization along the [010] direction. However, the \textit{relative} strength of the cusp to the subsequent drop and plateau does not change substantially. As we rotate out-of-plane towards the [001], the cusp is rapidly suppressed and replaced with a more prototypical ferromagnetic behavior. At this point, the cusp remains an outstanding point of research in both EuTi$_3$Bi$_4$ and the V$_3$Sb$_4$ cogener EuV$_3$Sb$_4$. Some other Eu-containing metals like EuCo$_2$As$_2$ and EuCo$_2$P$_2$ exhibit qualitatively similar magnetization results and subsequently manifest a helical magnetic ground state.\cite{kim2023evolution,ding2017nmr,sangeetha2016euco} However, additional work needs to be performed to rule out more mundane explanations (e.g. development of magnetic anisotropy). 

Up to this point, all members of the \textit{Ln}Ti$_3$Bi$_4$ family have exhibited a clear transition into a long-range ordered state. The astute observer will also note that all of the members of the \textit{Ln}Ti$_3$Bi$_4$ family up to this point were based on Kramers-ions (Ce, Nd, Sm, Gd, Eu$^{2+}$). PrTi$_3$Bi$_4$ is the first departure from this rule. As a non-Kramers ion, Pr$^{3+}$ has a fragile ground state doublet that can be heavily perturbed by crystal-field effects, disorder, and strain fields. Even in cases where a magnetic ground state can be preserved, the ground state doublet is often poorly isolated from the excited states, which can complicate analysis of Pr-containing compounds.

As in our prior examples, Figure \ref{fig:11}(a) shows the polar magnetization plot for single crystals of PrTi$_3$Bi$_4$. Before interpreting the results, please take note of Figure \ref{fig:11}(b). The temperature-dependent magnetization of PrTi$_3$Bi$_4$ shows no clear magnetic transition, with a broad increase in the magnetization upon cooling. We also tested crystals of PrTi$_3$Bi$_4$ oriented with $H\parallel [010]$ down to 60~mK, though no additional features were observed. Despite the lack for a clear ordering transition, magnetic anisotropy vaguely reminiscent of the SmTi$_3$Bi$_4$ and NdTi$_3$Bi$_4$ compounds can be observed as the sample rotates from $H \parallel r_\text{[010]}$ to $H \parallel r_\text{[001]}$. The system generally exhibits easy-plane anisotropy, with moments preferentially polarized in the \textit{a-b} plane. The magnetization rapidly drops to zero as the sample is rotated such that $H \parallel r_\text{[001]}$. Unlike all other compounds studied thus far, the difference between the various in-plane directions is weak. PrTi$_3$Bi$_4$ also does not exhibit a well-defined saturation magnetization up to 7~T. The isothermal magnetization at 1.8~K is shown in Figure \ref{fig:11}(c). By 7~T the ``easy-axis’’ [010] saturation magnetization is approximately 2.3$\mu_\text{B}$, approximately 70\% of the expected $gJ = 3.2\mu_\text{B}$ for Pr$^{3+}$. 

Analysis of the inverse susceptibility is similarly ambiguous. For PrTi$_3$Bi$_4$ we compare the in-plane $H \parallel r_\text{[010]}$ and $H \parallel r_\text{[100]}$, as the [001] magnetization trends towards zero and produces a non-physical inverse susceptibility. Both results result in similar results for the effective paramagnetic moment ($\mu_\text{Eff} = 3.9-4.0\mu_\text{B}$), which is slightly enhanced above the expected 3.57$\mu_\text{B}$ for Pr$^{3+}$. Conversely the $\theta_\text{CW}$ changes dramatically between between the two directions (8.5~K and $-$25~K). As PrTi$_3$Bi$_4$ shows no clear magnetic transition, it is somewhat difficult to interpret the difference. Under low fields, the moments clearly lie within the \textit{ab}-plane  and do not fully polarize by 7~T. One could imagine that there is an antiferromagnetic interaction along the [010] ``easy-axis’’ that reduces the saturation magnetization and results in the negative Weiss temperature within a system dominated by net ferromagnetic interactions. These considerations are pure speculation, however, and we require further investigations to unravel the ground state of PrTi$_3$Bi$_4$. 

Examining the heat capacity for crystals of PrTi$_3$Bi$_4$ corroborates the weak magnetism and lack of a well defined (long-range) ordered magnetic ground state. Figure \ref{fig:11}(e) reveals a broad, weak peak centered around 8.7~K. Coincidentally, this aligns with the Curie-Weiss temperature extracted when $H \parallel r_\text{[100]}$. The broad peak is wholly unaffected by the application of magnetic fields, though this matches the extraordinarily weak response in the isothermal magnetization (Figure \ref{fig:11}(c))for crystals with $H \parallel r_\text{[001]}$. One can also clearly see that magnetic fluctuations extend well beyond the broad 8.7~K peak. Figure \ref{fig:11}(f) shows the integrated entropy analysis for PrTi$_3$Bi$_4$, highlighting that the entropy release is a gradual continuous process under cooling. The entropy contribution primarily from the broad peak (1.8--20~K) approaches $R\ln 2$, which is initially surprising considering the lack of a clear ordering transition. 

We suspect that Pr may adopt a short-range ordered or spin glass-like state with moments that largely lie within the \textit{ab}-plane, but further measurements will be required to confirm. We remind the reader that the local coordination environment of Pr in PrTi$_3$Bi$_4$ is highly nonuniform. While we approximated the coordination as 9-fold (see Figure \ref{fig:2}), this requires a spread of \textit{Ln}-Bi bond lengths. The low symmetry of the \textit{Ln} coordination shell, combined with the sensitivity of non-Kramers Pr$^{3+}$ to crystal field effects makes the distorted coordination particularly impactful. More work will be required to illuminate the true nature of the ground state in PrTi$_3$Bi$_4$.

\subsubsection{Nonmagnetic LaTi$_3$Bi$_4$ and YbTi$_3$Bi$_4$}
For our final section, we briefly examine the properties of LaTi$_3$Bi$_4$ and YbTi$_3$Bi$_4$. We demonstrated previously that Yb adopts the nonmagnetic divalent Yb$^{2+}$ in crystals of \textit{Ln}V$_3$Sb$_4$. This also appears to be the case for the \textit{Ln}Ti$_3$Bi$_4$ family, though we also have LaTi$_3$Bi$_4$ as an example of the trivalent nonmagnetic rare-earth compound. Figure \ref{fig:12} provides magnetization, electrical resistivity, and heat capacity results for YbTi$_3$Bi$_4$ and LaTi$_3$Bi$_4$. Examining figure \ref{fig:12}(a,d) we can see that the temperature dependent magnetization of both compounds is exceedingly weak (note the scale multiplier 10$^{-3}$--10$^{-4}$). Between the two compounds, LaTi$_3$Bi$_4$ exhibits a substantially weaker response, and has been included as a reference trace along YbTi$_3$Bi$_4$ in Figure \ref{fig:12}(a). Even so, YbTi$_3$Bi$_4$ is decidedly nonmagnetic. \ref{fig:12}(a,inset) demonstrates a saturation magnetization of 0.01$\mu_\text{B}$ per Yb$^{2+}$, consistent with a small fraction of impurity spins or potentially trivalent Yb$^{3+}$. LaTi$_3$Bi$_4$ exhibits an even lower effective magnetization of 0.005$\mu_\text{B}$ per La$^{3+}$ by 7~T. 

Considering the nonmagnetic nature of LaTi$_3$Bi$_4$ and YbTi$_3$Bi$_4$, they were examined for superconductivity down to 1.8~K. A weak drop in the resistivity at base temperatures was noted for LaTi$_3$Bi$_4$, though subsequent magnetization measurements down to 60~mK did not indicate any clear signatures of superconductivity. No signatures of quantum oscillations in the magnetoresistance were observed up to 12~T. Figure \ref{fig:12}(b,e) show the electrical resistivity as a function of temperature under zero field conditions. Minimal differences were noted with the application of modest fields. A quadratic fit ($\rho=\rho_0+AT^2$) to the low-temperature regime provides values of $r_0$=7.06~$\mu$Ohm~cm and $A$=0.0021~$\mu$Ohm~cm~K$^{-2}$ for YbTi$_3$Bi$_4$ and $r_0$=67.1~$\mu$Ohm~cm and $A$=0.0036~$\mu$Ohm~cm~K$^{-2}$ for LaTi$_3$Bi$_4$.

Though used throughout this manuscript as the nonmagnetic lattice standards, the heat capacity results for LaTi$_3$Bi$_4$ and YbTi$_3$Bi$_4$ are shown in Figure \ref{fig:12}(c,f) for completeness. Insets to the heat capacity plots show an appreciated temperature range plotted as $C_\text{p}/T$ vs $T^2$. A Sommerfeld fit to the low-temperature regime provides values of $\gamma$=1.5~mJ~mol$^{-1}$~K$^{-2}$ and $A$=17.8~mJ~mol$^{-1}$~K$^{-4}$ for YbTi$_3$Bi$_4$ and $\gamma$=1.0~mJ~mol$^{-1}$~K$^{-2}$ and $A$=10.2~mJ~mol$^{-1}$~K$^{-4}$ for LaTi$_3$Bi$_4$. While LaTi$_3$Bi$_4$ and YbTi$_3$Bi$_4$ seem nearly identical to the eye, recall that the entropy analysis for the magnetic \textit{Ln}Ti$_3$Bi$_4$ compounds is most accurate when the reference sample has the same valence as the magnetic sample. This appears to be a consequence of the mechanical properties of divalent EuTi$_3$Bi$_4$ and YbTi$_3$Bi$_4$, which are substantially softer than the rest of the series.

\setlength{\tabcolsep}{7.5pt} 
\renewcommand{\arraystretch}{1.15} 
\begin{table}[]
\begin{tabular}{@{}llllll@{}}
\toprule
Compound & $T_\text{Order}$ & Order & $\mu_\text{Eff,CW}$   & $\theta_\text{CW}$     & $M_\text{sat}$  \\ \midrule
GdTi$_3$Bi$_4$ & 13K    & AFM           & 8.5$\mu_\text{B}$   & 10--13K      & 7.5$\mu_\text{B}$        \\
CeTi$_3$Bi$_4$ & 3K     & AFM           & 2.7$\mu_\text{B}$   & 3K          & 1.5$\mu_\text{B}$  \\
EuTi$_3$Bi$_4$ & 11K    & FM            & 7.9$\mu_\text{B}$   & 12--16K      & 7.9$\mu_\text{B}$          \\
SmTi$_3$Bi$_4$ & 23K    & FM            & 0.8$\mu_\text{B}$   & 24K         & 0.4$\mu_\text{B}$        \\
NdTi$_3$Bi$_4$ & 9K     & FM            & 3.4$\mu_\text{B}$   & 10K         & 2.5$\mu_\text{B}$  \\
PrTi$_3$Bi$_4$ & 9K*    & SG*           & 4.0$\mu_\text{B}$   & -25K, 9K     & 2.3$\mu_\text{B}$          \\
LaTi$_3$Bi$_4$ & N/A    & NM            & N/A     & N/A         & 0.01$\mu_\text{B}$         \\
YbTi$_3$Bi$_4$ & N/A    & NM            & N/A     & N/A         & 0.01$\mu_\text{B}$        \\ \bottomrule
\end{tabular}
\caption{\label{sumTable} Summary of the primary results for the \textit{Ln}Ti$_3$Bi$_4$ compounds investigated in this manuscript. The ordering temperature, type of order, and the associated Curie-Weiss fitting parameters are shown. The saturation magnetization obtained at the maximum field, for the highest magnetization direction is also provided. Note that PrTi$_3$Bi$_4$ remains ambiguous at this point, and values are approximate and subject to refinement.}
\end{table}

A brief summary of all magnetic properties uncovered in this work have been included in Table \ref{sumTable} for easy reference.

\section{Conclusion}
Weaving together the intrinsically interesting electronic structure of the kagome network and the chemical degrees of freedom offered by a magnetic rare-earth sublattice has the potential to create new and complex magnetic materials. In this manuscript we have introduced the \textit{Ln}Ti$_3$Bi$_4$ (\textit{Ln}: La...Gd$^{3+}$, Eu$^{2+}$, Yb$^{2+}$) with the hallmark Ti-based kagome motif and quasi-1D chains of rare-earth atoms. The inherent anisotropy of the zig-zag chains imparts a wide array of rich and complex magnetic ground states. While we have predominantly focused on the magnetic properties in this foundational study, our ARPES results have also shown that the electronic structure is densely populated with features arising from the Ti-based kagome nets. The highly exfoliatable nature of single crystals also highlights the admixing of dimensionality in these systems: quasi-2D crystal structure, isolated 2D kagome networks, quasi-1D zig-zag chains and makes them prime candidates for ARPES, STM, and device manufacturing. Ultimately this report serves as an anchor to explore a new direction of magnetic kagome metals with unique and complex ground states.

\section{Acknowledgments}
Research directed by B.R.O. and G.D.S. is sponsored by the Laboratory Directed Research and Development Program of Oak Ridge National Laboratory, managed by UT-Battelle, LLC, for the US Department of Energy. The work of H.M., F.Y., E.M.C., D.S.P., J.Y., A.F.M., and M.A.M. was supported by the U.S. Department of Energy (DOE), Office of Science, Basic Energy Sciences (BES), Materials Sciences and Engineering Division. We thank the X-ray laboratory of the Oak Ridge National Laboratory Spallation Neutron Source for use of their MWL120 Real-Time Back-Reflection Laue Camera System used to orient single crystals. This research utilized beamline 21-ID-1 of the National Synchrotron Light Source II, a U.S. Department of Energy (DOE) Office of Science User Facility operated for the DOE Office of Science by Brookhaven National Laboratory under Contract No. DE-SC0012704. We thank Pyeongjae Park, Andrew D. Christianson, and Denver Strong for their editing, proofreading, and support.

\section{Notes Added}
Alongside this manuscript, several related works were posted on arXiv within a short time period.\cite{chen2023134,guo2023134} While results are similar to ours, the authors do not address the rotational dependence of the in-plane/ out-of-plane magnetization and tend to focus on smaller subsets of the compounds. One work has rotation dependence, but examines only the Sm compound.\cite{chen2023sm134}.

\bibliography{LnTi3Bi4}

\providecommand{\noopsort}[1]{}\providecommand{\singleletter}[1]{#1}%
\begin{thebibliography}{55}%
\makeatletter
\providecommand \@ifxundefined [1]{%
 \@ifx{#1\undefined}
}%
\providecommand \@ifnum [1]{%
 \ifnum #1\expandafter \@firstoftwo
 \else \expandafter \@secondoftwo
 \fi
}%
\providecommand \@ifx [1]{%
 \ifx #1\expandafter \@firstoftwo
 \else \expandafter \@secondoftwo
 \fi
}%
\providecommand \natexlab [1]{#1}%
\providecommand \enquote  [1]{``#1''}%
\providecommand \bibnamefont  [1]{#1}%
\providecommand \bibfnamefont [1]{#1}%
\providecommand \citenamefont [1]{#1}%
\providecommand \href@noop [0]{\@secondoftwo}%
\providecommand \href [0]{\begingroup \@sanitize@url \@href}%
\providecommand \@href[1]{\@@startlink{#1}\@@href}%
\providecommand \@@href[1]{\endgroup#1\@@endlink}%
\providecommand \@sanitize@url [0]{\catcode `\\12\catcode `\$12\catcode
  `\&12\catcode `\#12\catcode `\^12\catcode `\_12\catcode `\%12\relax}%
\providecommand \@@startlink[1]{}%
\providecommand \@@endlink[0]{}%
\providecommand \url  [0]{\begingroup\@sanitize@url \@url }%
\providecommand \@url [1]{\endgroup\@href {#1}{\urlprefix }}%
\providecommand \urlprefix  [0]{URL }%
\providecommand \Eprint [0]{\href }%
\providecommand \doibase [0]{https://doi.org/}%
\providecommand \selectlanguage [0]{\@gobble}%
\providecommand \bibinfo  [0]{\@secondoftwo}%
\providecommand \bibfield  [0]{\@secondoftwo}%
\providecommand \translation [1]{[#1]}%
\providecommand \BibitemOpen [0]{}%
\providecommand \bibitemStop [0]{}%
\providecommand \bibitemNoStop [0]{.\EOS\space}%
\providecommand \EOS [0]{\spacefactor3000\relax}%
\providecommand \BibitemShut  [1]{\csname bibitem#1\endcsname}%
\let\auto@bib@innerbib\@empty
\bibitem [{\citenamefont {Balents}(2010)}]{balents2010spin}%
  \BibitemOpen
  \bibfield  {author} {\bibinfo {author} {\bibfnamefont {L.}~\bibnamefont
  {Balents}},\ }\bibfield  {title} {\bibinfo {title} {Spin liquids in
  frustrated magnets},\ }\href@noop {} {\bibfield  {journal} {\bibinfo
  {journal} {Nature}\ }\textbf {\bibinfo {volume} {464}},\ \bibinfo {pages}
  {199} (\bibinfo {year} {2010})}\BibitemShut {NoStop}%
\bibitem [{\citenamefont {Wulferding}\ \emph {et~al.}(2010)\citenamefont
  {Wulferding}, \citenamefont {Lemmens}, \citenamefont {Scheib}, \citenamefont
  {R{\"o}der}, \citenamefont {Mendels}, \citenamefont {Chu}, \citenamefont
  {Han},\ and\ \citenamefont {Lee}}]{wulferding2010interplay}%
  \BibitemOpen
  \bibfield  {author} {\bibinfo {author} {\bibfnamefont {D.}~\bibnamefont
  {Wulferding}}, \bibinfo {author} {\bibfnamefont {P.}~\bibnamefont {Lemmens}},
  \bibinfo {author} {\bibfnamefont {P.}~\bibnamefont {Scheib}}, \bibinfo
  {author} {\bibfnamefont {J.}~\bibnamefont {R{\"o}der}}, \bibinfo {author}
  {\bibfnamefont {P.}~\bibnamefont {Mendels}}, \bibinfo {author} {\bibfnamefont
  {S.}~\bibnamefont {Chu}}, \bibinfo {author} {\bibfnamefont {T.}~\bibnamefont
  {Han}},\ and\ \bibinfo {author} {\bibfnamefont {Y.~S.}\ \bibnamefont {Lee}},\
  }\bibfield  {title} {\bibinfo {title} {Interplay of thermal and quantum spin
  fluctuations in the kagome lattice compound herbertsmithite},\ }\href@noop {}
  {\bibfield  {journal} {\bibinfo  {journal} {Phys. Rev. B}\ }\textbf {\bibinfo
  {volume} {82}},\ \bibinfo {pages} {144412} (\bibinfo {year}
  {2010})}\BibitemShut {NoStop}%
\bibitem [{\citenamefont {Han}\ \emph {et~al.}(2012)\citenamefont {Han},
  \citenamefont {Helton}, \citenamefont {Chu}, \citenamefont {Nocera},
  \citenamefont {Rodriguez-Rivera}, \citenamefont {Broholm},\ and\
  \citenamefont {Lee}}]{han2012fractionalized}%
  \BibitemOpen
  \bibfield  {author} {\bibinfo {author} {\bibfnamefont {T.-H.}\ \bibnamefont
  {Han}}, \bibinfo {author} {\bibfnamefont {J.~S.}\ \bibnamefont {Helton}},
  \bibinfo {author} {\bibfnamefont {S.}~\bibnamefont {Chu}}, \bibinfo {author}
  {\bibfnamefont {D.~G.}\ \bibnamefont {Nocera}}, \bibinfo {author}
  {\bibfnamefont {J.~A.}\ \bibnamefont {Rodriguez-Rivera}}, \bibinfo {author}
  {\bibfnamefont {C.}~\bibnamefont {Broholm}},\ and\ \bibinfo {author}
  {\bibfnamefont {Y.~S.}\ \bibnamefont {Lee}},\ }\bibfield  {title} {\bibinfo
  {title} {Fractionalized excitations in the spin-liquid state of a
  kagome-lattice antiferromagnet},\ }\href@noop {} {\bibfield  {journal}
  {\bibinfo  {journal} {Nature}\ }\textbf {\bibinfo {volume} {492}},\ \bibinfo
  {pages} {406} (\bibinfo {year} {2012})}\BibitemShut {NoStop}%
\bibitem [{\citenamefont {Fu}\ \emph {et~al.}(2015)\citenamefont {Fu},
  \citenamefont {Imai}, \citenamefont {Han},\ and\ \citenamefont
  {Lee}}]{fu2015evidence}%
  \BibitemOpen
  \bibfield  {author} {\bibinfo {author} {\bibfnamefont {M.}~\bibnamefont
  {Fu}}, \bibinfo {author} {\bibfnamefont {T.}~\bibnamefont {Imai}}, \bibinfo
  {author} {\bibfnamefont {T.-H.}\ \bibnamefont {Han}},\ and\ \bibinfo {author}
  {\bibfnamefont {Y.~S.}\ \bibnamefont {Lee}},\ }\bibfield  {title} {\bibinfo
  {title} {Evidence for a gapped spin-liquid ground state in a kagome
  {H}eisenberg antiferromagnet},\ }\href@noop {} {\bibfield  {journal}
  {\bibinfo  {journal} {Science}\ }\textbf {\bibinfo {volume} {350}},\ \bibinfo
  {pages} {655} (\bibinfo {year} {2015})}\BibitemShut {NoStop}%
\bibitem [{\citenamefont {Freedman}\ \emph {et~al.}(2010)\citenamefont
  {Freedman}, \citenamefont {Han}, \citenamefont {Prodi}, \citenamefont
  {M\"{u}ller}, \citenamefont {Huang}, \citenamefont {Chen}, \citenamefont
  {Webb}, \citenamefont {Lee}, \citenamefont {McQueen},\ and\ \citenamefont
  {Nocera}}]{freedman2010site}%
  \BibitemOpen
  \bibfield  {author} {\bibinfo {author} {\bibfnamefont {D.~E.}\ \bibnamefont
  {Freedman}}, \bibinfo {author} {\bibfnamefont {T.~H.}\ \bibnamefont {Han}},
  \bibinfo {author} {\bibfnamefont {A.}~\bibnamefont {Prodi}}, \bibinfo
  {author} {\bibfnamefont {P.}~\bibnamefont {M\"{u}ller}}, \bibinfo {author}
  {\bibfnamefont {Q.-Z.}\ \bibnamefont {Huang}}, \bibinfo {author}
  {\bibfnamefont {Y.-S.}\ \bibnamefont {Chen}}, \bibinfo {author}
  {\bibfnamefont {S.~M.}\ \bibnamefont {Webb}}, \bibinfo {author}
  {\bibfnamefont {Y.~S.}\ \bibnamefont {Lee}}, \bibinfo {author} {\bibfnamefont
  {T.~M.}\ \bibnamefont {McQueen}},\ and\ \bibinfo {author} {\bibfnamefont
  {D.~G.}\ \bibnamefont {Nocera}},\ }\bibfield  {title} {\bibinfo {title} {Site
  specific x-ray anomalous dispersion of the geometrically frustrated kagome
  magnet, herbertsmithite, {ZnCu$_3$(OH)$_6$Cl$_2$}},\ }\href@noop {}
  {\bibfield  {journal} {\bibinfo  {journal} {J. Am. Chem. Soc.}\ }\textbf
  {\bibinfo {volume} {132}},\ \bibinfo {pages} {16185} (\bibinfo {year}
  {2010})}\BibitemShut {NoStop}%
\bibitem [{\citenamefont {Ortiz}\ \emph {et~al.}(2019)\citenamefont {Ortiz},
  \citenamefont {Gomes}, \citenamefont {Morey}, \citenamefont {Winiarski},
  \citenamefont {Bordelon}, \citenamefont {Mangum}, \citenamefont {Oswald},
  \citenamefont {Rodriguez-Rivera}, \citenamefont {Neilson}, \citenamefont
  {Wilson} \emph {et~al.}}]{ortiz2019new}%
  \BibitemOpen
  \bibfield  {author} {\bibinfo {author} {\bibfnamefont {B.~R.}\ \bibnamefont
  {Ortiz}}, \bibinfo {author} {\bibfnamefont {L.~C.}\ \bibnamefont {Gomes}},
  \bibinfo {author} {\bibfnamefont {J.~R.}\ \bibnamefont {Morey}}, \bibinfo
  {author} {\bibfnamefont {M.}~\bibnamefont {Winiarski}}, \bibinfo {author}
  {\bibfnamefont {M.}~\bibnamefont {Bordelon}}, \bibinfo {author}
  {\bibfnamefont {J.~S.}\ \bibnamefont {Mangum}}, \bibinfo {author}
  {\bibfnamefont {I.~W.}\ \bibnamefont {Oswald}}, \bibinfo {author}
  {\bibfnamefont {J.~A.}\ \bibnamefont {Rodriguez-Rivera}}, \bibinfo {author}
  {\bibfnamefont {J.~R.}\ \bibnamefont {Neilson}}, \bibinfo {author}
  {\bibfnamefont {S.~D.}\ \bibnamefont {Wilson}}, \emph {et~al.},\ }\bibfield
  {title} {\bibinfo {title} {{New kagome prototype materials: discovery of
  KV$_3$Sb$_5$, RbV$_3$Sb$_5$, and CsV$_3$Sb$_5$}},\ }\href@noop {} {\bibfield
  {journal} {\bibinfo  {journal} {Phys. Rev. Materials}\ }\textbf {\bibinfo
  {volume} {3}},\ \bibinfo {pages} {094407} (\bibinfo {year}
  {2019})}\BibitemShut {NoStop}%
\bibitem [{\citenamefont {Ortiz}\ \emph
  {et~al.}(2020{\natexlab{a}})\citenamefont {Ortiz}, \citenamefont {Teicher},
  \citenamefont {Hu}, \citenamefont {Zuo}, \citenamefont {Sarte}, \citenamefont
  {Schueller}, \citenamefont {Abeykoon}, \citenamefont {Krogstad},
  \citenamefont {Rosenkranz}, \citenamefont {Osborn}, \citenamefont {Seshadri},
  \citenamefont {Balents}, \citenamefont {He},\ and\ \citenamefont
  {Wilson}}]{ortizCsV3Sb5}%
  \BibitemOpen
  \bibfield  {author} {\bibinfo {author} {\bibfnamefont {B.~R.}\ \bibnamefont
  {Ortiz}}, \bibinfo {author} {\bibfnamefont {S.~M.}\ \bibnamefont {Teicher}},
  \bibinfo {author} {\bibfnamefont {Y.}~\bibnamefont {Hu}}, \bibinfo {author}
  {\bibfnamefont {J.~L.}\ \bibnamefont {Zuo}}, \bibinfo {author} {\bibfnamefont
  {P.~M.}\ \bibnamefont {Sarte}}, \bibinfo {author} {\bibfnamefont {E.~C.}\
  \bibnamefont {Schueller}}, \bibinfo {author} {\bibfnamefont {A.~M.}\
  \bibnamefont {Abeykoon}}, \bibinfo {author} {\bibfnamefont {M.~J.}\
  \bibnamefont {Krogstad}}, \bibinfo {author} {\bibfnamefont {S.}~\bibnamefont
  {Rosenkranz}}, \bibinfo {author} {\bibfnamefont {R.}~\bibnamefont {Osborn}},
  \bibinfo {author} {\bibfnamefont {R.}~\bibnamefont {Seshadri}}, \bibinfo
  {author} {\bibfnamefont {L.}~\bibnamefont {Balents}}, \bibinfo {author}
  {\bibfnamefont {J.}~\bibnamefont {He}},\ and\ \bibinfo {author}
  {\bibfnamefont {S.~D.}\ \bibnamefont {Wilson}},\ }\bibfield  {title}
  {\bibinfo {title} {{CsV$_3$Sb$_5$: a $\mathbb{Z}_2$ topological kagome metal
  with a superconducting ground state}},\ }\href@noop {} {\bibfield  {journal}
  {\bibinfo  {journal} {Phys. Rev. Lett.}\ }\textbf {\bibinfo {volume} {125}},\
  \bibinfo {pages} {247002} (\bibinfo {year} {2020}{\natexlab{a}})}\BibitemShut
  {NoStop}%
\bibitem [{\citenamefont {Ortiz}\ \emph
  {et~al.}(2020{\natexlab{b}})\citenamefont {Ortiz}, \citenamefont {Kenney},
  \citenamefont {Sarte}, \citenamefont {Teicher}, \citenamefont {Seshadri},
  \citenamefont {Graf},\ and\ \citenamefont {Wilson}}]{ortiz2020KV3Sb5}%
  \BibitemOpen
  \bibfield  {author} {\bibinfo {author} {\bibfnamefont {B.~R.}\ \bibnamefont
  {Ortiz}}, \bibinfo {author} {\bibfnamefont {E.}~\bibnamefont {Kenney}},
  \bibinfo {author} {\bibfnamefont {P.~M.}\ \bibnamefont {Sarte}}, \bibinfo
  {author} {\bibfnamefont {S.~M.}\ \bibnamefont {Teicher}}, \bibinfo {author}
  {\bibfnamefont {R.}~\bibnamefont {Seshadri}}, \bibinfo {author}
  {\bibfnamefont {M.~J.}\ \bibnamefont {Graf}},\ and\ \bibinfo {author}
  {\bibfnamefont {S.~D.}\ \bibnamefont {Wilson}},\ }\bibfield  {title}
  {\bibinfo {title} {{Superconductivity in the $\mathbb{Z}_2$ kagome metal
  KV$_3$Sb$_5$}},\ }\href@noop {} {\bibfield  {journal} {\bibinfo  {journal}
  {Phys. Rev. Mater.}\ }\textbf {\bibinfo {volume} {5}},\ \bibinfo {pages}
  {034801} (\bibinfo {year} {2020}{\natexlab{b}})}\BibitemShut {NoStop}%
\bibitem [{\citenamefont {Yin}\ \emph {et~al.}(2021)\citenamefont {Yin},
  \citenamefont {Tu}, \citenamefont {Gong}, \citenamefont {Fu}, \citenamefont
  {Yan},\ and\ \citenamefont {Lei}}]{RbV3Sb5SC}%
  \BibitemOpen
  \bibfield  {author} {\bibinfo {author} {\bibfnamefont {Q.}~\bibnamefont
  {Yin}}, \bibinfo {author} {\bibfnamefont {Z.}~\bibnamefont {Tu}}, \bibinfo
  {author} {\bibfnamefont {C.}~\bibnamefont {Gong}}, \bibinfo {author}
  {\bibfnamefont {Y.}~\bibnamefont {Fu}}, \bibinfo {author} {\bibfnamefont
  {S.}~\bibnamefont {Yan}},\ and\ \bibinfo {author} {\bibfnamefont
  {H.}~\bibnamefont {Lei}},\ }\bibfield  {title} {\bibinfo {title}
  {{Superconductivity and normal-state properties of kagome metal RbV$_3$Sb$_5$
  single crystals}},\ }\href@noop {} {\bibfield  {journal} {\bibinfo  {journal}
  {Chin. Phys. Lett.}\ }\textbf {\bibinfo {volume} {38}},\ \bibinfo {pages}
  {037403} (\bibinfo {year} {2021})}\BibitemShut {NoStop}%
\bibitem [{\citenamefont {Park}\ \emph {et~al.}(2021)\citenamefont {Park},
  \citenamefont {Ye},\ and\ \citenamefont {Balents}}]{park2021electronic}%
  \BibitemOpen
  \bibfield  {author} {\bibinfo {author} {\bibfnamefont {T.}~\bibnamefont
  {Park}}, \bibinfo {author} {\bibfnamefont {M.}~\bibnamefont {Ye}},\ and\
  \bibinfo {author} {\bibfnamefont {L.}~\bibnamefont {Balents}},\ }\bibfield
  {title} {\bibinfo {title} {{Electronic instabilities of kagome metals: saddle
  points and Landau theory}},\ }\href@noop {} {\bibfield  {journal} {\bibinfo
  {journal} {Phys. Rev. B}\ }\textbf {\bibinfo {volume} {104}},\ \bibinfo
  {pages} {035142} (\bibinfo {year} {2021})}\BibitemShut {NoStop}%
\bibitem [{\citenamefont {Wang}\ \emph {et~al.}(2013)\citenamefont {Wang},
  \citenamefont {Li}, \citenamefont {Xiang},\ and\ \citenamefont
  {Wang}}]{PhysRevB.87.115135}%
  \BibitemOpen
  \bibfield  {author} {\bibinfo {author} {\bibfnamefont {W.-S.}\ \bibnamefont
  {Wang}}, \bibinfo {author} {\bibfnamefont {Z.-Z.}\ \bibnamefont {Li}},
  \bibinfo {author} {\bibfnamefont {Y.-Y.}\ \bibnamefont {Xiang}},\ and\
  \bibinfo {author} {\bibfnamefont {Q.-H.}\ \bibnamefont {Wang}},\ }\bibfield
  {title} {\bibinfo {title} {Competing electronic orders on kagome lattices at
  van hove filling},\ }\href {https://doi.org/10.1103/PhysRevB.87.115135}
  {\bibfield  {journal} {\bibinfo  {journal} {Phys. Rev. B}\ }\textbf {\bibinfo
  {volume} {87}},\ \bibinfo {pages} {115135} (\bibinfo {year}
  {2013})}\BibitemShut {NoStop}%
\bibitem [{\citenamefont {Kiesel}\ \emph {et~al.}(2013)\citenamefont {Kiesel},
  \citenamefont {Platt},\ and\ \citenamefont
  {Thomale}}]{kiesel2013unconventional}%
  \BibitemOpen
  \bibfield  {author} {\bibinfo {author} {\bibfnamefont {M.~L.}\ \bibnamefont
  {Kiesel}}, \bibinfo {author} {\bibfnamefont {C.}~\bibnamefont {Platt}},\ and\
  \bibinfo {author} {\bibfnamefont {R.}~\bibnamefont {Thomale}},\ }\bibfield
  {title} {\bibinfo {title} {{Unconventional Fermi surface instabilities in the
  kagome Hubbard model}},\ }\href@noop {} {\bibfield  {journal} {\bibinfo
  {journal} {Phys. Rev. Lett.}\ }\textbf {\bibinfo {volume} {110}},\ \bibinfo
  {pages} {126405} (\bibinfo {year} {2013})}\BibitemShut {NoStop}%
\bibitem [{\citenamefont {Meier}\ \emph {et~al.}(2020)\citenamefont {Meier},
  \citenamefont {Du}, \citenamefont {Okamoto}, \citenamefont {Mohanta},
  \citenamefont {May}, \citenamefont {McGuire}, \citenamefont {Bridges},
  \citenamefont {Samolyuk},\ and\ \citenamefont {Sales}}]{meier2020flat}%
  \BibitemOpen
  \bibfield  {author} {\bibinfo {author} {\bibfnamefont {W.~R.}\ \bibnamefont
  {Meier}}, \bibinfo {author} {\bibfnamefont {M.-H.}\ \bibnamefont {Du}},
  \bibinfo {author} {\bibfnamefont {S.}~\bibnamefont {Okamoto}}, \bibinfo
  {author} {\bibfnamefont {N.}~\bibnamefont {Mohanta}}, \bibinfo {author}
  {\bibfnamefont {A.~F.}\ \bibnamefont {May}}, \bibinfo {author} {\bibfnamefont
  {M.~A.}\ \bibnamefont {McGuire}}, \bibinfo {author} {\bibfnamefont {C.~A.}\
  \bibnamefont {Bridges}}, \bibinfo {author} {\bibfnamefont {G.~D.}\
  \bibnamefont {Samolyuk}},\ and\ \bibinfo {author} {\bibfnamefont {B.~C.}\
  \bibnamefont {Sales}},\ }\bibfield  {title} {\bibinfo {title} {{Flat bands in
  the CoSn-type compounds}},\ }\href@noop {} {\bibfield  {journal} {\bibinfo
  {journal} {Phys. Rev. B}\ }\textbf {\bibinfo {volume} {102}},\ \bibinfo
  {pages} {075148} (\bibinfo {year} {2020})}\BibitemShut {NoStop}%
\bibitem [{\citenamefont {Isakov}\ \emph {et~al.}(2006)\citenamefont {Isakov},
  \citenamefont {Wessel}, \citenamefont {Melko}, \citenamefont {Sengupta},\
  and\ \citenamefont {Kim}}]{PhysRevLett.97.147202}%
  \BibitemOpen
  \bibfield  {author} {\bibinfo {author} {\bibfnamefont {S.~V.}\ \bibnamefont
  {Isakov}}, \bibinfo {author} {\bibfnamefont {S.}~\bibnamefont {Wessel}},
  \bibinfo {author} {\bibfnamefont {R.~G.}\ \bibnamefont {Melko}}, \bibinfo
  {author} {\bibfnamefont {K.}~\bibnamefont {Sengupta}},\ and\ \bibinfo
  {author} {\bibfnamefont {Y.~B.}\ \bibnamefont {Kim}},\ }\bibfield  {title}
  {\bibinfo {title} {Hard-core bosons on the kagome lattice: Valence-bond
  solids and their quantum melting},\ }\href
  {https://doi.org/10.1103/PhysRevLett.97.147202} {\bibfield  {journal}
  {\bibinfo  {journal} {Phys. Rev. Lett.}\ }\textbf {\bibinfo {volume} {97}},\
  \bibinfo {pages} {147202} (\bibinfo {year} {2006})}\BibitemShut {NoStop}%
\bibitem [{\citenamefont {O'Brien}\ \emph {et~al.}(2010)\citenamefont
  {O'Brien}, \citenamefont {Pollmann},\ and\ \citenamefont
  {Fulde}}]{PhysRevB.81.235115}%
  \BibitemOpen
  \bibfield  {author} {\bibinfo {author} {\bibfnamefont {A.}~\bibnamefont
  {O'Brien}}, \bibinfo {author} {\bibfnamefont {F.}~\bibnamefont {Pollmann}},\
  and\ \bibinfo {author} {\bibfnamefont {P.}~\bibnamefont {Fulde}},\ }\bibfield
   {title} {\bibinfo {title} {Strongly correlated fermions on a kagome
  lattice},\ }\href {https://doi.org/10.1103/PhysRevB.81.235115} {\bibfield
  {journal} {\bibinfo  {journal} {Phys. Rev. B}\ }\textbf {\bibinfo {volume}
  {81}},\ \bibinfo {pages} {235115} (\bibinfo {year} {2010})}\BibitemShut
  {NoStop}%
\bibitem [{\citenamefont {R\"{u}egg}\ and\ \citenamefont
  {Fiete}(2011)}]{PhysRevB.83.165118}%
  \BibitemOpen
  \bibfield  {author} {\bibinfo {author} {\bibfnamefont {A.}~\bibnamefont
  {R\"{u}egg}}\ and\ \bibinfo {author} {\bibfnamefont {G.~A.}\ \bibnamefont
  {Fiete}},\ }\bibfield  {title} {\bibinfo {title} {Fractionally charged
  topological point defects on the kagome lattice},\ }\href
  {https://doi.org/10.1103/PhysRevB.83.165118} {\bibfield  {journal} {\bibinfo
  {journal} {Phys. Rev. B}\ }\textbf {\bibinfo {volume} {83}},\ \bibinfo
  {pages} {165118} (\bibinfo {year} {2011})}\BibitemShut {NoStop}%
\bibitem [{\citenamefont {Guo}\ and\ \citenamefont
  {Franz}(2009)}]{PhysRevB.80.113102}%
  \BibitemOpen
  \bibfield  {author} {\bibinfo {author} {\bibfnamefont {H.-M.}\ \bibnamefont
  {Guo}}\ and\ \bibinfo {author} {\bibfnamefont {M.}~\bibnamefont {Franz}},\
  }\bibfield  {title} {\bibinfo {title} {Topological insulator on the kagome
  lattice},\ }\href {https://doi.org/10.1103/PhysRevB.80.113102} {\bibfield
  {journal} {\bibinfo  {journal} {Phys. Rev. B}\ }\textbf {\bibinfo {volume}
  {80}},\ \bibinfo {pages} {113102} (\bibinfo {year} {2009})}\BibitemShut
  {NoStop}%
\bibitem [{\citenamefont {Yu}\ and\ \citenamefont {Li}(2012)}]{yu2012chiral}%
  \BibitemOpen
  \bibfield  {author} {\bibinfo {author} {\bibfnamefont {S.-L.}\ \bibnamefont
  {Yu}}\ and\ \bibinfo {author} {\bibfnamefont {J.-X.}\ \bibnamefont {Li}},\
  }\bibfield  {title} {\bibinfo {title} {{Chiral superconducting phase and
  chiral spin-density-wave phase in a Hubbard model on the kagome lattice}},\
  }\href@noop {} {\bibfield  {journal} {\bibinfo  {journal} {Phys. Rev. B}\
  }\textbf {\bibinfo {volume} {85}},\ \bibinfo {pages} {144402} (\bibinfo
  {year} {2012})}\BibitemShut {NoStop}%
\bibitem [{\citenamefont {Ko}\ \emph {et~al.}(2009)\citenamefont {Ko},
  \citenamefont {Lee},\ and\ \citenamefont {Wen}}]{ko2009doped}%
  \BibitemOpen
  \bibfield  {author} {\bibinfo {author} {\bibfnamefont {W.-H.}\ \bibnamefont
  {Ko}}, \bibinfo {author} {\bibfnamefont {P.~A.}\ \bibnamefont {Lee}},\ and\
  \bibinfo {author} {\bibfnamefont {X.-G.}\ \bibnamefont {Wen}},\ }\bibfield
  {title} {\bibinfo {title} {{Doped kagome system as exotic superconductor}},\
  }\href@noop {} {\bibfield  {journal} {\bibinfo  {journal} {Phys. Rev. B}\
  }\textbf {\bibinfo {volume} {79}},\ \bibinfo {pages} {214502} (\bibinfo
  {year} {2009})}\BibitemShut {NoStop}%
\bibitem [{\citenamefont {Ortiz}\ \emph {et~al.}(2021)\citenamefont {Ortiz},
  \citenamefont {Teicher}, \citenamefont {Kautzsch}, \citenamefont {Sarte},
  \citenamefont {Ratcliff}, \citenamefont {Harter}, \citenamefont {Ruff},
  \citenamefont {Seshadri},\ and\ \citenamefont {Wilson}}]{ortiz2021fermi}%
  \BibitemOpen
  \bibfield  {author} {\bibinfo {author} {\bibfnamefont {B.~R.}\ \bibnamefont
  {Ortiz}}, \bibinfo {author} {\bibfnamefont {S.~M.}\ \bibnamefont {Teicher}},
  \bibinfo {author} {\bibfnamefont {L.}~\bibnamefont {Kautzsch}}, \bibinfo
  {author} {\bibfnamefont {P.~M.}\ \bibnamefont {Sarte}}, \bibinfo {author}
  {\bibfnamefont {N.}~\bibnamefont {Ratcliff}}, \bibinfo {author}
  {\bibfnamefont {J.}~\bibnamefont {Harter}}, \bibinfo {author} {\bibfnamefont
  {J.~P.}\ \bibnamefont {Ruff}}, \bibinfo {author} {\bibfnamefont
  {R.}~\bibnamefont {Seshadri}},\ and\ \bibinfo {author} {\bibfnamefont
  {S.~D.}\ \bibnamefont {Wilson}},\ }\bibfield  {title} {\bibinfo {title}
  {{Fermi surface mapping and the nature of charge-density-wave order in the
  kagome superconductor CsV$_3$Sb$_5$}},\ }\href@noop {} {\bibfield  {journal}
  {\bibinfo  {journal} {Phys. Rev. X}\ }\textbf {\bibinfo {volume} {11}},\
  \bibinfo {pages} {041030} (\bibinfo {year} {2021})}\BibitemShut {NoStop}%
\bibitem [{\citenamefont {Zhao}\ \emph {et~al.}(2021)\citenamefont {Zhao},
  \citenamefont {Li}, \citenamefont {Ortiz}, \citenamefont {Teicher},
  \citenamefont {Park}, \citenamefont {Ye}, \citenamefont {Wang}, \citenamefont
  {Balents}, \citenamefont {Wilson},\ and\ \citenamefont
  {Zeljkovic}}]{zhao2021cascade}%
  \BibitemOpen
  \bibfield  {author} {\bibinfo {author} {\bibfnamefont {H.}~\bibnamefont
  {Zhao}}, \bibinfo {author} {\bibfnamefont {H.}~\bibnamefont {Li}}, \bibinfo
  {author} {\bibfnamefont {B.~R.}\ \bibnamefont {Ortiz}}, \bibinfo {author}
  {\bibfnamefont {S.~M.}\ \bibnamefont {Teicher}}, \bibinfo {author}
  {\bibfnamefont {T.}~\bibnamefont {Park}}, \bibinfo {author} {\bibfnamefont
  {M.}~\bibnamefont {Ye}}, \bibinfo {author} {\bibfnamefont {Z.}~\bibnamefont
  {Wang}}, \bibinfo {author} {\bibfnamefont {L.}~\bibnamefont {Balents}},
  \bibinfo {author} {\bibfnamefont {S.~D.}\ \bibnamefont {Wilson}},\ and\
  \bibinfo {author} {\bibfnamefont {I.}~\bibnamefont {Zeljkovic}},\ }\bibfield
  {title} {\bibinfo {title} {{Cascade of correlated electron states in the
  kagome superconductor CsV$_3$Sb$_5$}},\ }\href@noop {} {\bibfield  {journal}
  {\bibinfo  {journal} {Nature}\ }\textbf {\bibinfo {volume} {599}},\ \bibinfo
  {pages} {216} (\bibinfo {year} {2021})}\BibitemShut {NoStop}%
\bibitem [{\citenamefont {Hu}\ \emph {et~al.}(2022)\citenamefont {Hu},
  \citenamefont {Wu}, \citenamefont {Ortiz}, \citenamefont {Han}, \citenamefont
  {Plumb}, \citenamefont {Wilson}, \citenamefont {Schnyder}, \citenamefont
  {Shi} \emph {et~al.}}]{hu2022coexistence}%
  \BibitemOpen
  \bibfield  {author} {\bibinfo {author} {\bibfnamefont {Y.}~\bibnamefont
  {Hu}}, \bibinfo {author} {\bibfnamefont {X.}~\bibnamefont {Wu}}, \bibinfo
  {author} {\bibfnamefont {B.~R.}\ \bibnamefont {Ortiz}}, \bibinfo {author}
  {\bibfnamefont {X.}~\bibnamefont {Han}}, \bibinfo {author} {\bibfnamefont
  {N.~C.}\ \bibnamefont {Plumb}}, \bibinfo {author} {\bibfnamefont {S.~D.}\
  \bibnamefont {Wilson}}, \bibinfo {author} {\bibfnamefont {A.~P.}\
  \bibnamefont {Schnyder}}, \bibinfo {author} {\bibfnamefont {M.}~\bibnamefont
  {Shi}}, \emph {et~al.},\ }\bibfield  {title} {\bibinfo {title} {{Coexistence
  of trihexagonal and star-of-David pattern in the charge density wave of the
  kagome superconductor \textit{A}V$_3$Sb$_5$}},\ }\href@noop {} {\bibfield
  {journal} {\bibinfo  {journal} {Phys. Rev. B}\ }\textbf {\bibinfo {volume}
  {106}},\ \bibinfo {pages} {L241106} (\bibinfo {year} {2022})}\BibitemShut
  {NoStop}%
\bibitem [{\citenamefont {Kang}\ \emph {et~al.}(2023)\citenamefont {Kang},
  \citenamefont {Fang}, \citenamefont {Yoo}, \citenamefont {Ortiz},
  \citenamefont {Oey}, \citenamefont {Choi}, \citenamefont {Ryu}, \citenamefont
  {Kim}, \citenamefont {Jozwiak}, \citenamefont {Bostwick} \emph
  {et~al.}}]{kang2022microscopic}%
  \BibitemOpen
  \bibfield  {author} {\bibinfo {author} {\bibfnamefont {M.}~\bibnamefont
  {Kang}}, \bibinfo {author} {\bibfnamefont {S.}~\bibnamefont {Fang}}, \bibinfo
  {author} {\bibfnamefont {J.}~\bibnamefont {Yoo}}, \bibinfo {author}
  {\bibfnamefont {B.~R.}\ \bibnamefont {Ortiz}}, \bibinfo {author}
  {\bibfnamefont {Y.~M.}\ \bibnamefont {Oey}}, \bibinfo {author} {\bibfnamefont
  {J.}~\bibnamefont {Choi}}, \bibinfo {author} {\bibfnamefont {S.~H.}\
  \bibnamefont {Ryu}}, \bibinfo {author} {\bibfnamefont {J.}~\bibnamefont
  {Kim}}, \bibinfo {author} {\bibfnamefont {C.}~\bibnamefont {Jozwiak}},
  \bibinfo {author} {\bibfnamefont {A.}~\bibnamefont {Bostwick}}, \emph
  {et~al.},\ }\bibfield  {title} {\bibinfo {title} {Charge order landscape and
  competition with superconductivity in kagome metals},\ }\href@noop {}
  {\bibfield  {journal} {\bibinfo  {journal} {Nat. Mater.}\ }\textbf {\bibinfo
  {volume} {22}},\ \bibinfo {pages} {186} (\bibinfo {year} {2023})}\BibitemShut
  {NoStop}%
\bibitem [{\citenamefont {Jiang}\ \emph {et~al.}(2021)\citenamefont {Jiang},
  \citenamefont {Yin}, \citenamefont {Denner}, \citenamefont {Shumiya},
  \citenamefont {Ortiz}, \citenamefont {Xu}, \citenamefont {Guguchia},
  \citenamefont {He}, \citenamefont {Hossain}, \citenamefont {Liu} \emph
  {et~al.}}]{jiang2021unconventional}%
  \BibitemOpen
  \bibfield  {author} {\bibinfo {author} {\bibfnamefont {Y.-X.}\ \bibnamefont
  {Jiang}}, \bibinfo {author} {\bibfnamefont {J.-X.}\ \bibnamefont {Yin}},
  \bibinfo {author} {\bibfnamefont {M.~M.}\ \bibnamefont {Denner}}, \bibinfo
  {author} {\bibfnamefont {N.}~\bibnamefont {Shumiya}}, \bibinfo {author}
  {\bibfnamefont {B.~R.}\ \bibnamefont {Ortiz}}, \bibinfo {author}
  {\bibfnamefont {G.}~\bibnamefont {Xu}}, \bibinfo {author} {\bibfnamefont
  {Z.}~\bibnamefont {Guguchia}}, \bibinfo {author} {\bibfnamefont
  {J.}~\bibnamefont {He}}, \bibinfo {author} {\bibfnamefont {M.~S.}\
  \bibnamefont {Hossain}}, \bibinfo {author} {\bibfnamefont {X.}~\bibnamefont
  {Liu}}, \emph {et~al.},\ }\bibfield  {title} {\bibinfo {title}
  {{Unconventional chiral charge order in kagome superconductor
  KV$_3$Sb$_5$}},\ }\href@noop {} {\bibfield  {journal} {\bibinfo  {journal}
  {Nat. Mater.}\ }\textbf {\bibinfo {volume} {20}},\ \bibinfo {pages} {1353}
  (\bibinfo {year} {2021})}\BibitemShut {NoStop}%
\bibitem [{\citenamefont {Peng}\ \emph {et~al.}(2021)\citenamefont {Peng},
  \citenamefont {Han}, \citenamefont {Pokharel}, \citenamefont {Shen},
  \citenamefont {Li}, \citenamefont {Hashimoto}, \citenamefont {Lu},
  \citenamefont {Ortiz}, \citenamefont {Luo}, \citenamefont {Li}, \citenamefont
  {Guo}, \citenamefont {Wang}, \citenamefont {Cui}, \citenamefont {Sun},
  \citenamefont {Qiao}, \citenamefont {Wilson},\ and\ \citenamefont
  {He}}]{PhysRevLett.127.266401}%
  \BibitemOpen
  \bibfield  {author} {\bibinfo {author} {\bibfnamefont {S.}~\bibnamefont
  {Peng}}, \bibinfo {author} {\bibfnamefont {Y.}~\bibnamefont {Han}}, \bibinfo
  {author} {\bibfnamefont {G.}~\bibnamefont {Pokharel}}, \bibinfo {author}
  {\bibfnamefont {J.}~\bibnamefont {Shen}}, \bibinfo {author} {\bibfnamefont
  {Z.}~\bibnamefont {Li}}, \bibinfo {author} {\bibfnamefont {M.}~\bibnamefont
  {Hashimoto}}, \bibinfo {author} {\bibfnamefont {D.}~\bibnamefont {Lu}},
  \bibinfo {author} {\bibfnamefont {B.~R.}\ \bibnamefont {Ortiz}}, \bibinfo
  {author} {\bibfnamefont {Y.}~\bibnamefont {Luo}}, \bibinfo {author}
  {\bibfnamefont {H.}~\bibnamefont {Li}}, \bibinfo {author} {\bibfnamefont
  {M.}~\bibnamefont {Guo}}, \bibinfo {author} {\bibfnamefont {B.}~\bibnamefont
  {Wang}}, \bibinfo {author} {\bibfnamefont {S.}~\bibnamefont {Cui}}, \bibinfo
  {author} {\bibfnamefont {Z.}~\bibnamefont {Sun}}, \bibinfo {author}
  {\bibfnamefont {Z.}~\bibnamefont {Qiao}}, \bibinfo {author} {\bibfnamefont
  {S.~D.}\ \bibnamefont {Wilson}},\ and\ \bibinfo {author} {\bibfnamefont
  {J.}~\bibnamefont {He}},\ }\bibfield  {title} {\bibinfo {title} {Realizing
  kagome band structure in two-dimensional kagome surface states of
  \text{$R{\mathrm{V}}_{6}{\mathrm{Sn}}_{6}$ ($R=\mathrm{Gd}$, Ho)}},\ }\href
  {https://doi.org/10.1103/PhysRevLett.127.266401} {\bibfield  {journal}
  {\bibinfo  {journal} {Phys. Rev. Lett.}\ }\textbf {\bibinfo {volume} {127}},\
  \bibinfo {pages} {266401} (\bibinfo {year} {2021})}\BibitemShut {NoStop}%
\bibitem [{\citenamefont {Wang}\ \emph {et~al.}(2021)\citenamefont {Wang},
  \citenamefont {Neubauer}, \citenamefont {Duan}, \citenamefont {Yin},
  \citenamefont {Fujitsu}, \citenamefont {Hosono}, \citenamefont {Ye},
  \citenamefont {Zhang}, \citenamefont {Chi}, \citenamefont {Krycka},
  \citenamefont {Lei},\ and\ \citenamefont {Dai}}]{PhysRevB.103.014416}%
  \BibitemOpen
  \bibfield  {author} {\bibinfo {author} {\bibfnamefont {Q.}~\bibnamefont
  {Wang}}, \bibinfo {author} {\bibfnamefont {K.~J.}\ \bibnamefont {Neubauer}},
  \bibinfo {author} {\bibfnamefont {C.}~\bibnamefont {Duan}}, \bibinfo {author}
  {\bibfnamefont {Q.}~\bibnamefont {Yin}}, \bibinfo {author} {\bibfnamefont
  {S.}~\bibnamefont {Fujitsu}}, \bibinfo {author} {\bibfnamefont
  {H.}~\bibnamefont {Hosono}}, \bibinfo {author} {\bibfnamefont
  {F.}~\bibnamefont {Ye}}, \bibinfo {author} {\bibfnamefont {R.}~\bibnamefont
  {Zhang}}, \bibinfo {author} {\bibfnamefont {S.}~\bibnamefont {Chi}}, \bibinfo
  {author} {\bibfnamefont {K.}~\bibnamefont {Krycka}}, \bibinfo {author}
  {\bibfnamefont {H.}~\bibnamefont {Lei}},\ and\ \bibinfo {author}
  {\bibfnamefont {P.}~\bibnamefont {Dai}},\ }\bibfield  {title} {\bibinfo
  {title} {{Field-induced topological Hall effect and double-fan spin structure
  with a $c$-axis component in the metallic kagome antiferromagnetic compound
  YMn$_6$Sn$_6$}},\ }\href {https://doi.org/10.1103/PhysRevB.103.014416}
  {\bibfield  {journal} {\bibinfo  {journal} {Phys. Rev. B}\ }\textbf {\bibinfo
  {volume} {103}},\ \bibinfo {pages} {014416} (\bibinfo {year}
  {2021})}\BibitemShut {NoStop}%
\bibitem [{\citenamefont {Pokharel}\ \emph {et~al.}(2021)\citenamefont
  {Pokharel}, \citenamefont {Teicher}, \citenamefont {Ortiz}, \citenamefont
  {Sarte}, \citenamefont {Wu}, \citenamefont {Peng}, \citenamefont {He},
  \citenamefont {Seshadri},\ and\ \citenamefont
  {Wilson}}]{PhysRevB.104.235139}%
  \BibitemOpen
  \bibfield  {author} {\bibinfo {author} {\bibfnamefont {G.}~\bibnamefont
  {Pokharel}}, \bibinfo {author} {\bibfnamefont {S.~M.~L.}\ \bibnamefont
  {Teicher}}, \bibinfo {author} {\bibfnamefont {B.~R.}\ \bibnamefont {Ortiz}},
  \bibinfo {author} {\bibfnamefont {P.~M.}\ \bibnamefont {Sarte}}, \bibinfo
  {author} {\bibfnamefont {G.}~\bibnamefont {Wu}}, \bibinfo {author}
  {\bibfnamefont {S.}~\bibnamefont {Peng}}, \bibinfo {author} {\bibfnamefont
  {J.}~\bibnamefont {He}}, \bibinfo {author} {\bibfnamefont {R.}~\bibnamefont
  {Seshadri}},\ and\ \bibinfo {author} {\bibfnamefont {S.~D.}\ \bibnamefont
  {Wilson}},\ }\bibfield  {title} {\bibinfo {title} {{Electronic properties of
  the topological kagome metals YV$_6$Sn$_6$ and GdV$_6$Sn$_6$}},\ }\href
  {https://doi.org/10.1103/PhysRevB.104.235139} {\bibfield  {journal} {\bibinfo
   {journal} {Phys. Rev. B}\ }\textbf {\bibinfo {volume} {104}},\ \bibinfo
  {pages} {235139} (\bibinfo {year} {2021})}\BibitemShut {NoStop}%
\bibitem [{\citenamefont {Pokharel}\ \emph {et~al.}(2022)\citenamefont
  {Pokharel}, \citenamefont {Ortiz}, \citenamefont {Chamorro}, \citenamefont
  {Sarte}, \citenamefont {Kautzsch}, \citenamefont {Wu}, \citenamefont {Ruff},\
  and\ \citenamefont {Wilson}}]{PhysRevMaterials104202}%
  \BibitemOpen
  \bibfield  {author} {\bibinfo {author} {\bibfnamefont {G.}~\bibnamefont
  {Pokharel}}, \bibinfo {author} {\bibfnamefont {B.}~\bibnamefont {Ortiz}},
  \bibinfo {author} {\bibfnamefont {J.}~\bibnamefont {Chamorro}}, \bibinfo
  {author} {\bibfnamefont {P.}~\bibnamefont {Sarte}}, \bibinfo {author}
  {\bibfnamefont {L.}~\bibnamefont {Kautzsch}}, \bibinfo {author}
  {\bibfnamefont {G.}~\bibnamefont {Wu}}, \bibinfo {author} {\bibfnamefont
  {J.}~\bibnamefont {Ruff}},\ and\ \bibinfo {author} {\bibfnamefont {S.~D.}\
  \bibnamefont {Wilson}},\ }\bibfield  {title} {\bibinfo {title} {{Highly
  anisotropic magnetism in the vanadium-based kagome metal TbV$_6$Sn$_6$}},\
  }\href {https://doi.org/10.1103/PhysRevMaterials.6.104202} {\bibfield
  {journal} {\bibinfo  {journal} {Phys. Rev. Mater.}\ }\textbf {\bibinfo
  {volume} {6}},\ \bibinfo {pages} {104202} (\bibinfo {year}
  {2022})}\BibitemShut {NoStop}%
\bibitem [{\citenamefont {Rosenberg}\ \emph {et~al.}(2022)\citenamefont
  {Rosenberg}, \citenamefont {DeStefano}, \citenamefont {Guo}, \citenamefont
  {Oh}, \citenamefont {Hashimoto}, \citenamefont {Lu}, \citenamefont
  {Birgeneau}, \citenamefont {Lee}, \citenamefont {Ke}, \citenamefont {Yi},\
  and\ \citenamefont {Chu}}]{PhysRevB.106.115139}%
  \BibitemOpen
  \bibfield  {author} {\bibinfo {author} {\bibfnamefont {E.}~\bibnamefont
  {Rosenberg}}, \bibinfo {author} {\bibfnamefont {J.~M.}\ \bibnamefont
  {DeStefano}}, \bibinfo {author} {\bibfnamefont {Y.}~\bibnamefont {Guo}},
  \bibinfo {author} {\bibfnamefont {J.~S.}\ \bibnamefont {Oh}}, \bibinfo
  {author} {\bibfnamefont {M.}~\bibnamefont {Hashimoto}}, \bibinfo {author}
  {\bibfnamefont {D.}~\bibnamefont {Lu}}, \bibinfo {author} {\bibfnamefont
  {R.~J.}\ \bibnamefont {Birgeneau}}, \bibinfo {author} {\bibfnamefont
  {Y.}~\bibnamefont {Lee}}, \bibinfo {author} {\bibfnamefont {L.}~\bibnamefont
  {Ke}}, \bibinfo {author} {\bibfnamefont {M.}~\bibnamefont {Yi}},\ and\
  \bibinfo {author} {\bibfnamefont {J.-H.}\ \bibnamefont {Chu}},\ }\bibfield
  {title} {\bibinfo {title} {{Uniaxial ferromagnetism in the kagome metal
  TbV$_6$Sn$_6$}},\ }\href {https://doi.org/10.1103/PhysRevB.106.115139}
  {\bibfield  {journal} {\bibinfo  {journal} {Phys. Rev. B}\ }\textbf {\bibinfo
  {volume} {106}},\ \bibinfo {pages} {115139} (\bibinfo {year}
  {2022})}\BibitemShut {NoStop}%
\bibitem [{\citenamefont {Ghimire}\ \emph {et~al.}(2020)\citenamefont
  {Ghimire}, \citenamefont {Dally}, \citenamefont {Poudel}, \citenamefont
  {Jones}, \citenamefont {Michel}, \citenamefont {Magar}, \citenamefont
  {Bleuel}, \citenamefont {McGuire}, \citenamefont {Jiang}, \citenamefont
  {Mitchell}, \citenamefont {Lynn},\ and\ \citenamefont
  {Mazin}}]{sciadv_abe2680}%
  \BibitemOpen
  \bibfield  {author} {\bibinfo {author} {\bibfnamefont {N.~J.}\ \bibnamefont
  {Ghimire}}, \bibinfo {author} {\bibfnamefont {R.~L.}\ \bibnamefont {Dally}},
  \bibinfo {author} {\bibfnamefont {L.}~\bibnamefont {Poudel}}, \bibinfo
  {author} {\bibfnamefont {D.~C.}\ \bibnamefont {Jones}}, \bibinfo {author}
  {\bibfnamefont {D.}~\bibnamefont {Michel}}, \bibinfo {author} {\bibfnamefont
  {N.~T.}\ \bibnamefont {Magar}}, \bibinfo {author} {\bibfnamefont
  {M.}~\bibnamefont {Bleuel}}, \bibinfo {author} {\bibfnamefont {M.~A.}\
  \bibnamefont {McGuire}}, \bibinfo {author} {\bibfnamefont {J.~S.}\
  \bibnamefont {Jiang}}, \bibinfo {author} {\bibfnamefont {J.~F.}\ \bibnamefont
  {Mitchell}}, \bibinfo {author} {\bibfnamefont {J.~W.}\ \bibnamefont {Lynn}},\
  and\ \bibinfo {author} {\bibfnamefont {I.~I.}\ \bibnamefont {Mazin}},\
  }\bibfield  {title} {\bibinfo {title} {Competing magnetic phases and
  fluctuation-driven scalar spin chirality in the kagome metal
  \text{YMn$_6$Sn$_6$}},\ }\href {https://doi.org/10.1126/sciadv.abe2680}
  {\bibfield  {journal} {\bibinfo  {journal} {Science Advances}\ }\textbf
  {\bibinfo {volume} {6}},\ \bibinfo {pages} {eabe2680} (\bibinfo {year}
  {2020})}\BibitemShut {NoStop}%
\bibitem [{\citenamefont {Arachchige}\ \emph {et~al.}(2022)\citenamefont
  {Arachchige}, \citenamefont {Meier}, \citenamefont {Marshall}, \citenamefont
  {Matsuoka}, \citenamefont {Xue}, \citenamefont {McGuire}, \citenamefont
  {Hermann}, \citenamefont {Cao},\ and\ \citenamefont
  {Mandrus}}]{PhysRevLett.129.216402}%
  \BibitemOpen
  \bibfield  {author} {\bibinfo {author} {\bibfnamefont {H.~W.~S.}\
  \bibnamefont {Arachchige}}, \bibinfo {author} {\bibfnamefont {W.~R.}\
  \bibnamefont {Meier}}, \bibinfo {author} {\bibfnamefont {M.}~\bibnamefont
  {Marshall}}, \bibinfo {author} {\bibfnamefont {T.}~\bibnamefont {Matsuoka}},
  \bibinfo {author} {\bibfnamefont {R.}~\bibnamefont {Xue}}, \bibinfo {author}
  {\bibfnamefont {M.~A.}\ \bibnamefont {McGuire}}, \bibinfo {author}
  {\bibfnamefont {R.~P.}\ \bibnamefont {Hermann}}, \bibinfo {author}
  {\bibfnamefont {H.}~\bibnamefont {Cao}},\ and\ \bibinfo {author}
  {\bibfnamefont {D.}~\bibnamefont {Mandrus}},\ }\bibfield  {title} {\bibinfo
  {title} {Charge density wave in kagome lattice intermetallic
  \text{${\mathrm{ScV}}_{6}{\mathrm{Sn}}_{6}$}},\ }\href
  {https://doi.org/10.1103/PhysRevLett.129.216402} {\bibfield  {journal}
  {\bibinfo  {journal} {Phys. Rev. Lett.}\ }\textbf {\bibinfo {volume} {129}},\
  \bibinfo {pages} {216402} (\bibinfo {year} {2022})}\BibitemShut {NoStop}%
\bibitem [{\citenamefont {Yin}\ \emph {et~al.}(2020)\citenamefont {Yin},
  \citenamefont {Ma}, \citenamefont {Cochran}, \citenamefont {Xu},
  \citenamefont {Zhang}, \citenamefont {Tien}, \citenamefont {Shumiya},
  \citenamefont {Cheng}, \citenamefont {Jiang}, \citenamefont {Lian},
  \citenamefont {Song}, \citenamefont {Chang}, \citenamefont {Belopolski},
  \citenamefont {Multer}, \citenamefont {Litskevich}, \citenamefont {Cheng},
  \citenamefont {Yang}, \citenamefont {Swidler}, \citenamefont {Zhou},
  \citenamefont {Lin}, \citenamefont {Neupert}, \citenamefont {Wang},
  \citenamefont {Yao}, \citenamefont {Chang}, \citenamefont {Jia},\ and\
  \citenamefont {Zahid~Hasan}}]{Yin_2020}%
  \BibitemOpen
  \bibfield  {author} {\bibinfo {author} {\bibfnamefont {J.-X.}\ \bibnamefont
  {Yin}}, \bibinfo {author} {\bibfnamefont {W.}~\bibnamefont {Ma}}, \bibinfo
  {author} {\bibfnamefont {T.~A.}\ \bibnamefont {Cochran}}, \bibinfo {author}
  {\bibfnamefont {X.}~\bibnamefont {Xu}}, \bibinfo {author} {\bibfnamefont
  {S.~S.}\ \bibnamefont {Zhang}}, \bibinfo {author} {\bibfnamefont {H.-J.}\
  \bibnamefont {Tien}}, \bibinfo {author} {\bibfnamefont {N.}~\bibnamefont
  {Shumiya}}, \bibinfo {author} {\bibfnamefont {G.}~\bibnamefont {Cheng}},
  \bibinfo {author} {\bibfnamefont {K.}~\bibnamefont {Jiang}}, \bibinfo
  {author} {\bibfnamefont {B.}~\bibnamefont {Lian}}, \bibinfo {author}
  {\bibfnamefont {Z.}~\bibnamefont {Song}}, \bibinfo {author} {\bibfnamefont
  {G.}~\bibnamefont {Chang}}, \bibinfo {author} {\bibfnamefont
  {I.}~\bibnamefont {Belopolski}}, \bibinfo {author} {\bibfnamefont
  {D.}~\bibnamefont {Multer}}, \bibinfo {author} {\bibfnamefont
  {M.}~\bibnamefont {Litskevich}}, \bibinfo {author} {\bibfnamefont {Z.-J.}\
  \bibnamefont {Cheng}}, \bibinfo {author} {\bibfnamefont {X.~P.}\ \bibnamefont
  {Yang}}, \bibinfo {author} {\bibfnamefont {B.}~\bibnamefont {Swidler}},
  \bibinfo {author} {\bibfnamefont {H.}~\bibnamefont {Zhou}}, \bibinfo {author}
  {\bibfnamefont {H.}~\bibnamefont {Lin}}, \bibinfo {author} {\bibfnamefont
  {T.}~\bibnamefont {Neupert}}, \bibinfo {author} {\bibfnamefont
  {Z.}~\bibnamefont {Wang}}, \bibinfo {author} {\bibfnamefont {N.}~\bibnamefont
  {Yao}}, \bibinfo {author} {\bibfnamefont {T.-R.}\ \bibnamefont {Chang}},
  \bibinfo {author} {\bibfnamefont {S.}~\bibnamefont {Jia}},\ and\ \bibinfo
  {author} {\bibfnamefont {M.}~\bibnamefont {Zahid~Hasan}},\ }\bibfield
  {title} {\bibinfo {title} {Quantum-limit chern topological magnetism in
  \text{TbMn$_6$Sn$_6$}},\ }\href {https://doi.org/10.1038/s41586-020-2482-7}
  {\bibfield  {journal} {\bibinfo  {journal} {Nature}\ }\textbf {\bibinfo
  {volume} {583}},\ \bibinfo {pages} {533} (\bibinfo {year}
  {2020})}\BibitemShut {NoStop}%
\bibitem [{\citenamefont {Zhang}\ \emph {et~al.}(2022)\citenamefont {Zhang},
  \citenamefont {Liu}, \citenamefont {Cui}, \citenamefont {Guo}, \citenamefont
  {Wang}, \citenamefont {Shi}, \citenamefont {Zhang}, \citenamefont {Wang},
  \citenamefont {Dong}, \citenamefont {Sun}, \citenamefont {Dun},\ and\
  \citenamefont {Cheng}}]{PhysRevMaterials.6.105001}%
  \BibitemOpen
  \bibfield  {author} {\bibinfo {author} {\bibfnamefont {X.}~\bibnamefont
  {Zhang}}, \bibinfo {author} {\bibfnamefont {Z.}~\bibnamefont {Liu}}, \bibinfo
  {author} {\bibfnamefont {Q.}~\bibnamefont {Cui}}, \bibinfo {author}
  {\bibfnamefont {Q.}~\bibnamefont {Guo}}, \bibinfo {author} {\bibfnamefont
  {N.}~\bibnamefont {Wang}}, \bibinfo {author} {\bibfnamefont {L.}~\bibnamefont
  {Shi}}, \bibinfo {author} {\bibfnamefont {H.}~\bibnamefont {Zhang}}, \bibinfo
  {author} {\bibfnamefont {W.}~\bibnamefont {Wang}}, \bibinfo {author}
  {\bibfnamefont {X.}~\bibnamefont {Dong}}, \bibinfo {author} {\bibfnamefont
  {J.}~\bibnamefont {Sun}}, \bibinfo {author} {\bibfnamefont {Z.}~\bibnamefont
  {Dun}},\ and\ \bibinfo {author} {\bibfnamefont {J.}~\bibnamefont {Cheng}},\
  }\bibfield  {title} {\bibinfo {title} {{Electronic and magnetic properties of
  intermetallic kagome magnets \textit{R}V$_6$Sn$_6$ (\textit{R}: Tb--Tm)}},\
  }\href {https://doi.org/10.1103/PhysRevMaterials.6.105001} {\bibfield
  {journal} {\bibinfo  {journal} {Phys. Rev. Mater.}\ }\textbf {\bibinfo
  {volume} {6}},\ \bibinfo {pages} {105001} (\bibinfo {year}
  {2022})}\BibitemShut {NoStop}%
\bibitem [{\citenamefont {Lee}\ and\ \citenamefont
  {Mun}(2022)}]{PhysRevMaterials.6.083401}%
  \BibitemOpen
  \bibfield  {author} {\bibinfo {author} {\bibfnamefont {J.}~\bibnamefont
  {Lee}}\ and\ \bibinfo {author} {\bibfnamefont {E.}~\bibnamefont {Mun}},\
  }\bibfield  {title} {\bibinfo {title} {Anisotropic magnetic property of
  single crystals \text{$R{\mathrm{V}}_{6}{\mathrm{Sn}}_{6}$ $(R=\mathrm{Y},
  \mathrm{Gd}\text{\ensuremath{-}}\mathrm{Tm}, \mathrm{Lu})$}},\ }\href
  {https://doi.org/10.1103/PhysRevMaterials.6.083401} {\bibfield  {journal}
  {\bibinfo  {journal} {Phys. Rev. Mater.}\ }\textbf {\bibinfo {volume} {6}},\
  \bibinfo {pages} {083401} (\bibinfo {year} {2022})}\BibitemShut {NoStop}%
\bibitem [{\citenamefont {Shao-ying}\ \emph {et~al.}(2001)\citenamefont
  {Shao-ying}, \citenamefont {Peng}, \citenamefont {Run-wei}, \citenamefont
  {Sun Ji-rong}, \citenamefont {Hong-wei},\ and\ \citenamefont
  {Bao-gen}}]{ZhangShao-ying_2001}%
  \BibitemOpen
  \bibfield  {author} {\bibinfo {author} {\bibfnamefont {Z.}~\bibnamefont
  {Shao-ying}}, \bibinfo {author} {\bibfnamefont {Z.}~\bibnamefont {Peng}},
  \bibinfo {author} {\bibfnamefont {L.}~\bibnamefont {Run-wei}}, \bibinfo
  {author} {\bibfnamefont {C.~Z.-h.}\ \bibnamefont {Sun Ji-rong}}, \bibinfo
  {author} {\bibfnamefont {Z.}~\bibnamefont {Hong-wei}},\ and\ \bibinfo
  {author} {\bibfnamefont {S.}~\bibnamefont {Bao-gen}},\ }\bibfield  {title}
  {\bibinfo {title} {Structure, magnetic properties and giant magnetoresistance
  of \text{YMn$_6$Sn$_{6-x}$Ga$_x$ (x = 0-0.6)} compounds},\ }\href
  {https://doi.org/10.1088/1009-1963/10/4/318} {\bibfield  {journal} {\bibinfo
  {journal} {Chin. Phys.}\ }\textbf {\bibinfo {volume} {10}},\ \bibinfo {pages}
  {345} (\bibinfo {year} {2001})}\BibitemShut {NoStop}%
\bibitem [{\citenamefont {Ma}\ \emph {et~al.}(2021)\citenamefont {Ma},
  \citenamefont {Xu}, \citenamefont {Yin}, \citenamefont {Yang}, \citenamefont
  {Zhou}, \citenamefont {Cheng}, \citenamefont {Huang}, \citenamefont {Qu},
  \citenamefont {Wang}, \citenamefont {Hasan},\ and\ \citenamefont
  {Jia}}]{PhysRevLett.126.246602}%
  \BibitemOpen
  \bibfield  {author} {\bibinfo {author} {\bibfnamefont {W.}~\bibnamefont
  {Ma}}, \bibinfo {author} {\bibfnamefont {X.}~\bibnamefont {Xu}}, \bibinfo
  {author} {\bibfnamefont {J.-X.}\ \bibnamefont {Yin}}, \bibinfo {author}
  {\bibfnamefont {H.}~\bibnamefont {Yang}}, \bibinfo {author} {\bibfnamefont
  {H.}~\bibnamefont {Zhou}}, \bibinfo {author} {\bibfnamefont {Z.-J.}\
  \bibnamefont {Cheng}}, \bibinfo {author} {\bibfnamefont {Y.}~\bibnamefont
  {Huang}}, \bibinfo {author} {\bibfnamefont {Z.}~\bibnamefont {Qu}}, \bibinfo
  {author} {\bibfnamefont {F.}~\bibnamefont {Wang}}, \bibinfo {author}
  {\bibfnamefont {M.~Z.}\ \bibnamefont {Hasan}},\ and\ \bibinfo {author}
  {\bibfnamefont {S.}~\bibnamefont {Jia}},\ }\bibfield  {title} {\bibinfo
  {title} {Rare earth engineering in
  \text{$R{\mathrm{Mn}}_{6}{\mathrm{Sn}}_{6}$
  ($R=\text{Gd}\text{\ensuremath{-}}\text{Tm}$, Lu)} topological kagome
  magnets},\ }\href {https://doi.org/10.1103/PhysRevLett.126.246602} {\bibfield
   {journal} {\bibinfo  {journal} {Phys. Rev. Lett.}\ }\textbf {\bibinfo
  {volume} {126}},\ \bibinfo {pages} {246602} (\bibinfo {year}
  {2021})}\BibitemShut {NoStop}%
\bibitem [{\citenamefont {Ortiz}\ \emph {et~al.}(2023)\citenamefont {Ortiz},
  \citenamefont {Pokharel}, \citenamefont {Gundayao}, \citenamefont {Li},
  \citenamefont {Kaboudvand}, \citenamefont {Kautzsch}, \citenamefont {Sarker},
  \citenamefont {Ruff}, \citenamefont {Hogan}, \citenamefont {Alvarado} \emph
  {et~al.}}]{ortiz2023ybv}%
  \BibitemOpen
  \bibfield  {author} {\bibinfo {author} {\bibfnamefont {B.~R.}\ \bibnamefont
  {Ortiz}}, \bibinfo {author} {\bibfnamefont {G.}~\bibnamefont {Pokharel}},
  \bibinfo {author} {\bibfnamefont {M.}~\bibnamefont {Gundayao}}, \bibinfo
  {author} {\bibfnamefont {H.}~\bibnamefont {Li}}, \bibinfo {author}
  {\bibfnamefont {F.}~\bibnamefont {Kaboudvand}}, \bibinfo {author}
  {\bibfnamefont {L.}~\bibnamefont {Kautzsch}}, \bibinfo {author}
  {\bibfnamefont {S.}~\bibnamefont {Sarker}}, \bibinfo {author} {\bibfnamefont
  {J.~P.}\ \bibnamefont {Ruff}}, \bibinfo {author} {\bibfnamefont
  {T.}~\bibnamefont {Hogan}}, \bibinfo {author} {\bibfnamefont {S.~J.~G.}\
  \bibnamefont {Alvarado}}, \emph {et~al.},\ }\bibfield  {title} {\bibinfo
  {title} {{YbV$_3$Sb$_4$ and EuV$_3$Sb$_4$ vanadium-based kagome metals with
  Yb$^{2+}$ and Eu$^{2+}$ zigzag chains}},\ }\href@noop {} {\bibfield
  {journal} {\bibinfo  {journal} {Phys. Rev. Mater.}\ }\textbf {\bibinfo
  {volume} {7}},\ \bibinfo {pages} {064201} (\bibinfo {year}
  {2023})}\BibitemShut {NoStop}%
\bibitem [{\citenamefont {Ovchinnikov}\ and\ \citenamefont
  {Bobev}(2018)}]{ovchinnikov2018synthesis}%
  \BibitemOpen
  \bibfield  {author} {\bibinfo {author} {\bibfnamefont {A.}~\bibnamefont
  {Ovchinnikov}}\ and\ \bibinfo {author} {\bibfnamefont {S.}~\bibnamefont
  {Bobev}},\ }\bibfield  {title} {\bibinfo {title} {{Synthesis, Crystal and
  Electronic Structure of the Titanium Bismuthides Sr$_5$Ti$_{12}$Bi$_{19+x}$,
  Ba$_5$Ti$_{12}$Bi$_{19+x}$, and
  Sr$_{5-\delta}$Eu$_\delta$Ti$_{12}$Bi$_{19+x}$ (x=0.5--1.0; $\delta$=2.4,
  4.0)}},\ }\href@noop {} {\bibfield  {journal} {\bibinfo  {journal} {Eur. J.
  Inorg. Chem.}\ }\textbf {\bibinfo {volume} {2018}},\ \bibinfo {pages} {1266}
  (\bibinfo {year} {2018})}\BibitemShut {NoStop}%
\bibitem [{\citenamefont {Ovchinnikov}\ and\ \citenamefont
  {Bobev}(2019)}]{ovchinnikov2019bismuth}%
  \BibitemOpen
  \bibfield  {author} {\bibinfo {author} {\bibfnamefont {A.}~\bibnamefont
  {Ovchinnikov}}\ and\ \bibinfo {author} {\bibfnamefont {S.}~\bibnamefont
  {Bobev}},\ }\bibfield  {title} {\bibinfo {title} {Bismuth as a reactive
  solvent in the synthesis of multicomponent transition-metal-bearing
  bismuthides},\ }\href@noop {} {\bibfield  {journal} {\bibinfo  {journal}
  {Inorg. Chem.}\ }\textbf {\bibinfo {volume} {59}},\ \bibinfo {pages} {3459}
  (\bibinfo {year} {2019})}\BibitemShut {NoStop}%
\bibitem [{\citenamefont {Motoyama}\ \emph {et~al.}(2018)\citenamefont
  {Motoyama}, \citenamefont {Sezaki}, \citenamefont {Gouchi}, \citenamefont
  {Miyoshi}, \citenamefont {Nishigori}, \citenamefont {Mutou}, \citenamefont
  {Fujiwara},\ and\ \citenamefont {Uwatoko}}]{motoyama2018magnetic}%
  \BibitemOpen
  \bibfield  {author} {\bibinfo {author} {\bibfnamefont {G.}~\bibnamefont
  {Motoyama}}, \bibinfo {author} {\bibfnamefont {M.}~\bibnamefont {Sezaki}},
  \bibinfo {author} {\bibfnamefont {J.}~\bibnamefont {Gouchi}}, \bibinfo
  {author} {\bibfnamefont {K.}~\bibnamefont {Miyoshi}}, \bibinfo {author}
  {\bibfnamefont {S.}~\bibnamefont {Nishigori}}, \bibinfo {author}
  {\bibfnamefont {T.}~\bibnamefont {Mutou}}, \bibinfo {author} {\bibfnamefont
  {K.}~\bibnamefont {Fujiwara}},\ and\ \bibinfo {author} {\bibfnamefont
  {Y.}~\bibnamefont {Uwatoko}},\ }\bibfield  {title} {\bibinfo {title}
  {{Magnetic properties of new antiferromagnetic heavy-fermion compounds,
  Ce$_3$TiBi$_5$ and CeTi$_3$Bi$_4$}},\ }\href@noop {} {\bibfield  {journal}
  {\bibinfo  {journal} {Physica B Condens.}\ }\textbf {\bibinfo {volume}
  {536}},\ \bibinfo {pages} {142} (\bibinfo {year} {2018})}\BibitemShut
  {NoStop}%
\bibitem [{\citenamefont {Canfield}\ \emph {et~al.}(2016)\citenamefont
  {Canfield}, \citenamefont {Kong}, \citenamefont {Kaluarachchi},\ and\
  \citenamefont {Jo}}]{canfield2016use}%
  \BibitemOpen
  \bibfield  {author} {\bibinfo {author} {\bibfnamefont {P.~C.}\ \bibnamefont
  {Canfield}}, \bibinfo {author} {\bibfnamefont {T.}~\bibnamefont {Kong}},
  \bibinfo {author} {\bibfnamefont {U.~S.}\ \bibnamefont {Kaluarachchi}},\ and\
  \bibinfo {author} {\bibfnamefont {N.~H.}\ \bibnamefont {Jo}},\ }\bibfield
  {title} {\bibinfo {title} {Use of frit-disc crucibles for routine and
  exploratory solution growth of single crystalline samples},\ }\href@noop {}
  {\bibfield  {journal} {\bibinfo  {journal} {Philosophical magazine}\ }\textbf
  {\bibinfo {volume} {96}},\ \bibinfo {pages} {84} (\bibinfo {year}
  {2016})}\BibitemShut {NoStop}%
\bibitem [{ESI()}]{ESI}%
  \BibitemOpen
  \href@noop {} {}\bibinfo {note} {See Supplemental Information for further
  details}\BibitemShut {NoStop}%
\bibitem [{\citenamefont {Blaha}\ \emph {et~al.}(2001)\citenamefont {Blaha},
  \citenamefont {Schwarz}, \citenamefont {Madsen}, \citenamefont {Kvasnicka},\
  and\ \citenamefont {Luitz}}]{blaha2001}%
  \BibitemOpen
  \bibfield  {author} {\bibinfo {author} {\bibfnamefont {P.}~\bibnamefont
  {Blaha}}, \bibinfo {author} {\bibfnamefont {K.}~\bibnamefont {Schwarz}},
  \bibinfo {author} {\bibfnamefont {G.~K.}\ \bibnamefont {Madsen}}, \bibinfo
  {author} {\bibfnamefont {D.}~\bibnamefont {Kvasnicka}},\ and\ \bibinfo
  {author} {\bibfnamefont {J.}~\bibnamefont {Luitz}},\ }\bibfield  {title}
  {\bibinfo {title} {Wien2k, an augmented plane wave+ local orbitals program
  for calculating crystal properties, edited by k},\ }\href@noop {} {\bibfield
  {journal} {\bibinfo  {journal} {Schwarz, Vienna University of Technology,
  Austria}\ } (\bibinfo {year} {2001})}\BibitemShut {NoStop}%
\bibitem [{\citenamefont {Perdew}\ \emph {et~al.}(1996)\citenamefont {Perdew},
  \citenamefont {Burke},\ and\ \citenamefont {Ernzerhof}}]{perdew1996}%
  \BibitemOpen
  \bibfield  {author} {\bibinfo {author} {\bibfnamefont {J.~P.}\ \bibnamefont
  {Perdew}}, \bibinfo {author} {\bibfnamefont {K.}~\bibnamefont {Burke}},\ and\
  \bibinfo {author} {\bibfnamefont {M.}~\bibnamefont {Ernzerhof}},\ }\bibfield
  {title} {\bibinfo {title} {Generalized gradient approximation made simple},\
  }\href@noop {} {\bibfield  {journal} {\bibinfo  {journal} {Phys. Rev. Lett.}\
  }\textbf {\bibinfo {volume} {77}},\ \bibinfo {pages} {3865} (\bibinfo {year}
  {1996})}\BibitemShut {NoStop}%
\bibitem [{\citenamefont {Yasui}\ \emph {et~al.}(2020)\citenamefont {Yasui},
  \citenamefont {Butler}, \citenamefont {Khanh}, \citenamefont {Hayami},
  \citenamefont {Nomoto}, \citenamefont {Hanaguri}, \citenamefont {Motome},
  \citenamefont {Arita}, \citenamefont {Arima}, \citenamefont {Tokura} \emph
  {et~al.}}]{yasui2020imaging}%
  \BibitemOpen
  \bibfield  {author} {\bibinfo {author} {\bibfnamefont {Y.}~\bibnamefont
  {Yasui}}, \bibinfo {author} {\bibfnamefont {C.~J.}\ \bibnamefont {Butler}},
  \bibinfo {author} {\bibfnamefont {N.~D.}\ \bibnamefont {Khanh}}, \bibinfo
  {author} {\bibfnamefont {S.}~\bibnamefont {Hayami}}, \bibinfo {author}
  {\bibfnamefont {T.}~\bibnamefont {Nomoto}}, \bibinfo {author} {\bibfnamefont
  {T.}~\bibnamefont {Hanaguri}}, \bibinfo {author} {\bibfnamefont
  {Y.}~\bibnamefont {Motome}}, \bibinfo {author} {\bibfnamefont
  {R.}~\bibnamefont {Arita}}, \bibinfo {author} {\bibfnamefont {T.-h.}\
  \bibnamefont {Arima}}, \bibinfo {author} {\bibfnamefont {Y.}~\bibnamefont
  {Tokura}}, \emph {et~al.},\ }\bibfield  {title} {\bibinfo {title} {{Imaging
  the coupling between itinerant electrons and localised moments in the
  centrosymmetric skyrmion magnet GdRu$_2$Si$_2$}},\ }\href@noop {} {\bibfield
  {journal} {\bibinfo  {journal} {Nature communications}\ }\textbf {\bibinfo
  {volume} {11}},\ \bibinfo {pages} {5925} (\bibinfo {year}
  {2020})}\BibitemShut {NoStop}%
\bibitem [{\citenamefont {Kurumaji}\ \emph {et~al.}(2019)\citenamefont
  {Kurumaji}, \citenamefont {Nakajima}, \citenamefont {Hirschberger},
  \citenamefont {Kikkawa}, \citenamefont {Yamasaki}, \citenamefont {Sagayama},
  \citenamefont {Nakao}, \citenamefont {Taguchi}, \citenamefont {Arima},\ and\
  \citenamefont {Tokura}}]{kurumaji2019skyrmion}%
  \BibitemOpen
  \bibfield  {author} {\bibinfo {author} {\bibfnamefont {T.}~\bibnamefont
  {Kurumaji}}, \bibinfo {author} {\bibfnamefont {T.}~\bibnamefont {Nakajima}},
  \bibinfo {author} {\bibfnamefont {M.}~\bibnamefont {Hirschberger}}, \bibinfo
  {author} {\bibfnamefont {A.}~\bibnamefont {Kikkawa}}, \bibinfo {author}
  {\bibfnamefont {Y.}~\bibnamefont {Yamasaki}}, \bibinfo {author}
  {\bibfnamefont {H.}~\bibnamefont {Sagayama}}, \bibinfo {author}
  {\bibfnamefont {H.}~\bibnamefont {Nakao}}, \bibinfo {author} {\bibfnamefont
  {Y.}~\bibnamefont {Taguchi}}, \bibinfo {author} {\bibfnamefont {T.-h.}\
  \bibnamefont {Arima}},\ and\ \bibinfo {author} {\bibfnamefont
  {Y.}~\bibnamefont {Tokura}},\ }\bibfield  {title} {\bibinfo {title} {Skyrmion
  lattice with a giant topological hall effect in a frustrated
  triangular-lattice magnet},\ }\href@noop {} {\bibfield  {journal} {\bibinfo
  {journal} {Science}\ }\textbf {\bibinfo {volume} {365}},\ \bibinfo {pages}
  {914} (\bibinfo {year} {2019})}\BibitemShut {NoStop}%
\bibitem [{\citenamefont {Hirschberger}\ \emph {et~al.}(2019)\citenamefont
  {Hirschberger}, \citenamefont {Nakajima}, \citenamefont {Gao}, \citenamefont
  {Peng}, \citenamefont {Kikkawa}, \citenamefont {Kurumaji}, \citenamefont
  {Kriener}, \citenamefont {Yamasaki}, \citenamefont {Sagayama}, \citenamefont
  {Nakao} \emph {et~al.}}]{hirschberger2019skyrmion}%
  \BibitemOpen
  \bibfield  {author} {\bibinfo {author} {\bibfnamefont {M.}~\bibnamefont
  {Hirschberger}}, \bibinfo {author} {\bibfnamefont {T.}~\bibnamefont
  {Nakajima}}, \bibinfo {author} {\bibfnamefont {S.}~\bibnamefont {Gao}},
  \bibinfo {author} {\bibfnamefont {L.}~\bibnamefont {Peng}}, \bibinfo {author}
  {\bibfnamefont {A.}~\bibnamefont {Kikkawa}}, \bibinfo {author} {\bibfnamefont
  {T.}~\bibnamefont {Kurumaji}}, \bibinfo {author} {\bibfnamefont
  {M.}~\bibnamefont {Kriener}}, \bibinfo {author} {\bibfnamefont
  {Y.}~\bibnamefont {Yamasaki}}, \bibinfo {author} {\bibfnamefont
  {H.}~\bibnamefont {Sagayama}}, \bibinfo {author} {\bibfnamefont
  {H.}~\bibnamefont {Nakao}}, \emph {et~al.},\ }\bibfield  {title} {\bibinfo
  {title} {Skyrmion phase and competing magnetic orders on a breathing
  kagom{\'e} lattice},\ }\href@noop {} {\bibfield  {journal} {\bibinfo
  {journal} {Nature communications}\ }\textbf {\bibinfo {volume} {10}},\
  \bibinfo {pages} {5831} (\bibinfo {year} {2019})}\BibitemShut {NoStop}%
\bibitem [{\citenamefont {White}\ and\ \citenamefont
  {Van~Vleck}(1961)}]{white1961sign}%
  \BibitemOpen
  \bibfield  {author} {\bibinfo {author} {\bibfnamefont {J.}~\bibnamefont
  {White}}\ and\ \bibinfo {author} {\bibfnamefont {J.}~\bibnamefont
  {Van~Vleck}},\ }\bibfield  {title} {\bibinfo {title} {Sign of knight shift in
  samarium intermetallic compounds},\ }\href@noop {} {\bibfield  {journal}
  {\bibinfo  {journal} {Physical Review Letters}\ }\textbf {\bibinfo {volume}
  {6}},\ \bibinfo {pages} {412} (\bibinfo {year} {1961})}\BibitemShut {NoStop}%
\bibitem [{\citenamefont {Malik}\ and\ \citenamefont
  {Vijayaraghavan}(1974)}]{malik1974crystal}%
  \BibitemOpen
  \bibfield  {author} {\bibinfo {author} {\bibfnamefont {S.}~\bibnamefont
  {Malik}}\ and\ \bibinfo {author} {\bibfnamefont {R.}~\bibnamefont
  {Vijayaraghavan}},\ }\bibfield  {title} {\bibinfo {title} {{Crystal field
  effects on the saturation magnetic moment of Sm$^{3+}$ ion in ferronagnetic
  samarium compounds}},\ }\href@noop {} {\bibfield  {journal} {\bibinfo
  {journal} {Pramana}\ }\textbf {\bibinfo {volume} {3}},\ \bibinfo {pages}
  {122} (\bibinfo {year} {1974})}\BibitemShut {NoStop}%
\bibitem [{\citenamefont {Kim}\ \emph {et~al.}(2023)\citenamefont {Kim},
  \citenamefont {Kim}, \citenamefont {Hong}, \citenamefont {Shin},
  \citenamefont {Jeong}, \citenamefont {Kim}, \citenamefont {Moon},
  \citenamefont {Lee},\ and\ \citenamefont {Choi}}]{kim2023evolution}%
  \BibitemOpen
  \bibfield  {author} {\bibinfo {author} {\bibfnamefont {J.~H.}\ \bibnamefont
  {Kim}}, \bibinfo {author} {\bibfnamefont {M.~K.}\ \bibnamefont {Kim}},
  \bibinfo {author} {\bibfnamefont {J.~M.}\ \bibnamefont {Hong}}, \bibinfo
  {author} {\bibfnamefont {H.~J.}\ \bibnamefont {Shin}}, \bibinfo {author}
  {\bibfnamefont {K.~W.}\ \bibnamefont {Jeong}}, \bibinfo {author}
  {\bibfnamefont {J.~S.}\ \bibnamefont {Kim}}, \bibinfo {author} {\bibfnamefont
  {K.}~\bibnamefont {Moon}}, \bibinfo {author} {\bibfnamefont {N.}~\bibnamefont
  {Lee}},\ and\ \bibinfo {author} {\bibfnamefont {Y.~J.}\ \bibnamefont
  {Choi}},\ }\bibfield  {title} {\bibinfo {title} {{Evolution of anisotropic
  magnetic properties through helix-to-fan transition in helical
  antiferromagnetic EuCo$_2$As$_2$}},\ }\href@noop {} {\bibfield  {journal}
  {\bibinfo  {journal} {Commun. Phys.}\ }\textbf {\bibinfo {volume} {6}},\
  \bibinfo {pages} {20} (\bibinfo {year} {2023})}\BibitemShut {NoStop}%
\bibitem [{\citenamefont {Ding}\ \emph {et~al.}(2017)\citenamefont {Ding},
  \citenamefont {Higa}, \citenamefont {Sangeetha}, \citenamefont {Johnston},\
  and\ \citenamefont {Furukawa}}]{ding2017nmr}%
  \BibitemOpen
  \bibfield  {author} {\bibinfo {author} {\bibfnamefont {Q.-P.}\ \bibnamefont
  {Ding}}, \bibinfo {author} {\bibfnamefont {N.}~\bibnamefont {Higa}}, \bibinfo
  {author} {\bibfnamefont {N.}~\bibnamefont {Sangeetha}}, \bibinfo {author}
  {\bibfnamefont {D.}~\bibnamefont {Johnston}},\ and\ \bibinfo {author}
  {\bibfnamefont {Y.}~\bibnamefont {Furukawa}},\ }\bibfield  {title} {\bibinfo
  {title} {{NMR determination of an incommensurate helical antiferromagnetic
  structure in EuCo$_2$As$_2$}},\ }\href@noop {} {\bibfield  {journal}
  {\bibinfo  {journal} {Phys. Rev. B}\ }\textbf {\bibinfo {volume} {95}},\
  \bibinfo {pages} {184404} (\bibinfo {year} {2017})}\BibitemShut {NoStop}%
\bibitem [{\citenamefont {Sangeetha}\ \emph {et~al.}(2016)\citenamefont
  {Sangeetha}, \citenamefont {Cuervo-Reyes}, \citenamefont {Pandey},\ and\
  \citenamefont {Johnston}}]{sangeetha2016euco}%
  \BibitemOpen
  \bibfield  {author} {\bibinfo {author} {\bibfnamefont {N.}~\bibnamefont
  {Sangeetha}}, \bibinfo {author} {\bibfnamefont {E.}~\bibnamefont
  {Cuervo-Reyes}}, \bibinfo {author} {\bibfnamefont {A.}~\bibnamefont
  {Pandey}},\ and\ \bibinfo {author} {\bibfnamefont {D.}~\bibnamefont
  {Johnston}},\ }\bibfield  {title} {\bibinfo {title} {{EuCo$_2$P$_2$: A model
  molecular-field helical Heisenberg antiferromagnet}},\ }\href@noop {}
  {\bibfield  {journal} {\bibinfo  {journal} {Phys. Rev. B}\ }\textbf {\bibinfo
  {volume} {94}},\ \bibinfo {pages} {014422} (\bibinfo {year}
  {2016})}\BibitemShut {NoStop}%
\bibitem [{\citenamefont {Chen}\ \emph
  {et~al.}(2023{\natexlab{a}})\citenamefont {Chen}, \citenamefont {Zhou},
  \citenamefont {Zhang}, \citenamefont {Ji}, \citenamefont {Liao},
  \citenamefont {Ji}, \citenamefont {Li}, \citenamefont {Guo}, \citenamefont
  {Shen}, \citenamefont {Yu} \emph {et~al.}}]{chen2023134}%
  \BibitemOpen
  \bibfield  {author} {\bibinfo {author} {\bibfnamefont {L.}~\bibnamefont
  {Chen}}, \bibinfo {author} {\bibfnamefont {Y.}~\bibnamefont {Zhou}}, \bibinfo
  {author} {\bibfnamefont {H.}~\bibnamefont {Zhang}}, \bibinfo {author}
  {\bibfnamefont {X.}~\bibnamefont {Ji}}, \bibinfo {author} {\bibfnamefont
  {K.}~\bibnamefont {Liao}}, \bibinfo {author} {\bibfnamefont {Y.}~\bibnamefont
  {Ji}}, \bibinfo {author} {\bibfnamefont {Y.}~\bibnamefont {Li}}, \bibinfo
  {author} {\bibfnamefont {Z.}~\bibnamefont {Guo}}, \bibinfo {author}
  {\bibfnamefont {X.}~\bibnamefont {Shen}}, \bibinfo {author} {\bibfnamefont
  {R.}~\bibnamefont {Yu}}, \emph {et~al.},\ }\bibfield  {title} {\bibinfo
  {title} {{Tunable magnetism and electron correlation in Titanium-based Kagome
  metals RETi$_3$Bi$_4$ (RE= Yb, Pr, and Nd) by rare-earth engineering}},\
  }\href@noop {} {\bibfield  {journal} {\bibinfo  {journal} {arXiv preprint
  arXiv:2307.02942}\ } (\bibinfo {year} {2023}{\natexlab{a}})}\BibitemShut
  {NoStop}%
\bibitem [{\citenamefont {Guo}\ \emph {et~al.}(2023)\citenamefont {Guo},
  \citenamefont {Zhou}, \citenamefont {Ding}, \citenamefont {Qu}, \citenamefont
  {Liu}, \citenamefont {Du}, \citenamefont {Zhang}, \citenamefont {Li},
  \citenamefont {Zhang}, \citenamefont {Zhou}, \citenamefont {Qi},
  \citenamefont {Guo}, \citenamefont {Wang}, \citenamefont {Fei}, \citenamefont
  {Huang}, \citenamefont {Qian}, \citenamefont {Shen}, \citenamefont {Weng},\
  and\ \citenamefont {Song}}]{guo2023134}%
  \BibitemOpen
  \bibfield  {author} {\bibinfo {author} {\bibfnamefont {J.}~\bibnamefont
  {Guo}}, \bibinfo {author} {\bibfnamefont {L.}~\bibnamefont {Zhou}}, \bibinfo
  {author} {\bibfnamefont {J.}~\bibnamefont {Ding}}, \bibinfo {author}
  {\bibfnamefont {G.}~\bibnamefont {Qu}}, \bibinfo {author} {\bibfnamefont
  {Z.}~\bibnamefont {Liu}}, \bibinfo {author} {\bibfnamefont {Y.}~\bibnamefont
  {Du}}, \bibinfo {author} {\bibfnamefont {H.}~\bibnamefont {Zhang}}, \bibinfo
  {author} {\bibfnamefont {J.}~\bibnamefont {Li}}, \bibinfo {author}
  {\bibfnamefont {Y.}~\bibnamefont {Zhang}}, \bibinfo {author} {\bibfnamefont
  {F.}~\bibnamefont {Zhou}}, \bibinfo {author} {\bibfnamefont {W.}~\bibnamefont
  {Qi}}, \bibinfo {author} {\bibfnamefont {F.}~\bibnamefont {Guo}}, \bibinfo
  {author} {\bibfnamefont {T.}~\bibnamefont {Wang}}, \bibinfo {author}
  {\bibfnamefont {F.}~\bibnamefont {Fei}}, \bibinfo {author} {\bibfnamefont
  {Y.}~\bibnamefont {Huang}}, \bibinfo {author} {\bibfnamefont
  {T.}~\bibnamefont {Qian}}, \bibinfo {author} {\bibfnamefont {D.}~\bibnamefont
  {Shen}}, \bibinfo {author} {\bibfnamefont {H.}~\bibnamefont {Weng}},\ and\
  \bibinfo {author} {\bibfnamefont {F.}~\bibnamefont {Song}},\ }\bibfield
  {title} {\bibinfo {title} {Magnetic kagome materials reti3bi4 family with
  weak interlayer interactions},\ }\href@noop {} {\bibfield  {journal}
  {\bibinfo  {journal} {arXiv preprint arXiv:2308.14509}\ } (\bibinfo {year}
  {2023})}\BibitemShut {NoStop}%
\bibitem [{\citenamefont {Chen}\ \emph
  {et~al.}(2023{\natexlab{b}})\citenamefont {Chen}, \citenamefont {Zhou},
  \citenamefont {Zhang}, \citenamefont {Guo}, \citenamefont {Yu},\ and\
  \citenamefont {Wang}}]{chen2023sm134}%
  \BibitemOpen
  \bibfield  {author} {\bibinfo {author} {\bibfnamefont {L.}~\bibnamefont
  {Chen}}, \bibinfo {author} {\bibfnamefont {Y.}~\bibnamefont {Zhou}}, \bibinfo
  {author} {\bibfnamefont {H.}~\bibnamefont {Zhang}}, \bibinfo {author}
  {\bibfnamefont {Z.}~\bibnamefont {Guo}}, \bibinfo {author} {\bibfnamefont
  {X.}~\bibnamefont {Yu}},\ and\ \bibinfo {author} {\bibfnamefont
  {G.}~\bibnamefont {Wang}},\ }\bibfield  {title} {\bibinfo {title}
  {{Anisotropic and pressure tunable magnetism of titanium-based Kagome
  ferromagnet SmTi$_3$Bi$_4$}},\ }\href@noop {} {\bibfield  {journal} {\bibinfo
   {journal} {arXiv preprint arXiv:2308.14349}\ } (\bibinfo {year}
  {2023}{\natexlab{b}})}\BibitemShut {NoStop}%
\end{thebibliography}%

\end{document}